\documentclass[12pt]{article}
\usepackage{latexsym,graphicx,color,amsmath}
\usepackage{amssymb,mathrsfs}
\DeclareFontFamily{OT1}{rsfs10}{}
\DeclareFontShape{OT1}{rsfs10}{m}{n}{ <-> rsfs10 }{}
\DeclareMathAlphabet{\mathscript}{OT1}{rsfs10}{m}{n}

 \parskip=\medskipamount                 
 \thicklines                         

\def\spin{\mathfrak{spin}}
\def\e{\epsilon}
\def\ve{\varepsilon}

\def\bpl{\Big(}
\def\bpr{\Big)}
\def\ve{\varepsilon}
\def\der{\partial}
\def\brr{\begin{eqnarray}}
\def\err{\end{eqnarray}}
\def\ba{\left(\begin{array}}
\def\ea{\end{array}\right)}
\def\lf{\left.\begin{array}{c}}
\def\rf{\end{array}\right.}
\def\red{\color{red}}
\def\blue{\color{blue}}

\def\dslash{\hbox{\ooalign{$\displaystyle\partial$\cr$/$}}}

\def\Dslash{\hbox{\ooalign{$\displaystyle D$\cr$\hspace{.03in}/$}}}

\def\rlx{\relax\leavevmode}
\def\Ione{\rlx\hbox{\rm 1\kern-3pt l}}
\newcommand{\fr}[2]{{\textstyle\frac{#1}{#2}}}

\renewcommand{\theequation}{\arabic{section}.\arabic{equation}}

\begin{document}

 {\hbox to\hsize{\today \hfill {SUNY-O/1601}}}

 \vspace{1.1in}

 \thispagestyle{empty}

 \begin{center}
 {\Large\bf The Conformal Hyperplet}
 \\[.25in]
 {Michael Faux}\\[5mm]
 {\small\it
 Department of Physics,
      State University of New York, Oneonta, NY 13820\\[-1mm]
  {\tt  fauxmg@oneonta.edu}
 }\\[.6in]
 {\bf ABSTRACT}\\[.3in]
 \parbox{5in}{We introduce a finite off-shell hypermultiplet with no
 off-shell central charge.  This requires 192+192 degrees of freedom, all but 8+8 of
 which are auxiliary or gauge.
 In the absence of supergravity, the model has a saddle-point vacuum instability implying
 ghost-like propagators.  These are cured by realizing the model superconformally, such
 that the erstwhile ghosts are realized as compensators.
 Gauge fixing these links the physical hypermultiplets to supergravity.
 This evokes the prospect of realizing ${\cal N}=4$ Super Yang-Mills
 theory off-shell.}
 \end{center}

 \pagebreak

\setcounter{equation}{0}
\section{Introduction}
In the standard presentation of four-dimensional ${\cal N}=4$ Super Yang-Mills theory \cite{SYM}, the supersymmetry algebra closes on the
fermion component fields only when Euler-Lagrange field equations are satisfied.  In this sense, the theory exhibits
merely {\it on-shell} supersymmetry.  For a variety of reasons, it would be desirable
to have an {\it off-shell} description, in which the algebra closes on all component fields
without recourse to dynamics.  This would require a judicious complement of
non-propagating fields involving judicious transformation rules.  Such off-shell completions
are well known for interesting ${\cal N}=1$ and ${\cal N}=2$ theories.  But the case ${\cal N}=4$ exhibits an
ostensible obstruction, complicating prospects for an analogous development.
In fact, a simple counting argument arguably suggests that the search for an
auxiliary field structure for ${\cal N}=4$ SYM is quixotic.\footnote{One version of this argument was presented a long time ago, in \cite{Rocek}.}
In this paper we discuss this problem in detail, we identify the ${\cal N}=2$ antecedent to this problem,
and we introduce a field theoretic resolution in that latter context.

For ${\cal N}=2$ supersymmetry the problem lies with hypermultiplets, and derives from the
feature that the known off-shell descriptions of these have awkward traits, such as off-shell central charges
and/or an infinitude of auxiliary fields.  These issues preclude the amalgamation of these
with vector multiplets to form a desired off-shell ${\cal N}=4$ multiplet.  The key result of this paper is an existence proof
of a hypermultiplet with finite degrees of freedom and a vanishing central charge.  Preliminary work on this construction
appeared in \cite{RETM}, which introduced, as an epilog, a curious model describing exactly two
hypermultiplets, no off-shell central charge, and a finite number of auxiliary fields, but with the seeming flaw of
a saddle-point vacuum instability and ghostlike propagators.  This paper reconsiders that model and explains
how the apparent ghosts can be inoculated by a supergravity background.  This suggests that off-shell ${\cal N}=4$ Super Yang-Mills
theory has an inextricable link to supergravity.

Generalized hypermultiplets have been previously addressed in the context of projective superspace, as reviewed in
\cite{Kuzenko}.  Concerning these, there is likely a close relationship between the multiplets discussed in this paper and the so-called
arctic, antarctic, and tropical supermultiplets introduced in \cite{R2}.  In that reference, it is acknowledged 
that those models exhibit ghostlike propagators, although the problematic nature of that fact is not emphasized. 
The component approach in this paper provides a
complementary perspective, and notably also provides a novel resolution the ghost problem. 

Poincar{\'e} supergravity models are elegantly constructed as gauge-fixed superconformal models \cite{KTN,dw1,dw2}.
In these constructions, certain multiplets are introduced as conformal compensators.  This means that they contain
fields which are pure gauge under the extra generators of the larger superconformal symmetry.
These fields can be fixed according to desired criteria, breaking the symmetry.  This is practical because
the superconformal Weyl multiplets, which contain the graviton and the gravitino fields, enable relatively economical
control over the coupling between supergravity with matter and gauge fields.  Supergravity theories with different
auxiliary field content are obtained by using different conformal compensator multiplets.  For example, the old-miminal
\cite{FV1,SW1} and new-minimal \cite{SW2} versions of ${\cal N}=1$ supergravity are distinguished because the first of these uses a Chiral multiplet
as a compensator whereas the latter uses a Linear multiplet.

Minimal ${\cal N}=2$ supergravity \cite{structure,Cremmer} may be constructed using a single vector multiplet and
a single hypermultiplet as compensators.  The compensating hypermultiplet enters the action with ``wrong sign"
kinetic terms, differentiating it from physical hypermultiplets that can also couple
to the supergravity background.  This circumstance is dictated by the requirement that the Einstein kinetic term,
proportional to the scalar curvature, have
the correct sign, enabling that the energy be bounded from below.  The ghost-like sign on the compensating
hypermultiplet is not problematic because, owing to the superconformal symmetry,
all of its component fields are pure gauge or auxiliary. Analogously, the ghost-like sector of the
model introduced in \cite{RETM}, which has a pairing of a ``healthy" hypermultiplet and a ghost-like hypermultiplet,
is rendered non-problematic by generalizing that construction into
a superconformal model, such that its ghost-like sector can be gauge-fixed, leaving a single physical
hypermultiplet coupled to Poincar{\'e} supergravity.

In Section 2 we explain why an off-shell context for ${\cal N}=4$ Super Yang-Mills theory
has proved elusive.  We do this using a simple counting argument based on a few stated premises.  In Section 3 we
restrict the on-shell ${\cal N}=4$ theory to its ${\cal N}=2$ constituents,
review known off-shell completions of these, and explain why these cannot re-amalgamate into
an ${\cal N}=4$ model.
In Section 4 we methodically develop a pair of ${\cal N}=2$ multiplets with a total of 192+192 off-shell degrees of
freedom having the on-shell field content of exactly two hypermultiplets, but with no off-shell central charge.
This is done using a variant of the relaxation paradigm introduced in \cite{HST}.  The pair of multiplets
we introduce are ``dynamically linked" in the sense that an action exists only if both of these
are employed.  There is an unavoidable ghost-like sector. We refer to this model as the Hyperplet.

In Section 5 we explain how the Hyperplet presents a seeming obstacle to superconformal generalization,
and we explain how that obstacle can be overcome if there is an extra (hidden) $SU(2)$ isometry.
We then show that the desired isometry does exist, and we re-cast the model in a way that renders
all of its isometries manifest.  Using these results, we develop the conformal version of the Hyperplet
in Section 6.  Section 7 comments on these results, and explains how the existence of the conformal
Hyperplet resolves the hypermultiplet off-shell problem at the level of multiplet structures.

We include eight Appendices, which are are an important part of this paper. They have detailed explanations of
subtle points, notational issues, and pertinent side calculations.  Notably, Appendix A provides a very brief
reference concerning the structure of ${\cal N}=4$ Super Yang-Mills theory.  Appendix D presents
an infinite class of ${\cal N}=2$ supermultiplets involving useful $SU(2)_R\times SU(2)_H$ tensor
structures employed systematically in the main text. Spinor indices are suppressed throughout this paper.

\setcounter{equation}{0}
\section{Off-shell counting}
\label{cntg}
In this section we provide a state counting argument, based on a few preliminary assumptions,
which seemingly negates the possibility for an off-shell description of ${\cal N}=4$ SYM theory.
The balance of the paper supplies evidence for a loophole, enabled by
new couplings to a supergravity background.

The on-shell ${\cal N}=4$ SYM theory includes gauge potentials $A_a^I$,
massless chiral gaugino fields $\chi_i^I=-\gamma_5\,\chi_i^I$, and
massless complex scalars $\phi_{ij}^I=\phi_{[ij]}^I$, subject to
the constraint $\phi_{ij}^I=\fr12\,\ve_{ijkl}\,\phi^{I\,kl}$,
where $\ve_{ijkl}$ is the four-dimensional Levi-Civita tensor.
Mid-alphabet Latin indices $i$ and $j$ are fundamental
${\bf 4}$ indices under $SU(4)_R$, and are raised or lowered by complex
conjugation. The index $I$ is an adjoint index under the
gauge group ${\cal G}$. On-shell, for each value of $I$ the
boson fields $\phi_{ij}^I$ and $A_a^I$ describe respectively six and two degrees of freedom,
while the fermion fields $\chi_i^I$ describe eight.  Thus, for each value of $I$ there
are 8+8 propagating degrees of freedom, exhibiting the balance emblematic
of supersymmetry.  The transformation rules and action are exhibited in \ref{n4elements}.

Off-shell, for each value of $I$ the boson fields $\phi_{ij}^I$ and $A_a^I$ involve
respectively six and three degrees of freedom, while the fermion fields $\chi_i^I$  involve
sixteen.  Accordingly, an off-shell completion requires seven more extra bosonic degrees of freedom than extra fermionic degrees of freedom
per unit dimension of ${\cal G}$.

We assume, preliminarily, that all extra fields needed to close the supersymmetry algebra off-shell
have algebraic rather than differential field equations, rendering them auxiliary using common parlance.  Such fields do not describe
independent propagating
degrees of freedom, and do not correspond to particles.  We assume that there is a supersymmetric action
with derivative-free quadratic terms involving each auxiliary field.
We assume that each bosonic auxiliary field appears linearly in a term in the supersymmetry transformation rule for a fermionic field.
We assume there is no mass
parameter in the theory.  Finally, we assume that the only fields which are not auxiliary are the propagating fields described by the state counting
enumerated above.  We proceed to show that this set of constraints are not mutually tenable. Later in this paper we explain how this
circumstance is ameliorated by allowing extra fields which are not auxiliary in the sense defined above, but which my be removed
by gauge transformations.

In four spacetime dimensions, a generic propagating scalar has
dimension one and a generic propagating spinor has dimension three-halves.
\footnote{We work in units where $\hbar=c=1$, so that distances (and times) have units of ${\rm (mass)}^{-1}$,
identifying $[x^a]=-1$, whereupon $[\der_a]=1$ and $[\,\Box]=2$.
Any term in the Lagrangian density ${\cal L}$ must have dimension 4, so that the
action $S=\int d^4x\,{\cal L}$ is properly dimensionless.  A generic scalar $\Phi$ has kinetic term
proportional to $\Phi\,\Box\Phi$ which has dimension 4 only if $[\Phi]=1$.  A generic
Weyl spinor $\Psi$ has kinetic term proportional to $\bar{\Psi}\,\dslash\Psi$, and this has
dimension 4 only if $[\Psi]=3/2$.}  In general, bosonic fields have integer dimension and
fermionic fields have half-integer dimension. Thus, it is possible to have dimension two
bosonic auxiliary fields, such as the field usually called $F$ in the ${\cal  N}=1$ Chiral multiplet or the field
usually called $D$ in the ${\cal N}=1$ Vector multiplet, which can
appear in dimension four terms $F^2$ or $D^2$.   Any other auxiliary fields, including any fermionic
auxiliary fields, must come in pairs such that the sum of their dimensions is four.
Otherwise, we could not form any dimension four quadratic terms involving these fields without
derivatives.

Since the physical bosons in ${\cal N}=4$ SYM carry an even number of fundamental $SU(4)_R$ indices,
and since the supercharge carries one such index, it follows that any fermion fields in any multiplet
containing the physical fields must carry an odd number of fundamental $SU(4)_R$ indices.  Any other
fully-auxiliary multiplet that couples quadratically to the physical fields must have a similar structure
in order that the action be $SU(4)_R$ invariant.  Accordingly,
any auxiliary fermion in an off-shell version of ${\cal N}=4$ SYM conforming to our preliminary criteria must transform as a rank $k$ tensor under
$SU(4)_R$, where $k$ is odd.  Every such representation has a dimensionality which is doubly-even, {\it i.e.}
an integral multiple of four.  This fact is proved in \ref{pcount}.

Any off-shell ${\cal N}=4$ supermultiplet must have an integral multiple of 128 bosonic degrees of freedom matched
by the same number of fermionic degrees of freedom.  The number 128 may be ascertained by considering
a (shadow) one-dimensional theory obtained by dimensional reduction onto a 0-brane.  In this case the
supersymmetry reduces to $N=16$ in one-dimension, where the representation theory can be tractably discerned using
methods involving graphs and codes, as described in \cite{adinkras,DFGHIL_codes,worldline,doubly}.  An
overview of this technology is given in \ref{adinkrareview}.

The auxiliary fermions in any off-shell version of ${\cal N}=4$ SYM
must describe, in total, an integral multiple of $2\cdot 4\cdot 4=32$ degrees of freedom, where the factor of two reflects that
auxiliary fermions come paired as explained above, the leading factor of four reflects that fermions assemble as spinors,
which in four dimension carry a multiple of four degrees of freedom off-shell, and the second factor of four reflects that
the auxiliary spinors must assemble as odd-rank tensor representations of $SU(4)_R$, which have a dimensionality of 4 mod 4,
as explained above and proved in \ref{pcount}.
Per unit dimension of ${\cal G}$, the total number of fermionic degrees of freedom in the off-shell theory includes 16 from the propagating gaugino fields
$\chi_i^I$ plus $32\,n$ from auxiliary fermionic fields.
Since the total number of fermionic degrees of freedom must be an integral multiple of 128,
it follows that
\brr 16+32\,n &=& 128\,m \,,
\label{countnm}\err
where $n$ counts the auxiliary fermions and $m$ counts the minimal 1D shadow
multiplets per unit dimension of ${\cal G}$ describing the ``skeleton" of the theory. Both $n$ and $m$ are necessarily integers.
Equation (\ref{countnm}) has no solution for integer $n$ and integer $m$ unless $n=0$.  We proceed to explain why $n$ cannot be
zero in an off-shell theory consistent with our preliminary assumptions.
Taken together, these two conclusions imply an algebraic impasse precluding an off shell
description of ${\cal N}=4$ SYM unless we relax one or more elements of our preliminary assumptions.

If there are no fermionic auxiliary fields, {\it i.e.} if $n=0$ in (\ref{countnm}), then
the foregoing discussion implies that
the physical boson fields $\phi_{ij}^I$ and the physical fermion fields $\chi_i^I$
must transform in a common multiplet, with respective dimensions one and three-halves, and that exactly
seven extra boson auxiliary degrees of freedom per unit dimension of ${\cal G}$ must transform in that same multiplet,
and must have dimension two.

If there are no auxiliary fermions then the auxiliary bosons must transform in
the adjoint of ${\cal G}$ and span a seven dimensional
representation of $\spin(1,3)\times SU(4)_R$.
Moreover, these new fields must appear in the supersymmetry transformation rule for $\chi_i^I$
in terms commensurate with the Clifford algebra $CL_{1,3}({\mathbb R})_{\mathbb C}$, of which spinors are elements.
The most general possibility consistent with the stated symmetry and field dimension requirements
would be
\brr \delta_Q\chi_i^I &=& \cdots
     +X_1^I\,\e_i
     +X^I_{2\,ab}\,\gamma^{ab}\,\e_i
     +X^I_{3\,a\,(ij)}\,\gamma^a\,\e^j
     +X^I_{4\,a\,[ij]}\,\gamma^a\,\e^j
       \,,
\label{dc7}\err
where the ellipsis includes terms involving the physical boson fields $\phi_{ij}^I$ and $A_a^I$, as exhibited in
(\ref{n4tran}).
The four explicit terms in (\ref{dc7}) include possible auxiliary bosonic fields $X^I_{1\cdots 4}$ transforming in
distinct irreducible representations of $\spin(1,3)\times SU(4)_R$, as manifested by the index structures.
For example $X^I_{3\,a\,(ij)}$ would transform as a vector ${\bf 4}$ under $\spin(1,3)$, manifested by the
index $a$, and as
a symmetric rank-2 tensor {\bf 10} under $SU(4)_R$, manifested by the symmetrized index pair $(ij)$.
There are no other possible linear terms consistent with the chirality of $\chi_i^I$ and with $\spin(1,3)\times SU(4)_R$
covariance.

If there are no auxiliary fermion fields, then the
dimension two auxiliary bosons must describe exactly seven degrees of freedom for each value of the index $I$.
This restricts the possibilities in (\ref{dc7}).  For example,
$X_{3\,a\,(ij)}^I$ involves $4\times 10=40$ degrees of freedom
and $X_{4\,a\,[ij]}$ involves $4\times 6=24$ degrees of freedom
for each value of $I$.  Thus, neither of these can be included supersymmetrically without
also including auxiliary fermions.

The remaining possible auxiliary bosons contribute the requisite seven off-shell degrees of freedom
per unit dimension of ${\cal G}$ only if the singlet $X_1^I$ and
the two-form $X_{2\,ab}^I$, are constrained to be real.
In this case, the supersymmetry transformation rules
must include
\brr \delta_Q\,\chi_i^I &=& \cdots
     +X_1^I\,\e_i
     +X_{2\,ab}^I\,\gamma^{ab}\,\e_i
     \nonumber\\[.1in]
     \delta_Q\,X_1^I &=&
     \fr12\,i\,\bar{\e}^i\,\dslash \chi_i^I
     +\fr12\,i\,\bar{\e}_i\,\dslash\chi^{I\,i}
     \nonumber\\[.1in]
     \delta_Q\,X_{2\,ab}^I &=&
     i\,\bar{\e}^i\,(\,{\red\nu_1}\,\gamma_{ab}\dslash+{\red\nu_2}\,\gamma_{[a}\der_{b]}\,)\,\chi_i^I
     +{\rm h.c.}
\label{tranXX}\err
plus non-linear terms which covariantize the derivatives, where ${\red\nu_1}$ and ${\red\nu_2}$ are
real coefficients.  The ellipsis
in the first line of (\ref{tranXX}) corresponds to the terms involving the physical fields
$\phi_{ij}^I$ and $A_a^I$, exhibited in
(\ref{n4tran}). There are no other
covariant linear terms involving the relevant fields which can augment these
transformation rules.  The coefficients in the first line of (\ref{tranXX})
may be fixed by scaling $X_1^I$ and $X_{2\,ab}^I$.  The coefficients
in the second line are determined by requiring that two supersymmetry transformations
applied to $X_1^I$ commute into a derivative of $X_1^I$ according to the ${\cal N}=4$ supersymmetry algebra,
exhibited in (\ref{qqcom}).

Immediately, we encounter two problems.
First, using the linear terms exibited above, we compute
\brr [\,\delta_Q(\e_1)\,,\,\delta_Q(\e_2)\,]\,X_1^I
     &=& 2\,i\,\bar{\e}_{[2}^i\,\gamma^a\,\e_{1]\,i}\,(\,
     \der_a X_1^I+2\,\der^b X_{2\,ab}^I\,) \,.
\label{anom4}\err
The first term on the right-hand side is the linear contribution appropriate to this commutator.
But the second term, in which the field $X_{2\,ab}^I$ appears, is an unwanted anomaly, spoiling the
requirements of the supersymmetry algebra.
Second, and worse, it is not possible to choose the coefficients ${\red\nu_1}$ and ${\red\nu_2}$ so as
to obtain the requisite $X_{2\,ab}^I$ contribution to the $[\delta_Q,\delta_Q]$ commutator applied
to $X_{2\,ab}^I$.
\footnote{Superficially, the defining commutator (\ref{qqcom}) is satisfied
if  $(\,{\red\nu_1}\,,\,{\red\nu_2}\,)=(\,0\,,-\fr12\,)$ provided $\der_{[a}X_{2\,bc]}^I=0$.
But such a Bianchi identity is incommensurate with the requirements
that $X_{2\,ab}^I$ be a dimension two auxiliary field with six degrees of freedom per unit dimension of ${\cal G}$, off-shell.
In fact, the properly covariant version of this identity this would render $X_{2\,ab}^I$ indistinguishable
from the Yang-Mills field strength ${\cal F}_{ab}^I$.}

A natural remedy to the anomaly encountered in the previous paragraph would be to include
new fermion fields with dimension 5/2 which transform under supersymmetry into the auxiliary bosons
and into which the auxiliary bosons transform.  This could synergistically supply new terms to (\ref{anom4})
which could cancel the anomaly.  However, this would violate our restriction to the case $n=0$ since this would
involve extra fermions.  What we have shown by this is that, modulo our stated assumptions, it is not possible to close the
${\cal N}=4$ supersymmetry algebra off-shell without including auxiliary fermions.  But we have already proved that,
modulo our stated assumptions, it is not possible to close the ${\cal N}=4$ supersymmetry algebra if we have
any auxiliary fermions.  Thus, unless we relax one or more of our assumptions, we are stymied.

\setcounter{equation}{0}
\section{The ${\cal N}=2$ context for ${\cal N}=4$ SYM}
The essence of the problem described above may be clarified by examining the
${\cal N}=4$ SYM multiplet in terms of its ${\cal N}=2$ constituents,
which are a vector multiplet and a hypermultiplet.
The former has a well known and perfectly acceptable off-shell extension.
But the only known extensions of the latter have awkward features,
such as an off-shell central charge or an infinite number of auxiliary fields. In this section we
explicitly restrict the on-shell ${\cal N}=4$ SYM multiplet to its ${\cal N}=2$ constituents,
we augment these with ``standard" auxiliary fields, and we explain how the awkward features of the
traditional hypermultiplets obstruct an off-shell ${\cal N}=2\to {\cal N}=4$ amalgamation.  This motivates subsequent sections
in which we engage a systematic search for an alternative off-shell hypermultiplet.

When we restrict ${\cal N}=4$ to an ${\cal N}=2$ sub-algebra,
the $R$ symmetry restricts as $SU(4)_R\to SU(2)_R\times SU(2)_H$, where the first factor on the right-hand side is
the residual $R$ symmetry.  The second factor $SU(2)_H\subset SU(4)_R$ commutes with
$SU(2)_R\subset SU(4)_R$.  We implement this restriction by identifying an $SU(2)_R$ transformation
with the first two $SU(4)_R$ tensor index values, $i=1,2$.  The group $SU(2)_H$ acts on the remaining two indices; we
re-designate these, using a hat, by writing $(\,T_3\,,\,T_4\,)=(\,T^{\hat{1}}\,,\,-T^{\hat{2}}\,)$.
In other words, we write the latter $SU(4)_R$ indices $i=3,4$ respectfully as $SU(2)_H$ indices $\hat{\imath}=\hat{1},\hat{2}$
and we raise these using the invariant tensor $\ve^{\hat{\imath}\hat{\jmath}}$.

Henceforth, we suppress the adjoint ${\cal G}$ index $I$
and we omit non-linear contributions proportional to the structure constant $f_{IJ}\,^K$.  This restricts
attention to the abelian case.  This restriction is suitable for our purposes; any extra subtleties attending
the non-abelian generalization can be addressed, in future work, once the off-shell problem has been
adequately addressed in this elemental context.

The ${\cal N}=4$ Lorentz scalars $\phi_{ij}=\fr12\,\ve_{ijkl}\,\phi^{kl}$, describe three complex fields,
$\phi_{12}=\phi^{34}$, $\phi_{13}=\phi^{42}$, and $\phi_{23}=\phi^{14}$.
The first of these is a singlet under $SU(2)_R$ and $SU(2)_H$, and can be
re-written as $\phi_{12}\equiv\ve_{12}\,X$,
or $\phi^{34}\equiv\ve_{\hat{1}\hat{2}}\,X$, noting that $\ve_{12}=\ve_{\hat{1}\hat{2}}=1$.
More succinctly, $X\equiv \phi_{12}$. The remaining Lorentz scalars
structure as
\brr \ba{cc}\phi_{13}&\phi_{14}\\\phi_{23}&\phi_{24}\ea &=&
     \ba{cc}\phi_1\,^{\hat{1}}&-\phi_1\,^{\hat{2}}\\\phi_2\,^{\hat{1}}&-\phi_2\,^{\hat{2}}\ea  \,.
\err
The constraints $\phi_{13}=\phi^{42}$, and $\phi_{23}=\phi^{14}$ translate as $\phi_i\,^{\hat{\imath}}=\ve_{ij}\,\ve^{\hat{\imath}\hat{\jmath}}\,\phi^j\,_{\hat{\jmath}}$,
which is manifestly $SU(2)_R\times SU(2)_H$ covariant. In the ${\cal N}=2$ context, indices $i$ or $\hat{\imath}$
implicitly assume only two values. Collectively, the four degrees of freedom
described by $\phi_i\,^{\hat{\imath}}$ form a quaternion.\footnote{This is because the constraint
$\phi_i\,^{\hat{\imath}}=\ve_{ij}\,\ve^{\hat{\imath}\hat{\jmath}}\,\phi^j\,_{\hat{\jmath}}$ is
solved uniquely by $\phi_i\,^{\hat{\imath}}=(\,a^0\,\Ione+i\,a^I\,\sigma_I\,)_i\,^{\hat{\imath}}$,
where $\sigma_I$ are the Pauli matrices and $a^{0,1,2,3}$ are real fields.}
The ${\cal N}=4$ fermions $\chi_i$ parse as two pairs, $(\,\chi_1\,,\,\chi_2\,)\equiv (\,-\Omega^2\,,\,\Omega^1\,)$
and $(\,\chi_3\,,\,\chi_4\,)\equiv(\xi^{\hat{1}}\,,\,-\xi^{\hat{2}}\,)$.
Under $SU(2)_R\times SU(2)_H$ the first pair, $\chi_i$, transforms as ${\bf 2}\times {\bf 1}$
and the second pair, $\xi^{\hat{\imath}}$, transforms as
${\bf 1}\times {\bf 2}$. \footnote{\label{ftp}For product groups ${\cal G}\times \tilde{\cal G}$,
we write tensor products of representations of either group factor as ${\bf R}_1\otimes {\bf R}_2$.
(These are ``intra-group" representations.) Tensor products of representations between
group factors are written ${\bf R}\times \tilde{\bf R}$.  (These are ```inter-group" representations.)
In more standard mathematical notation both kinds of tensor products would be written using $\otimes$.}

We have defined $(\,\phi^{12}\,,\,\chi^i\,)\equiv(\,X\,,\,-\ve^{ij}\,\Omega_j\,)$, so that
$(\,X\,,\,\Omega_i\,)=(\,\fr12\,\ve_{ij}\,\phi^{ij}\,,\,\ve_{ij}\,\chi^j\,)$.
Along with the field strength $F_{ab}$,
these transform under an ${\cal N}=2$ supersymmetry transformation, parameterized by
$\e^{i=1,2}$, according to
\brr \delta_Q\,X &=& i\,\bar{\e}^i\,\Omega_i
     \nonumber\\[.1in]
     \delta_Q\,\Omega_i &=& \dslash X\,\e_i
     +\fr14\,\ve_{ij}\,F_{ab}\,\gamma^{ab}\,\e^j
     {\blue +Y_{ij}\,\e^j\,}
     \nonumber\\[.1in]
     \delta_Q\,F_{ab} &=& 2\,i\,\ve^{ij}\,\bar{\e}_i\,\gamma_{[a}\,\der_{b]}\chi_j
     +{\rm h.c.}
     \nonumber\\[.1in]
     {\blue\delta_Q\,Y_{ij}} &{\blue =}& {\blue i\,\bar{\e}_{(i}\,\dslash\Omega_{j)}
     +{\rm h.c.}}
\label{v2rules}\err
where the terms involving $(\,X\,,\,\Omega_i\,,\,F_{ab}\,)$ follow directly from the restriction of the ${\cal N}=4$ rules, and the terms
involving $Y_{ij}$ have been added in.  These latter terms represent the well-known off-shell completion
comprising the off-shell ${\cal N}=2$ Vector multiplet. The auxiliary field $Y_{ij}=Y_{(ij)}=\ve_{ik}\,\ve_{jl}\,Y^{kl}$ describes three additional
off-shell degrees of freedom.

The balance of the ${\cal  N}=4$ SYM components transform as follows,
\brr \delta_Q\,\phi_i\,^{\hat{\imath}} &=&
     i\,\bar{\e}_i\,\zeta^{\hat{\imath}}
     +i\,\ve^{\hat{\imath}\hat{\jmath}}\,\ve_{ij}\,\bar{\e}^j\,\zeta_{\hat{\jmath}}
     \nonumber\\[.1in]
     \delta_Q\,\zeta^{\hat{\imath}} &=&
     \dslash\phi_i\,^{\hat{\imath}}\,\e^i
     {\red +\ve^{ij}\,\phi_i^{({\rm z})\,\hat{\imath}}\,\e_j}
     \nonumber\\[.1in]
     {\red \delta_Q\,\phi^{\rm (z)}_i\,^{\hat{\imath}}} &{\red =}&
     {\red i\,\bar{\e}_i\,\zeta^{({\rm z})\,\hat{\imath}}
     +i\,\ve^{\hat{\imath}\hat{\jmath}}\,\ve_{ij}\,\bar{\e}^j\,\zeta^{(\rm z)}_{\hat{\jmath}}}
\label{fhrules}\err
where the terms involving $(\,\phi_i\,^{\hat{\imath}}\,,\,\xi^{\hat{\imath}}\,)$  are the direct translation of the restricted ${\cal N}=4$ transformation rules,
and the terms involving $\phi^{\rm (z)}_i\,^{\hat{\imath}}$ describe a well-known augmentation which enables off-shell closure with a caveat:
this enlarged multiplet respects a modified ${\cal N}=2$ supersymmetry algebra, including a non-trivial off-shell central charge.
The auxiliary field $\phi^{\rm (z)}_i\,^{\hat{\imath}}$ describes four additional off--shell degrees of freedom.  This version is
the Fayet Hypermultiplet \cite{Fayet}.  It satisfies the algebra
\brr [\,\delta_Q(\e_1)\,,\,\delta_Q(\e_2)\,] &=&
     2\,i\,\bar{\e}_{[2}^i\,\dslash\,\e_{1]\,i}+\delta_Z({\rm z})
\err
where the second term is a central charge transformation, which acts as
$\delta_Z({\rm z})={\rm z}\,Z$, where $Z\,\phi_i\,^{\hat{\imath}}=\phi^{\rm (z)}_i\,^{\hat{\imath}}$
and $Z\,\xi^{\hat{\imath}}=\xi^{{\rm (z)}\,\hat{\imath}}$, and where the parameter is given by
${\rm z}=i\,\ve_{ij}\,\bar{\e}_1^i\,\e_2^j+{\rm h.c.}$
The field $\phi^{\rm (z)}_i\,^{\hat{\imath}}$ is an independent auxiliary quaternion.  But
$\xi^{{\rm (z)}\,\hat{\imath}}$ is not an independent field.  Instead, this is defined
by a constraint given below.
In fact, the central charge generates semi-infinite sequences
\brr \phi_i\,^{\hat{\imath}} &\to&
     \phi^{\rm (z)}_i\,^{\hat{\imath}} \,\,\to\,\,
     \phi^{\rm (zz)}_i\,^{\hat{\imath}} \,\,\to\,\,\cdots
     \nonumber\\[.1in]
     \xi^{\hat{\imath}} &\to& \xi^{{\rm (z)}\,\hat{\imath}}
     \,\,\to\,\,\xi^{{\rm (zz)}\,\hat{\imath}}  \,\,\to\,\,\cdots
\err
where all fields except the elemental components $(\,\phi_i\,^{\hat{\imath}}\,,\,\xi^{\hat{\imath}}\,,\,\phi^{\rm (z)}_i\,^{\hat{\imath}}\,)$,
are defined by the constraint $\xi^{{\rm (z)}\,\hat{\imath}}=-\ve^{\hat{\imath}\hat{\jmath}}\,\dslash\xi^{\hat{\jmath}}$,
the constraint $\phi^{\rm (zz)}_i\,^{\hat{\imath}}=\Box\phi_i\,^{\hat{\imath}}$,
and analogs of these generated by appending additional ${\rm z}$s to the superscripts on both sides
of either of these equations.
For any non-negative integer $n$, the result of the $n$-fold iteration $Z^n$ applied to the elemental fields is
$(\,\phi^{\rm (n)}_i\,^{\hat{\imath}}\,,\,\xi^{{\rm (n)}\hat{\imath}}\,,\,\phi^{\rm (n+1)}_i\,^{\hat{\imath}}\,)$,
where the superscript $(n)$ denotes a string $({\rm z}\cdots{\rm z})$ with $n$ elements.  Each such set has transformation
rules similar to (\ref{fhrules}).  The constraints imply $Z^2=\Box$.

The off-shell multiplets (\ref{v2rules}) and (\ref{fhrules}) play pivotal roles in the
${\cal N}=2$ supersymmetry literature: in the development of ${\cal N}=2$ supergravity \cite{dw1,dw2,structure},
special geometry \cite{SpecG,SpecG1,SpecG2}, non-linear sigma models \cite{Sigmas_03,Sigmas_02,Sigmas_01}, Seiberg-Witten theory
\cite{SWitt1,SWitt2}, and ruminations about ${\cal N}=2$ phenomenology.
In much of that work, the auxiliary fields $Y_{ij}$ and $\phi^{\rm (z)}_i\,^{\hat{\imath}}$ are important ingredients in the
construction of supersymmetric densities, and are therefore relevant to the physical content.  However, the ``traditional"
constructions (\ref{v2rules}) and (\ref{fhrules}) prove inadequate as a basis for an off-shell completion of ${\cal N}=4$
SYM theory, as we explain presently.

The two multiplets described by (\ref{v2rules}) and (\ref{fhrules}) respect distinct
algebras off-shell: the Fayet Hypermultiplet has a non-trivial central charge in contradistinction to the off-shell Vector multiplet.
An amalgamation of these two into a single ${\cal N}=4$ multiplet would require
intertwining supersymmetry transformations involving the parameters $\e_{\hat{\imath}}$, which we switched off to expose the
${\cal N}=2$ underpinning.  These would connect the Vector multiplet fields with the Hypermultiplet fields and vice-versa.
The fermions $\chi_i$ and $\xi^{\hat{\imath}}$ would combine as a vector ${\bf 4}$ representation of $SU(4)_R$.
As a result, the complex boson $X$ would transform under $\e_{\hat{\imath}}$ into $\xi^{\hat{\imath}}$.
However, since $Z\,\xi^{\hat{\imath}}\ne 0$ while $Z\,X=0$, it follows that the central charge could not commute with all of the ${\cal N}=4$
supercharges.  This is inconsistent with the super-conformal algebra, which limits the
bosonic charges which do not commute with the supercharges to the Poincar{\'e} generators, the $SU(4)_R\times U(1)_R$
generators, dilatations, and special conformal transformations, none of which correspond to $Z$.  In this way, the
desired amalgamation is precluded by the superconformal Jacobi identities.

The auxiliary fields $Y_{ij}$ and $\phi^{\rm (z)}_i\,^{\hat{\imath}}$ transform under
$SU(2)_R\times SU(2)_H$ as ${\bf 3}\times{\bf 1}$ and ${\bf 2}\times {\bf 2}$.
If the two multiplets  (\ref{v2rules}) and (\ref{fhrules}) can be amalgamated off-shell,
then these must combine as a seven-dimensional representation of
$SU(4)_R$.  However, there is no seven-dimensional representation
of $SU(4)_R$ which branches into $(\,{\bf 3}\times {\bf 1}\,)\oplus (\,{\bf 2}\times {\bf 2}\,)$
when $SU(4)_R\to SU(2)_R\times SU(2)_H$.  This provides another obstruction precluding the desired amalgamation.

Thus, at least one of the two multiplets
(\ref{v2rules}) and (\ref{fhrules}) requires an alternate off-shell description,
involving different auxiliary fields, to provide any prospect of amalgamation.
Toward that end, it is natural to postulate an off-shell hypermultiplet respecting
the ${\cal N}=2$ supersymmetry algebra with a trivial central charge off-shell.
However, there is an apparent obstruction based on a simple counting argument.
In the denouement of this section, we discuss this
matter and propose a resolution.

Any auxiliary fermions in a possible off-shell hypermultiplet must describe
an integer multiple of $2\cdot 4\cdot 2$ degrees of freedom.  In this accounting, the leading factor of
two reflects that auxiliary fermions come paired, the factor of four reflects that 4D spinors
have four degrees of freedom off-shell, and the second factor of two reflects that all fields
transform as doublets under $SU(2)_H$.  Since the physical fermions $\xi^{\hat{\imath}}$ involve
eight degrees of freedom, the total number of off-shell fermionic degrees of freedom
would be $8+16\,n$, where $n$ is the number of minimal auxiliary spinor pairs.
However, every irreducible ${\cal N}=2$ multiplet contains an integer multiple
of eight degrees of freedom, as explained in \ref{adinkrareview}.  Thus, the ${\cal N}=2$ analog of the
${\cal N}=4$ counting restriction (\ref{countnm}) is
\brr 8+16\,n &=& 8\,m \,,
\label{counth2}\err
where $n$ counts the auxiliary fermions and $m$ counts the minimal 1D shadow multiplets describing the ``skeleton"
of the hypermultiplet.  There is no solution to (\ref{counth2}) for integer $n$ and integer $m$ unless $n=0$.
Since any alternative hypermultiplet necessarily requires extra fermions, we are forced to reconsider our
initial assumptions, to determine which may be revised.

 A systematic approach, described in the next section, yields an
off-shell multiplet involving 192+192 degrees of freedom, in which all but 16+16 of these
are auxiliary.  The non-auxiliary degrees of freedom properly describe the field content of two
 hypermultiplets: two quaternions and two spinor doublets transforming under $SU(2)_R\times SU(2)_H$
 respectively as ${\bf 2}\times {\bf 2}$ and ${\bf 1}\times{\bf 2}$, {\it i.e.}, properly.   This construction, which we deem the
``Hyperplet",  was discovered nine years ago, and was described in \cite{RETM}.  However,
 its two hypermultiplet sectors contribute with opposite signs to the Hamiltonian.
Accordingly, the Hyperplet seems to exhibit a vacuum instability and to possess inconsistent ghost-like propagators.
This apparent problem relegated the Hyperplet to the status of a merely amusing curiosity.
However, the following observation suggests a means to a resolution.

An ostensibly ghost-like Fayet Hypermultiplet provides a conformal compensator in a well-known approach
to realising ${\cal N}=2$ Poincar{\'e} supergravity as a gauge fixed conformal supergravity theory.  In this approach,
a theory with $N_H$ hypermultiplets is built as a theory involving $N_H+1$ hypermultiplets
coupled to conformal supergravity along with $N_V+1$ abelian Vector multiplets.  One of the hypermultiplets enters
the action with ``wrong sign" kinetic terms, and that multiplet is pure gauge aside from its auxiliary degrees of freedom.
The pure gauge fields, including the ghost-like quaternion and its fermion doublet superpartner, provide conformal compensators,
which fix the background to Poincar{\'e} supergravity.

We propose that a conformal analog of the Hyperplet is suited for a similar purpose: to provide a novel
comformal compensator, enabling a new off-shell description of ${\cal N}=2$ Poincar{\'e} supergravity.
What is attractive about this proposition is
that it elegantly resolves the problem associated with the obstruction indicated by (\ref{counth2})
and, at the same time, resolves the issue of ghosts in the Hyperplet.  It does this
by including one extra $SU(2)_H$ spinor doublet which is pure gauge --- a kind of field not considered in our
preliminary assumptions, but which is enabled if we realize the multiplet superconformally.  This extra spinor doublet
included in the ghosty half of the Hyperplet modifies
(\ref{counth2}) into
\brr 8+8+16\,n &=& 8\,m \,,
\err
where the three terms on the left-hand side correspond to physical fermions, pure-gauge fermions, and auxiliary fermions.
The second of these is a new addition enabled by the new context.  Now the equation does admit solutions for non-zero
$n$.  Fixing the gauge we obtain a model with a physical hypermultiplet coupled to Poincar{\'e} supergravity,
for which the supersymmetry algebra closes off shell with a vanishing central charge.  The price we pay
is that the coupling to ${\cal N}=2$ supergravity is essential.

\setcounter{equation}{0}
\section{The 192+192 Hyperplet}
By definition, a hypermultiplet is an ${\cal N}=2$ supermultiplet supporting a propagating quaternion $\phi_i\,^{\hat{\imath}}=\ve_{ij}\,\ve^{\hat{\imath}\hat{\jmath}}\,\phi^j\,_{\hat{\jmath}}$ and a
doublet of propagating spinors $\xi^{\hat{\imath}}=-\gamma_5\,\xi^{\hat{\imath}}$, with the following properties.  These fields transform under $SU(2)_R\times SU(2)_H$
in the manner prescribed by the specified indices, namely as ${\bf 2}\times {\bf 2}$ and ${\bf 1}\times {\bf 2}$, respectively.
A supersymmetric action should exist, and, aside from the possibility of background supergravity,
there should be no other propagating fields than the 4+4 on-shell physical
degrees of freedom already specified. \footnote{While the fields $\phi_i\,^{\hat{\imath}}$ and $\xi^{\hat{\imath}}$ account
for 4+4 degrees of freedom on-shell, these correspond to 8+8 degrees of freedom off-shell.}  In other words all other degrees of freedom should be auxiliary,
in the sense defined above, or be pure gauge.   We seek to construct a hypermultiplet with finite degrees of freedom and a
trivial central charge off-shell. We deem such a construction a hyperplet.

In \ref{multclass} we present an infinite palette of multiplets with a variety of $SU(2)_R\times SU(2)_H$
tensor structures, all with trivial central charge off-shell.  In a quest to build a hyperplet, one might
choose from that class, as a starting point, the Dual Sextet multiplet $\widetilde{\cal M}_6\,^{\hat{2}}$
because its sole lowest component is a quaternion
$\Phi_i\,^{\hat{\imath}}$.  But, the sole fermion at the next level in that multiplet is $\Psi_{ij}\,^{\hat{\imath}}$, which
transforms irreducibly as ${\bf 3}\times{\bf 2}$ rather than as ${\bf 1}\times{\bf 2}$.  In our hyperplet theory, any unwanted
dimension 3/2 fermions need to be ``switched off" by coupling them without derivatives to a dimension 5/2
fermion with a commensurate $SU(2)_R\times SU(2)_H$ tensor structure.  So, in order to succeed with
$\widetilde{\cal M}_6\,^{\hat{2}}$, we must identify additional multiplets to supply the requisite
auxiliary fermion to switch off $\Psi_{ij}\,^{\hat{\imath}}$, and to supply the desired
fermion $\xi^{\hat{\imath}}$.  All other fields that come along for the ride must
conspire to be non-propagating. Moreover, we need to construct a supersymmetric action to
accomplish these tasks.  We could proceed to seek such a resolution.  However, we shall first
present two other possible starting points. Ultimately, we will find that all three starting points
lead to the same solution.

As a second possible starting point, we could consider the only other multiplet from the class described
in \ref{multclass} with a sole quaternion as its lowest component, namely $\widetilde{\cal M}_5\,^{2,\hat{2}}$.
In this case, the fermions at the second level
transform as
\footnote{See footnote \ref{ftp} for clarification about the tensor product notation.}
\brr (\,{\bf 2}\otimes {\bf 2}\,)\times{\bf 2} &=& (\,{\bf 1}\oplus{\bf 3}\,)\times {\bf 2}
     \nonumber\\[.1in]
     &=& (\,{\bf 1}\times{\bf 2}\,)\oplus (\,{\bf 3}\times {\bf 2}\,) \,.
\err
In terms of specific fields, this decomposition can be written as
$\Psi_i\,^{k\hat{\imath}}=\delta_i\,^k\,\xi^{\hat{\imath}}+\ve^{kj}\,\psi_{ij}\,^{\hat{\imath}}$.
Thus, we find a suitable fermion $\xi^{\hat{\imath}}$, but we
again have a rogue fermion $\psi_{ij}\,^{\hat{\imath}}=\psi_{(ij)}\,^{\hat{\imath}}$ which needs attention,
as explained above.

There is another multiplet in our palette which has a quaternion at its lowest level,
namely ${\cal M}_3\,^{2,\hat{2}}$.  We call this the Extended Tensor Multiplet,
because this is obtained from the Tensor multiplet ${\cal M}_3$ via extension, by adding new
indices $(\cdot)^{i\,\hat{\imath}}$.  This is our third (of three) logical starting points.
\footnote{The multiplet ${\cal M}_3\,^{2,\hat{2}}$ is the simplest in the class
of ``extended $p$-plets" ${\cal M}_p\,^{(p-1),\hat{2}}$, each of which has a quaternion, along with
rogue bosons, at its lowest level.
The cases $p>3$ result in relatively complicated models analogous to the ones we are discussing.
These have similar fields to the Hyperplet, but have ``echoes" involving higher-rank tensor components.}
In this case, the bosons at the lowest level transform as
\brr (\,{\bf 3}\otimes{\bf 2}\,)\times{\bf 2} &=&
     (\,{\bf 2}\oplus{\bf 4}\,)\times {\bf 2}
     \nonumber\\[.1in]
     &=& (\,{\bf 2}\times{\bf 2}\,)\oplus (\,{\bf 4}\times{\bf 2}\,) \,,
\err
where the first term in the final line is the quaternion.
In terms of specific fields, this decomposition can be written as
$\Phi_{ij}\,^{l\hat{\imath}}=\delta_{(i}\,^l\,\phi_{j)}\,^{\hat{\imath}}+\ve^{jk}\,u_{ijk}\,^{\hat{\imath}}$.
Now we have a rogue boson $u_{ijk}\,^{\hat{\imath}}=u_{(ijk)}\,^{\hat{\imath}}$ at the lowest level.  This field has dimension one
since it sits at the same level as the physical quaternion field.  In order to ``switch off" this field
({\it i.e.}, render it non-dynamical) we need to couple it to a dimension three auxiliary boson (Lagrange multiplier) with commensurate
tensor structure.  We proceed to work with this case.

We use diagrams to helpfully visualize field content and to guide our reasoning.  Accordingly,
we represent the Extended Tensor Multiplet ${\cal M}_3\,^{2,\hat{2}}$ as
\brr
\includegraphics[width=1.7in]{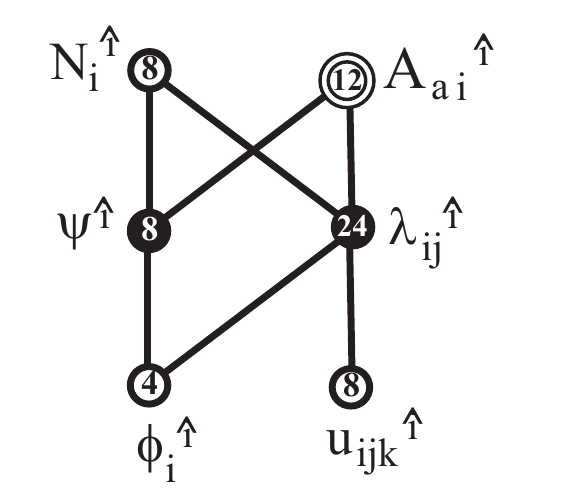} \,.
\label{et1}\err
This graph depicts the component fields and their supersymmetry transformations
using conventions explained in \ref{diappen}.
\footnote{The precise coefficients of the transformation rules are unimportant
for our discussion at this stage.  These are straightforward to work out, given the terms implied by the diagram
and given the supersymmetry algebra.  Such diagrams facilitate helpfully intuitive
speculating about supersymmetric models.}
The components in (\ref{et1}) are expressed using irreducible
representations of $SU(2)_R\times SU(2)_H$, as explained in the previous paragraph.   This multiplet
has hypermultiplet fields $\phi_i\,^{\hat{\imath}}$ and $\xi^{\hat{\imath}}$, along with a dimension two complex
auxiliary field $N_i\,^{\hat{\imath}}$.  But the fermion $\lambda_{ij}\,^{\hat{\imath}}$ and the
other bosonic fields $A_{a\,i}\,^{\hat{\imath}}$ and $u_{ijk}\,^{\hat{\imath}}$ each require attention.
We shall address these problems systematically.

First of all, the divergence-free vector is the Hodge dual of the field strength of a tensor potential,
$A_{a\,i}\,^{\hat{\imath}}=\ve_a\,^{bcd}\,\der_b B_{cd\,i}\,^{\hat{\imath}}$. As such, this this represents an unwanted
propagating field that we should neutralize, {\it i.e.}, render auxiliary.  We can do this by ``relaxing" the constraint by
supplying the divergence $\der^a A_{a\,i}\,^{\hat{\imath}}$ using another supermultiplet.  Since that divergence has dimension
three and is a Lorentz scalar, we need a multiplet which has a dimension three field transforming as $X_i\,^{\hat{\imath}}$,
as its highest-weight component.  A natural choice is the Sextet multiplet ${\cal M}_6\,^{\hat{2}}$,
\brr \includegraphics[width=1.6in]{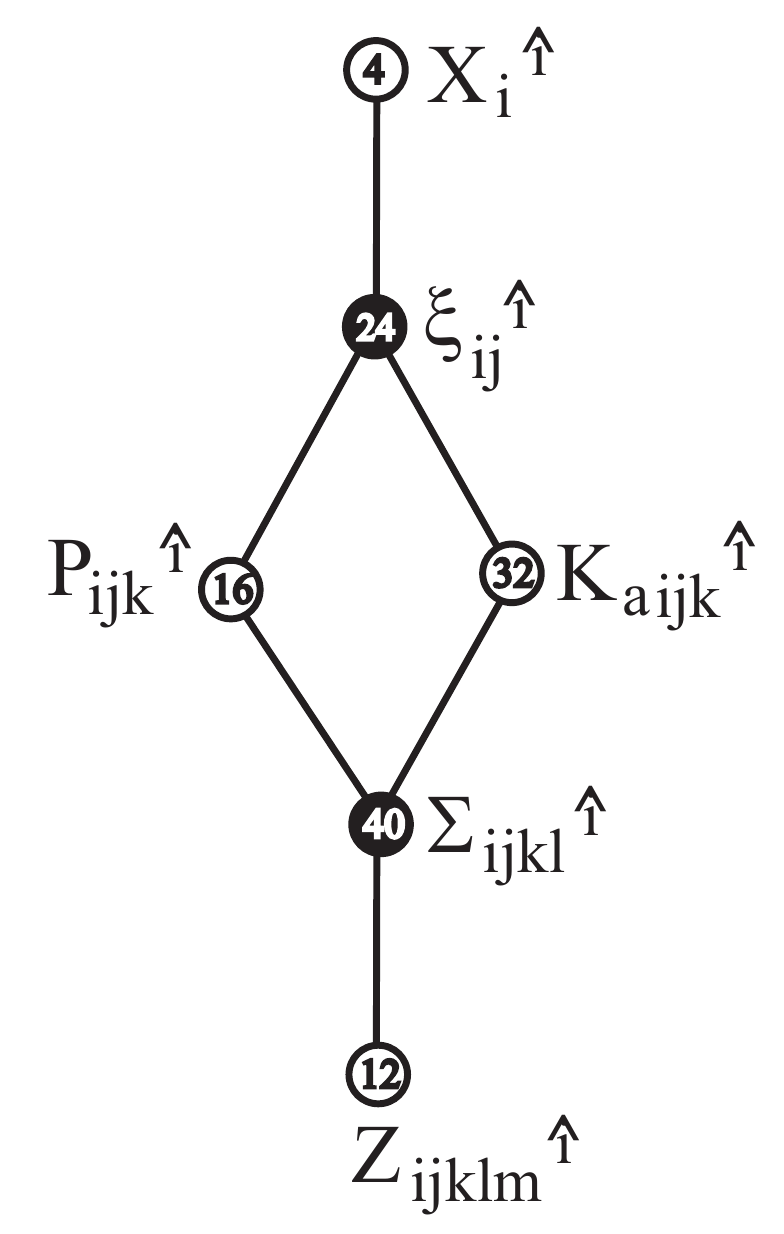}
\err
Given this multiplet along with (\ref{et1}), we can identify $X_i\,^{\hat{\imath}}=\der^a A_{a\,i}\,^{\hat{\imath}}$, provided we add
new terms to the transformation rules, guided by supersymmetry.  In particular,
since $X_i\,^{\hat{\imath}}$ transforms into $\xi_{ij}\,^{\hat{\imath}}$,
the algebra can close only if the rule $\delta_Q A_{a\,i}\,^{\hat{\imath}}$ is
augmented by suitable $\xi_{ij}\,^{\hat{\imath}}$ dependent terms.  And, since the commutator $\delta_Q^2 N_i\,^{\hat{\imath}}$
is proportional to $\der^aA_{a\,i}\,^{\hat{\imath}}$, we need to add $\xi_{ij}\,^{\hat{\imath}}$ dependent terms to $\delta_Q N_i\,^{\hat{\imath}}$
too.
\footnote{The notation $\delta_Q^2$ is shorthand
 for the commutator $[\,\delta_Q(\e_1)\,,\,\delta_Q(\e_2)\,]$.}
 This process continues to other components.  With patience, we find a complete set of needed terms by imposing
proper closure on all components.  What results is
\brr \includegraphics[width=3in]{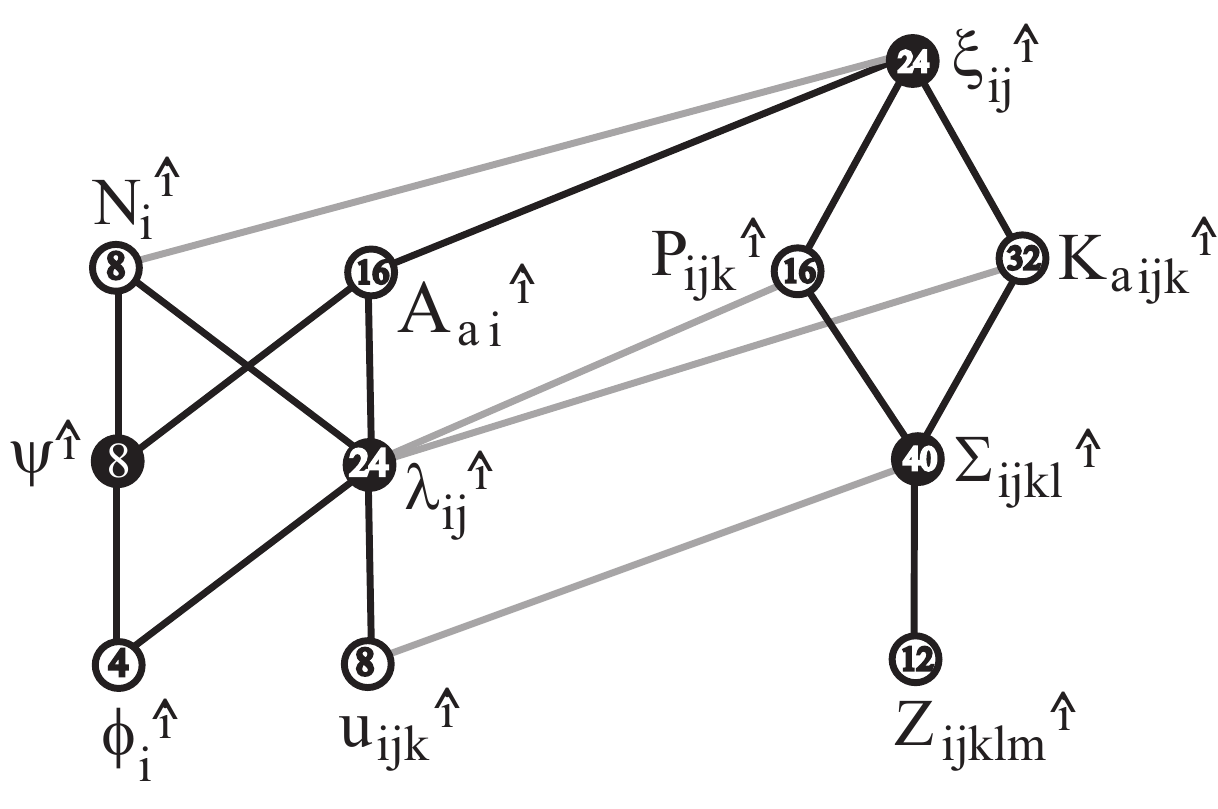} \,,
\label{retmadinkra}\err
where the grey edges are explained in \ref{diappen}.
We call (\ref{retmadinkra}) the Relaxed Extended Tensor Multiplet, or the RETM for short.

We have managed to render $A_{a\,i}\,^{\hat{\imath}}$ as an unconstrained dimension two auxiliary field.  As an extra bonus,
we obtained the dimension 5/2 spinor field $\xi_{ij}\,^{\hat{\imath}}$ which pairs with the dimension 3/2 spinor field
$\lambda_{ij}\,^{\hat{\imath}}$, to ``switch it off". Thus, these two fermion fields form an auxiliary pair.
We have also added two new ``innocuous" auxiliary fields $P_{ijk}\,^{\hat{\imath}}$ and $K_{a\,ijk}\,^{\hat{\imath}}$,
along with a problematic fermion $\Sigma_{ijkl}\,^{\hat{\imath}}$ and a new problematic boson,
$Z_{ijklm}\,^{\hat{\imath}}$.

We can neutralize the dimension one rogue boson fields $u_{ijk}\,^{\hat{\imath}}$ and $Z_{ijklm}\,^{\hat{\imath}}$
by coupling these to commensurate dimension three fields, provided these are situated as highest-weight components in
other multiplets.  This requirement is satisfied by the Dual Sextet $\widetilde{\cal M}_6\,^{\hat{2}}$
and the Dual Quadruplet $\widetilde{\cal M}_4\,^{\hat{2}}$,
given respectively by
\brr \includegraphics[width=3in]{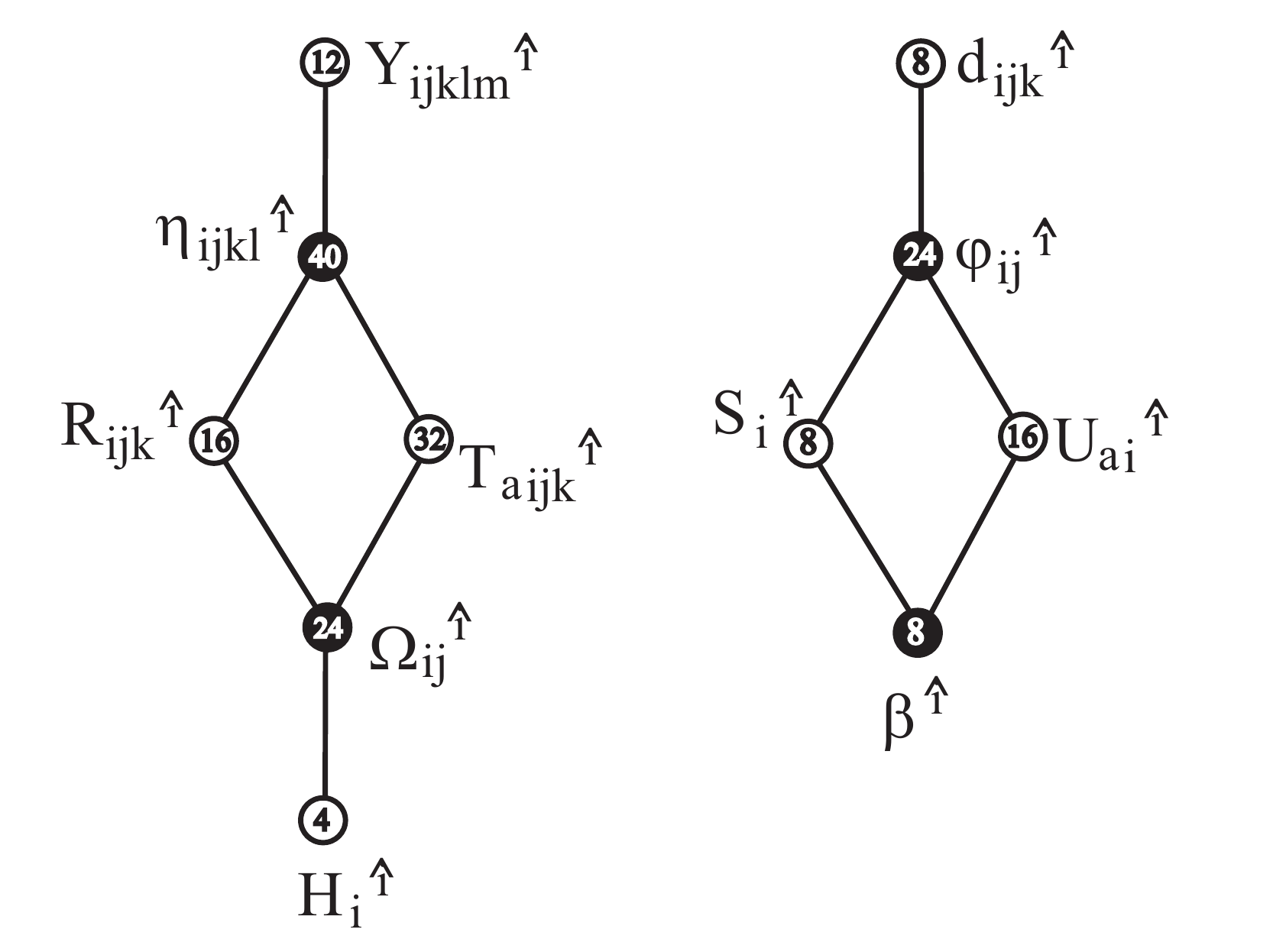} \,.
\label{dsdqul}\err
Notice that the Dual Quadruplet only spans four distinct height assignments, and is therefore ``shorter"
by one field compared to the Dual Sextet.  We have assigned the height (scaling weight) of the Dual
Quadruplet relative to the Dual Sextet, so that the highest weight fields in each case have dimension three.
This ``lifts" the Dual Quadruplet up, so that it contributes no new fields at dimension one --- a satisfying
and promising circumstance.

The multiplets in (\ref{dsdqul}) include the two fields $d_{ijk}\,^{\hat{\imath}}$ and $Y_{ijklm}\,^{\hat{\imath}}$
which, by design, combine as auxiliary pairs
with the RETM fields $u_{ijk}\,^{\hat{\imath}}$ and $Z_{ijklm}\,^{\hat{\imath}}$.  They include
four innocuous auxiliary bosons, $R_{ijk}\,^{\hat{\imath}}$, $T_{a\,ijk}\,^{\hat{\imath}}$,
$S_i\,^{\hat{\imath}}$, and $U_{a\,i}\,^{\hat{\imath}}$. They include a dimension 5/2 spinor
$\eta_{ijkl}\,^{\hat{\imath}}$, which can combine as an auxiliary pair with the RETM spinor $\Sigma_{ijkl}\,^{\hat{\imath}}$.
And they include two other spinors, $\Omega_{ij}\,^{\hat{\imath}}$ and $\varphi_{ij}\,^{\hat{\imath}}$, which
together form an auxiliary fermion pair.  All that remains are two fields, $H_i\,^{\hat{\imath}}$
and $\beta^{\hat{\imath}}$ which comprise the field content of a second hypermultiplet.

We seek an invariant action involving the component fields in (\ref{retmadinkra})
and (\ref{dsdqul}), using the following straightforward algorithm. First, we collect every possible dimension four, real, scalar,
bilinear term, such as $\phi_i\,^{\hat{\imath}}\,\Box\,\phi^i\,_{\hat{\imath}}$, $Z_{ijklm}\,^{\hat{\imath}}\,Y^{ijklm}\,_{\hat{\imath}}$,
$i\,\lambda_{ij}\,^{\hat{\imath}}\,\dslash\Omega^{ij}\,_{\hat{\imath}}$, and so forth. We then vary these under supersymmetry and impose the
criteria that some linear combination of the variations is a total derivative.  If there is a choice of coefficients which
enables that criterion, then this identifies a supersymmetric action.  If there is no such choice, then the
multiplets cannot ``assemble", as is, to provide an action.

Given the multiplets in (\ref{retmadinkra}) and (\ref{dsdqul}),
there is no supersymmetric action.  However, this circumstance is repaired by adding new terms to the transformation rules for the
Dual Sextet involving the fields of the Dual Quadruplet.
Interestingly, it is possible to do this in a manner consistent with
the supersymmetry.  The new terms  correspond diagrammatically to grey edges,
as explained in \ref{diappen}, resulting in the following modified multiplet
\brr \includegraphics[width=3in]{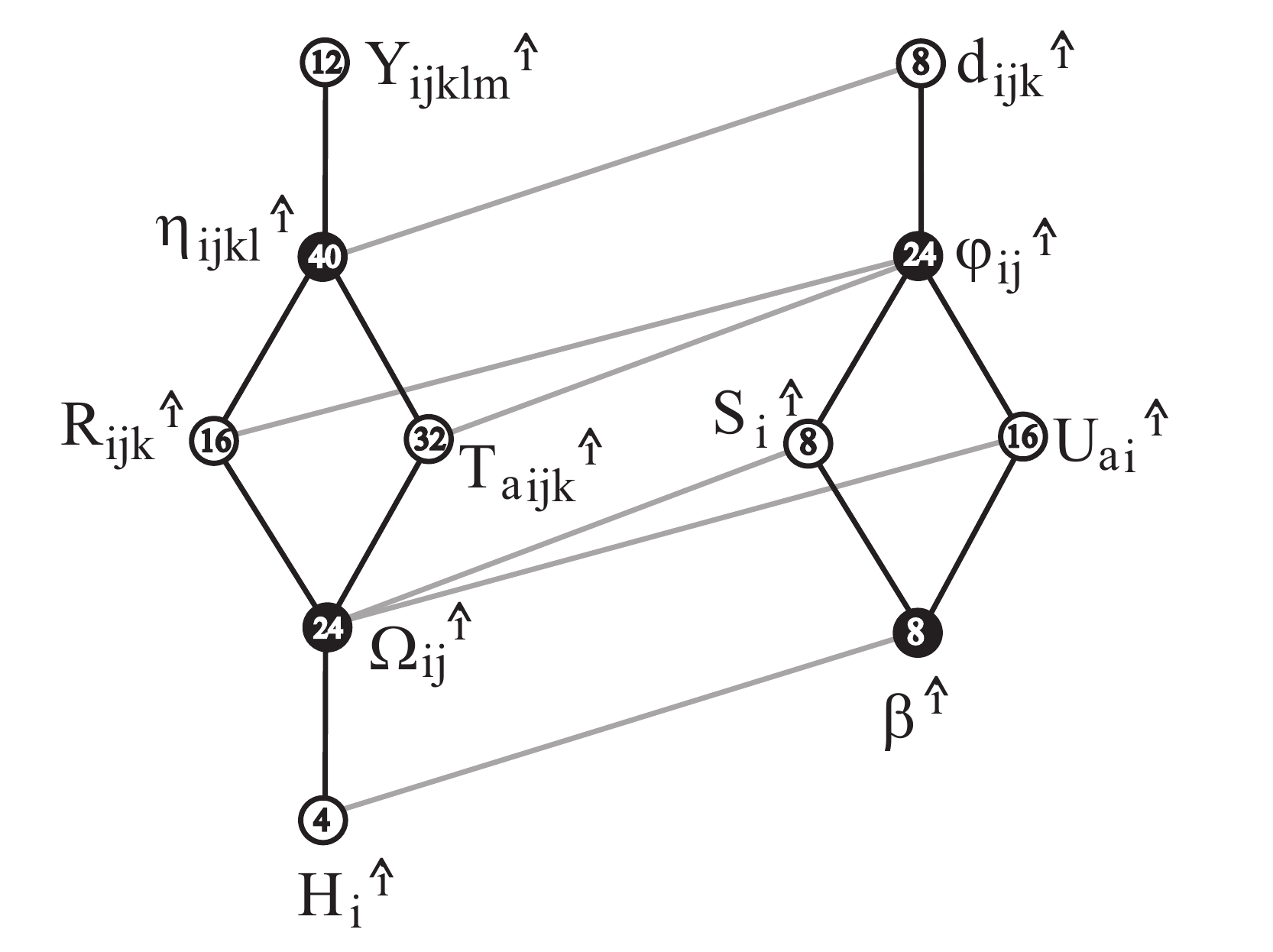} \,.
\label{dsdqtethered}\err
It is instructive to compare (\ref{dsdqul}) with (\ref{dsdqtethered}), and to ponder the difference between these.
The latter involves one-way links which, collectively, contribute nothing to any of the supersymmetry commutators
applied to any of the components.  This is accomplished
by conspiracies between different two-edge ``routes" on the graph, connecting one vertex to another.
For example, two $Q^i$ transformations map the field $R_{ijk}\,^{\hat{\imath}}$ into the field $d_{ijk}\,^{\hat{\imath}}$
via two different intermediary fermions --- one way using $\eta_{ijkl}\,^{\hat{\imath}}$ and the other way using
$\varphi_{ij}\,^{\hat{\imath}}$.  These two contributions to
$\delta_Q^2\,R_{ijk}\,^{\hat{\imath}}$ cancel, as do the contributions to the commutator applied to any
of the Dual Sextet fields involving any of the Dual Quadruplet fields.
 This process creates a new (reducible) supermultiplet by ``tethering" together two irreducible parts.
\footnote{Conversely, if a given multiplet admits a frame, obtained by field redefinitions, for which its Adinkra
has separate parts that could be disconnected by a single scissor cut through any number of grey edges,
then this indicates that the multiplet is reducible.  This concept is explained in \cite{frames}.}
In (\ref{dsdqtethered}) we have obtained an agglomerate of a Dual Sextet (DS) and a Dual Quadruplet
(DQ), which we call the DSDQ multiplet.

At this point, we may collect the field content of our model.  We have the fields of the
RETM and the fields of the DSDQ, given by
\brr \includegraphics[width=5in]{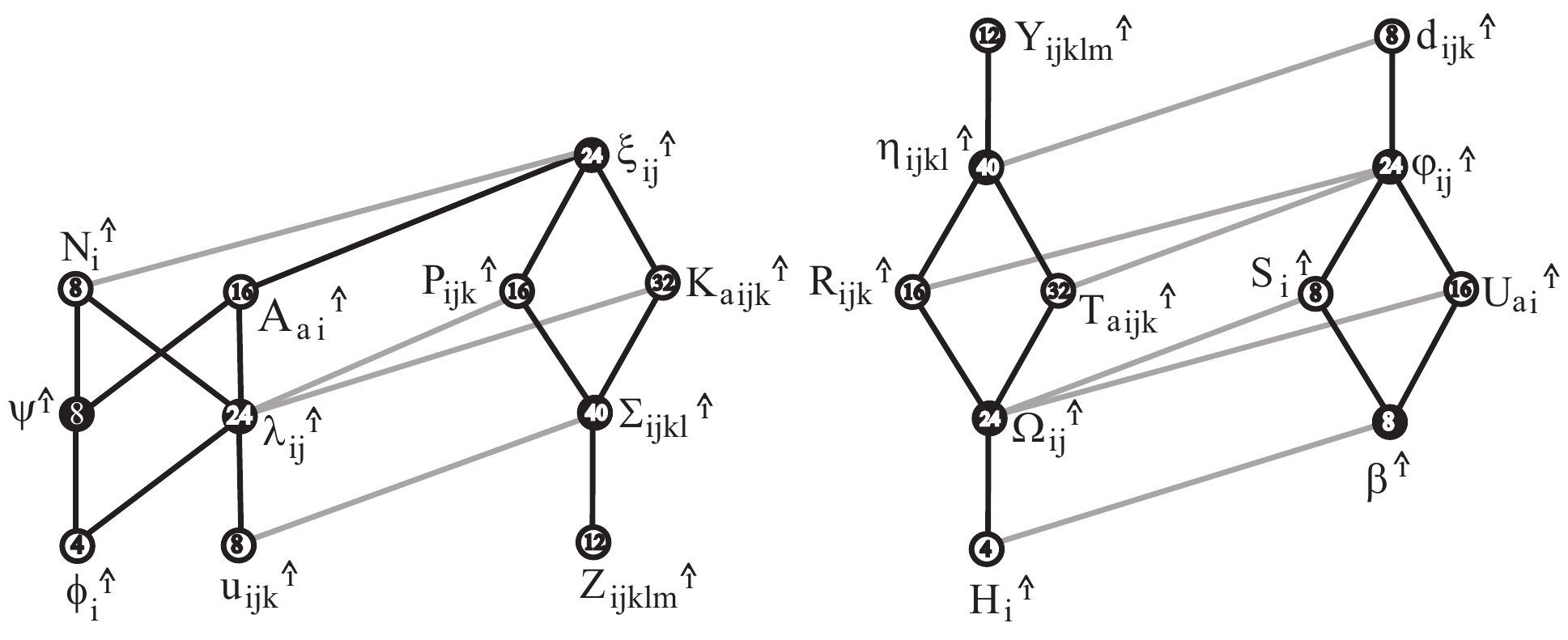}
\label{retmdsdq}\err
By virtue of judicious pairings, all component fields in (\ref{retmdsdq})
are auxiliary except for two hypermultiplet-like pairings,
$(\,\phi_i\,^{\hat{\imath}}\,,\,\xi^{\hat{\imath}}\,)$ and
$(\,H_i\,^{\hat{\imath}}\,,\,\beta^{\hat{\imath}}\,)$.
A supersymmetric action exists which facilitates this.
The transformation rules and the full supersymmetric action
are exhibited in \ref{hpac}.

A canonical form for the action, with diagonalized kinetic terms, is obtained by
reorganizing (or shifting) the component fields using linear re-definitions.  The precise
shifts and the complete form of the final action are also exhibited in \ref{hpac}.
The relevant part for our discussion is
\brr S &=& \int d^4x\,\bpl\,\fr12\,\phi_i\,^{\hat{\imath}}\,\Box\,\phi^i\,_{\hat{\imath}}
       -\fr12\,H_i\,^{\hat{\imath}}\,\Box H^i\,_{\hat{\imath}}
       +i\,\bar{\psi}^{\imath}\,\dslash\psi_{\hat{\imath}}
       -i\,\bar{\beta}^{\hat{\imath}}\,\dslash\beta_{\hat{\imath}}
       +\cdots \bpr
\label{theaction}\err
The ellipsis includes auxilary fields quadratically paired without any derivatives.
It is remarkable that an
action functional exists involving only hypermultiplet sectors, exhibiting supersymmetry with no
off-shell central charge.  It is curious that this result seems to require as Byzantine a construction as
that exhibited in (\ref{retmdsdq}).  Indeed, the RETM+DSDQ auxiliary field
complex is elaborate.  This might suggest an interesting complexity in generalized models involving
higher-order or non-linear couplings, as might be enabled by involving background multiplets of
the sort described below.

The most important aspect of (\ref{theaction}) is that the two hypermultiplet sectors appear
with opposite signs on their kinetic terms.  This seems to imply a saddle-point vacuum instability, a bottomless classical
energy spectra, negative norm states, or ghost-like propagators in the quantum theory.   Thus, unless we can
adequately address this matter, the RETM+DSDQ complex describes merely an appealing curiosity.

Similar anomalous signs have been harbingers of deeper physics in other contexts.  Most famous, perhaps,
is Dirac's resolution of negative energy solutions as a prediction of antimatter.  More recently,
instabilities similar to those found in (\ref{theaction}) were addressed in the context of cosmological models
\cite{G1,O1,O2},
where higher-order and higher-derivative interactions, motivated by Galilean invariance, conspire to dynamically
stabilize the vacuum against small perturbations.
\footnote{We have scrutinized generalized Hyperplet models using the most general fourth-order Galilean-invariant higher-derivative
interactions, and have not found evidence for an analogous stabilization mechanism. (This was done
collaboratively with Kevin Iga.)}

The most appealing observation concerning the ghost-like sectors in (\ref{theaction}) is facilitated
by comparison with Fayet Hypermultiplet models.  As explained above, a well-known way to couple
a hypermultiplet to ${\cal N}=2$ Poincar{\'e} supergravity is to formulate an analogous superconformal theory with two superconformal
Fayet Hypermultiplets, including one with wrong-sign kinetic terms, analogous
to (\ref{theaction}).  In that context, the dilatation symmetry, the $SU(2)_R$ symmetry, and the gauged $S$-supersymmetry
collectively inoculate the wrong-sign hypermultiplet since both the ghostlike quaternion and
its super partner fermion doublet become pure gauge. A similar mechanism
resolves the sign in the Hyperplet model described above.  With this in mind, it becomes incumbent on us to
develop a superconformal generalization of the Hyperplet. This endeavor occupies the
balance of this paper.

We should mention that in much earlier work \cite{HST}, the multiplet ${\cal M}_5$ was used to relax ${\cal M}_3$, and then
$\widetilde{\cal M}_5$ was coupled to provide relevant auxiliary fields.  Thus, the authors of that paper
used a Relaxed Tensor Multiplet (RTM) rather than a Relaxed Extended Tensor Multiplet (RETM) as a starting point. What results in
their case are propagating fields that configure as $(\,L_i\,^j\,,\,\psi_i\,)$, meaning that the bosons transform under
$SU(2)_R$ as ${\bf 2}\otimes {\bf 2}={\bf 1}\oplus{\bf 3}$, and the fermions transform as ${\bf 2}$.
There is no attendant $SU(2)_H$ symmetry.  Because of the symmetry discrepancy, that earlier work did
not describe bona fide hypermultiplets.  However, that work originally introduced the relaxation paradigm as
a means to evade off-shell central charges.
By contrast, we have used ${\cal M}_6\,^{\hat{1}}$ to relax ${\cal M}_3\,^{1,\hat{1}}$, and then
added $\widetilde{\cal M}_4\,^{\hat{1}}$ tethered to ${\cal M}_6\,^{\hat{1}}$ to provide extra
auxiliary fields.  Our approach has obtained proper hypermultiplet fields, albeit doubled, and
presenting an ostensible vacuum instability.  Our next goal is to attempt to reconcile the latter using
superconformal gravity as a mediator for absorbing the ghosts.

\setcounter{equation}{0}
\section{Hidden Isometries}
\label{hidis}
To construct a superconformal analog of the Hyperplet model described
above, we need, as a first step, to augment the
transformation rules for the RETM and the DSDQ to accommodate
the larger ${\cal N}=2$ superconformal algebra.   We should enhance all
derivatives to be covariant with respect to the full superconformal
algebra, using gauge potentials present in the ${\cal N}=2$ Weyl multiplet,
for example. And we should identify the proper scaling weights and the $S$ supersymmetry rules,
and augment the $Q$ supersymmetry rules to involve background
supergravity matter fields where required.
Moreover, we should make sure we properly understand the $U(1)_R\times SU(2)_R$
representations of the fields.  This last point supplies something of a surprise
in this analysis.

The superconformal algebra is generated by general coordinate transformations, along with
the Lorentz, dilatation, and special conformal generators $(\,M_{ab}\,,\,D\,,\,K_a\,)$,
supersymmetry generators $Q^i$ and $S^i$, and the $U(1)_R\times SU(2)_R$,
generators, specified as ${\cal A}$ and $G^I$, where $I=1,2,3$.
The conformal $S$ supersymmetry is characterized its
anti-commutator with $Q$ supersymmetry --- the superconformal algebra requires
that $\{Q,S\}$ is proportional to a Lorentz transformation plus a dilatation
plus a $U(1)_R$ and an $SU(2)_R$ transformation.  The generator $Q$ has
dimension one-half.  Thus, on an Adinkra, $Q$ generates upward maps and also downward maps
enabled by derivatives.  But $S$ has dimension minus one-half, and therefore only
generates downward maps.

Consider the field $Z_{ijklm}\,^{\hat{\imath}}$, located at the bottom right of the RETM Adinkra,
shown as the left diagram in (\ref{retmdsdq}).  Since the transformation we have called $SU(2)_R$ acts non-trivially on this
field, and since $Q$ maps $Z_{ijklm}\,^{\hat{\imath}}$ only to $\Sigma_{ijkl}\,^{\hat{\imath}}$,
it is essential that the spinor $\Sigma_{ijkl}\,^{\hat{\imath}}$ has an $S$-transformation
involving $Z_{ijklm}\,^{\hat{\imath}}$.  Since $u_{ijk}\,^{\hat{\imath}}$ transforms under $Q$
into $\Sigma_{ijkl}\,^{\hat{\imath}}$, it follows that $\{Q,S\}$ acting on $u_{ijk}\,^{\hat{\imath}}$
will produce a term involving $Z_{ijklm}\,^{\hat{\imath}}$.
This circumstance is puzzling, however, because, evidently, none of the operators
$M_{ab}$, $D$, $A$, or $G^I$ map $u_{ijk}\,^{\hat{\imath}}$ onto
$Z_{ijklm}\,^{\hat{\imath}}$.  Specifically, since $u_{ijk}\,^{\hat{\imath}}$ is a real Lorentz
scalar, it follows that $M_{ab}\,\phi={\cal A}\,\phi=0$.
Under a dilatation this field transforms as $\delta_D\,u_{ijk}\,^{\hat{\imath}}=w\,\Lambda_D\,u_{ijk}\,^{\hat{\imath}}$, where $w$ is
scaling weight of the RETM and $\Lambda_D$ is a dimensionless parameter.  Moreover, we have assumed that $u_{ijk}\,^{\hat{\imath}}$ transforms as
a symmetric rank three tensor ${\bf 4}$ under $SU(2)_R$, which would tell us
$\delta_R(\Theta)\,u_{ijk}\,^{\hat{\imath}}=-3\,\Theta^I\,(\,T_I\,)_{(i}\,^p\,u_{jk)p}\,^{\hat{\imath}}$,
using notation explained in \ref{su2rules}.  So there seems to be no way to assign $S$ transformations to the
component fields in a manner consistent with the superconformal Jacobi identities.

The way out of this quandary is to seek a non-manifest (hidden) isometry $SU(2)_A\times SU(2)_B$
associated with the RETM and the DSDQ, such that the group erstwhile called $SU(2)_R$ is a subgroup
of $SU(2)_A\times SU(2)_B$, and that $SU(2)_A$ is, in fact, the {\it rightful} $R$ symmetry.  (The rightful $R$ symmetry is the transformation
which appears in the $\{S,Q\}$ anticommutator, and which acts
on the supercharges $Q^i$ and $S^i$, such that these transform as doublets.) The factor $SU(2)_B$ is then a ``spectator" which commutes
with all superconformal generators.  Under this scenario, the isometry group associated with the Hyperplet
would be $U(1)_R\times SU(2)_A\times SU(2)_B\times SU(2)_H$.  This proposition can be tested as follows.

If there is a hidden (spectator) isometry $SU(2)_B$, then this must mix the component fields at each level
in (\ref{retmdsdq}).  For example, the fields
 $(\,\phi_i\,^{\hat{\imath}}\,,\,u_{ijk}\,^{\hat{\imath}}\,,\,Z_{ijklm}\,^{\hat{\imath}}\,)$
would transform amongst each other in a non-diagonal representation that respects the $SU(2)$
algebra and commutes with the supercharges.  By writing generic transformations as sensible generic
covariant linear combinations, this becomes a straightforward linear algebra problem.  Gratifyingly,
a unique non-trivial solution exists, which we exhibit presently.

\subsection{The $SU(2)_A\times SU(2)_B$ Automorphism of the RETM}
Using the RETM  transformation rules (\ref{retmrules}) and using the algorithm described above,
we find the following unique solution for the hidden $SU(2)_B$ automorphism.
The lowest-weight bosons transform as
\brr \delta_B\,\phi_i\,^{\hat{\imath}} &=&
      -\fr56\,\Theta_B^I\,(\,T_I\,)_l\,^k\,\bpl\,
      2\,\delta_i\,^l\,\phi_k\,^{\hat{\imath}}
      +3\,\ve^{lj}\,u_{ijk}\,^{\hat{\imath}}\,\bpr
      \nonumber\\[.1in]
      \delta_B\,u_{ijk}\,^{\hat{\imath}} &=&
      \fr{1}{15}\,\Theta_B^I\,(\,T_I\,)_p\,^m\,\bpl\,
      10\,\delta_{(i}\,^p\,\phi_j\,^{\hat{\imath}}\,\ve_{k)m}
      -33\,\delta_{(i}\,^p\,u_{jk)m}\,^{\hat{\imath}}
      -48\,\ve^{pl}\,Z_{ijkml}\,^{\hat{\imath}}\,\bpr
      \nonumber\\[.1in]
      \delta_B\,Z_{ijklm}\,^{\hat{\imath}} &=&
      \fr38\,\Theta_B^I\,(\,T_I\,)_{(i}\,^n\,\bpl\,
      u_{jkl}\,^{\hat{\imath}}\,\ve_{m)n}
      -8\,Z_{jklm)n}\,^{\hat{\imath}}\,\bpr \,,
 \err
 which describes the representation ${\bf 4}\oplus{\bf 4}\oplus{\bf 4}$.
 The level-2 fermions transform as
 \brr \delta_B\,\psi^{\hat{\imath}} &=&
      -2\,\Theta_B^I\,\ve^{ij}\,(\,T_I\,)_i\,^k\,\lambda_{jk}\,^{\hat{\imath}}
      \nonumber\\[.1in]
      \delta_B\,\lambda_{ij}\,^{\hat{\imath}} &=&
      -\fr13\,\Theta_B^I\,(\,T_I\,)_m\,^l\,\bpl\,
      2\,\ve_{l(i}\,\delta_{j)}\,^m\,\psi^{\hat{\imath}}
      +3\,\delta_{(i}\,^m\,\lambda_{j)l}\,^{\hat{\imath}}
      +8\,\ve^{mk}\,\Sigma^*_{ijkl}\,^{\hat{\imath}}\,\bpr
      \nonumber\\[.1in]
      \delta_B\,\Sigma^*_{ijkl}\,^{\hat{\imath}} &=&
      \fr18\,\Theta_B^I\,(\,T_I\,)_{(i}\,^m\,\bpl\,
      3\,\lambda_{jk}\,^{\hat{\imath}}\,\ve_{l)m}
      -16\,\Sigma^*_{jkl)m}\,^{\hat{\imath}}\,\bpr \,,
 \err
 which describes a ${\bf 3}\oplus{\bf 3}\oplus{\bf 3}$ representation.
 \footnote{Since complex conjugation toggles index placement, we define
 a space-saving ``star" notation typified by the expression $\Sigma^*_{ijkl}\,^{\hat{\imath}}=\ve_{im}\,\ve_{jn}\,\ve_{jp}\,\ve_{kq}\,\ve^{\hat{\imath}\hat{\jmath}}\,\Sigma^{mnpq}\,_{\hat{\jmath}}$.
 in general, a star is a complex conjugation with all indices contracted with Levi-Civita tensors.}
 The level-3 bosons transform as
 \brr \delta_B\,N_i\,^{\hat{\imath}} &=&
      \fr13\,\Theta_B^I\,(\,T_I\,)_l\,^j\,\bpl\,
      \delta_i\,^l\,N_j\,^{\hat{\imath}}
      +8\,\ve^{lk}\,P^*_{ijk}\,^{\hat{\imath}}\,\bpr
      \nonumber\\[.1in]
      \delta_B\,P^*_{ijk}\,^{\hat{\imath}} &=&
      -\fr14\,\Theta_B^I\,(\,T_I\,)_{(i}\,^l\,\bpl\,
      N_j\,^{\hat{\imath}}\,\ve_{k)l}
      +4\,P^*_{jk)l}\,^{\hat{\imath}} \,\bpr
 \err
 which describes a ${\bf 2}\oplus{\bf 2}\oplus{\bf 2}$ representation, and
 \brr \delta_B\,A_{a\,i}\,^{\hat{\imath}} &=&
      \fr13\,\Theta_B^I\,(\,T_I\,)_l\,^j\,\bpl\,
      \delta_i\,^l\,A_{a\,j}\,^{\hat{\imath}}
      -8\,\ve^{lk}\,K_{a\,ijk}\,^{\hat{\imath}}\,\bpr
      \nonumber\\[.1in]
      \delta_B\,K_{a\,ijk}\,^{\hat{\imath}} &=&
      \fr14\,\Theta_B^I\,(\,T_I\,)_{(i}\,^l\,\bpl\,
      A_{a\,j}\,^{\hat{\imath}}\,\ve_{k)l}
      -4\,K_{a\,jk)l}\,^{\hat{\imath}} \,\bpr
 \err
 which also describes a ${\bf 2}+{\bf 2}+{\bf 2}$ representation.
 Finally, the level-4 fermions are $SU(2)_B$ singlets,
 \brr \delta_B\,\xi^*_{ij}\,^{\hat{\imath}} &=& 0 \,,
 \label{dbxi}\err
 which describes a ${\bf 1}\oplus{\bf 1}\oplus{\bf 1}$ representation under $SU(2)_B$.
 The specified representation assignments are readily determined by translating any one of the three
$SU(2)_B$ generators into a square matrix that acts on a column comprised of each component.
The eigenvalues of that matrix unambiguously parse as a group of sets $\{-J,...,+J\}$ where $J$ is
an integer or a half-integer. The representation is a direct sum of sub-representations, one associated
with each eigenvalue set. Those with integer $J$ correspond a dimension $2\,J+1$ representaton, while
those with half-integer $J$ correspond to a dimension $2\,J$ representation.

Since the isometry previously called $SU(2)_R$, which acts diagonally, {\it i.e.} as a tensor operation
on the components as shown in (\ref{retmdsdq}), is not the proper $R$ symmetry in the superconformal
context, we will refer to that subgroup henceforth as $SU(2)_C$.
Using the $SU(2)_B$ transformation rules shown above, we find that the transformation
$\delta_C$ does not commute with $\delta_B$.  Instead,
$[\,\delta_C(\theta)\,,\,\delta_B(\theta)\,]$ is a $\delta_B$ transformation, such that
the difference $\delta_C-\delta_B$ commutes with $\delta_B$ when acting on
any of the components.  Thus, the ``diagonal"
$SU(2)$ transformations are
\brr \delta_A(\theta) &=& \delta_C(\theta)-\delta_B(\theta)
     \nonumber\\[.1in]
     \delta_B(\theta_B) &=& \delta_B(\theta_B) \,.
\err
and we identify $SU(2)_A$ as the proper $R$-symmetry. It is now straightforward to
re-structure the RETM so that all of its isometries become manifest.
We sometimes refer to $SU(2)_A$ as $SU(2)_{\cal R}$, where the calligraphic subscript
distinguishes this subgroup from the erstwhile $R$ symmetry $SU(2)_R\cong SU(2)_C$.

Since each of the $SU(2)_B$ representations spanned by RETM components are triple sums, and since
$SU(2)_A$ commutes with $SU(2)_B$, we conclude that the
$SU(2)_A\times SU(2)_B$ representations, and their branching
under $SU(2)_A\times SU(2)_B\to SU(2)_C$ are
\brr {\bf 1}\otimes{\bf 3} &\to& {\bf 3}
     \hspace{.9in}
     \xi_{ij}\,^{\hat{\imath}}
     \nonumber\\[.1in]
     {\bf 2}\otimes{\bf 3} &\to& {\bf 2}\oplus{\bf 4}
     \hspace{.6in}
     (\,N_i\,^{\hat{\imath}}\,,\,P_{ijk}\,^{\hat{\imath}}\,) \,\,{\rm and}\,\,
     (\,A_{a\,i}\,^{\hat{\imath}}\,,\,K_{a\,ijk}\,^{\hat{\imath}}\,)
     \nonumber\\[.1in]
     {\bf 3}\otimes{\bf 3} &\to& {\bf 1}\oplus{\bf 3}\oplus{\bf 5}
     \hspace{.3in}
     (\,\psi^{\hat{\imath}} \,,\,\lambda_{ij}\,^{\hat{\imath}}\,,\,\Sigma_{ijkl}\,^{\hat{\imath}}\,)
     \nonumber\\[.1in]
     {\bf 4}\otimes{\bf 3} &\to& {\bf 2}\oplus{\bf 4}\oplus{\bf 6}
     \hspace{.3in}
     (\,\phi_i\,^{\hat{\imath}}\,,\,u_{ijk}\,^{\hat{\imath}}\,,\,Z_{ijklm}\,^{\hat{\imath}}\,)
\label{rbranch}\err
where the relevant components are identified to the right, in the frame where
the $SU(2)_C$ representation is manifest but the $SU(2)_A\times SU(2)_B$ representation is
not.  With (\ref{rbranch}) as a guide, we can diagrammatically
re-configure the RETM as
\begin{center}
\includegraphics[width=2in]{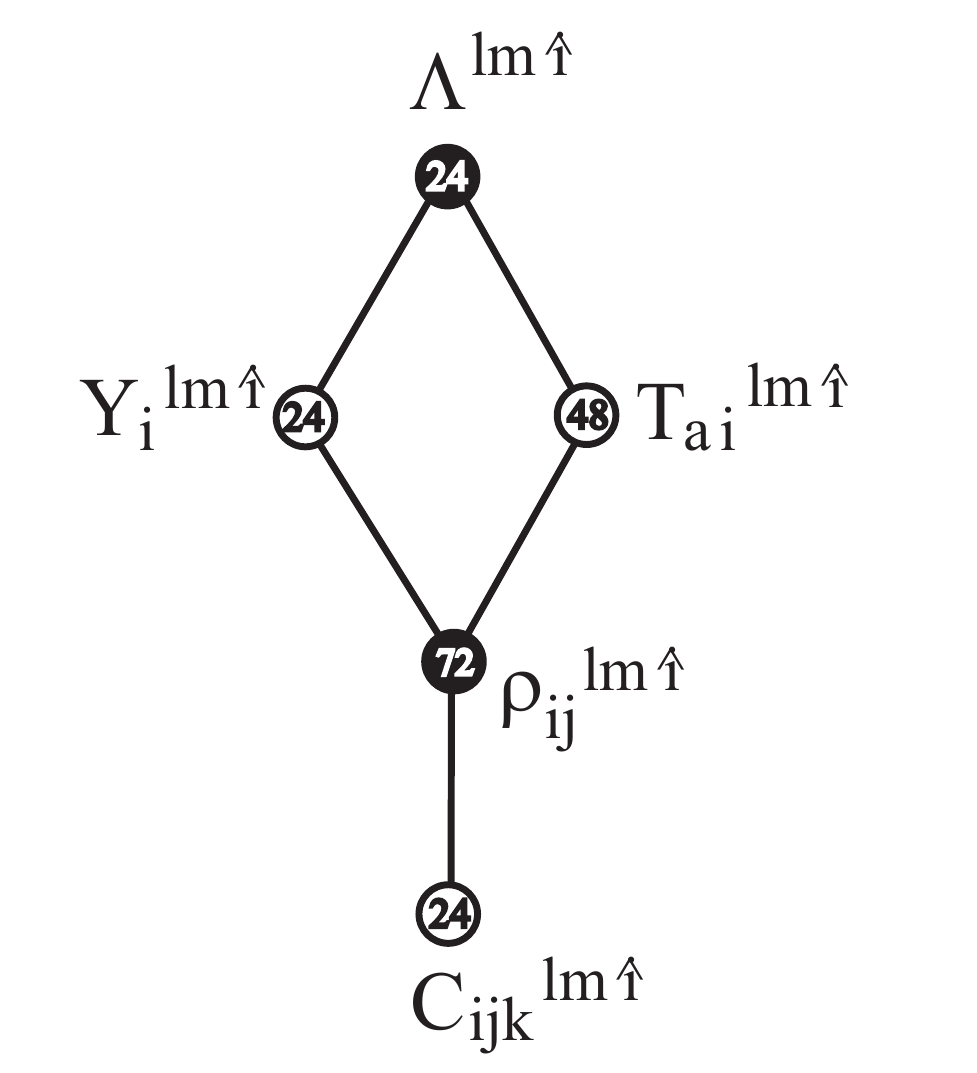}
\label{retmcoz}\end{center}
where the new component names manifestly exhibit all of the isometries according to the following scheme.
The group $SU(2)_A$ acts only on the lower indices while the group $SU(2)_B$ acts only on the
undecorated upper indices.  For example, this means that
\brr  \delta_A(\Theta_A)\,C_{ijk}\,^{lm\,\hat{\imath}} &=&
     -3\,\Theta^I_A\,(\,T_I\,)_{(i}\,^p\,C_{jk)p}\,^{lm\,\hat{\imath}}
     \nonumber\\[.1in]
     \delta_B(\Theta_B)\,C_{ijk}\,^{lm\,\hat{\imath}} &=&
     +2\,\Theta^I_B\,(\,T_I\,)_p\,^{(l}\,C_{ijk}\,^{m)p\,\hat{\imath}} \,.
\label{abtrans}\err
The group $SU(2)_C$ is the diagonal subgroup, which acts on {\it all} indices.
\footnote{We could have distinguished the upper $SU(2)_B$ indices using a new diacritical mark, for example by writing
$C_{ijk}\,^{\check{l}\check{m}\,\hat{\imath}}$.  But that notation would be cumbersome
and unnecessary.  Moreover, as written the components are proper $SU(2)_C$ tensors,
and the $SU(2)_A\times SU(2)_B$ structure is easy to read.  Note that $SU(2)_B$ indices
are always on the same level as hatted $SU(2)_H$ indices, and $SU(2)_A$ indices are always on
the opposite level compared to hatted $SU(2)_H$ indices.}
Thus, we have $\delta_C(\theta_C)=\delta_A(\theta_C)+\delta_B(\theta_C)$,
so that
\brr  \delta_C(\Theta_C)\,C_{ijk}\,^{lm\,\hat{\imath}} &=&
     -3\,\Theta^I_C\,(\,T_I\,)_{(i}\,^p\,C_{jk)p}\,^{lm\,\hat{\imath}}
     +2\,\Theta^I_C\,(\,T_I\,)_p\,^{(l}\,C_{ijk}\,^{m)p\,\hat{\imath}} \,.
\err
The relationship between the new RETM guise and the old RETM guise is clarified
by the following decompositions
\brr C_{ijk}\,^{lm\hat{\imath}} &=&
     \delta_{(i}\,^l\,\delta_j\,^m\,\phi_{k)}\,^{\hat{\imath}}
     +\ve^{p(l}\,\delta_{(i}\,^{m)}\,u_{jk)p}\,^{\hat{\imath}}
     +\ve^{lp}\,\ve^{mq}\,Z_{ijkpq}\,^{\hat{\imath}}
     \nonumber\\[.1in]
     \rho_{ij}\,^{kl\hat{\imath}} &=&
     \delta_{(i}\,^k\,\delta_{j)}\,^l\,\psi^{\hat{\imath}}
     +\ve^{m(k}\,\delta_{(i}\,^{l)}\,\lambda_{j)m}\,^{\hat{\imath}}
     +\ve^{km}\,\ve^{ln}\,\Sigma^*_{ijmn}\,^{\hat{\imath}}
     \nonumber\\[.1in]
     Y_i\,^{jk\hat{\imath}} &=&
     \ve^{m(j}\,\delta_i\,^{k)}\,N_m\,^{\hat{\imath}}
     +\ve^{jm}\,\ve^{kn}\,P^*_{imn}\,^{\hat{\imath}}
     \nonumber\\[.1in]
     T_{a\,i}\,^{jk\hat{\imath}} &=&
     \ve^{m(i}\,\delta_i\,^{k)}\,A_{a\,m}\,^{\hat{\imath}}
     +\ve^{jm}\,\ve^{kn}\,K_{a\,imn}\,^{\hat{\imath}}
     \nonumber\\[.1in]
     \Lambda^{ij\hat{\imath}} &=&
     \ve^{im}\,\ve^{jn}\,\xi_{mn}\,^{\hat{\imath}} \,.
\label{retmdecom}\err
The natural tensor structures exposed by this analysis indicate that the RETM, which we originally constructed as a
tethered combination of two multiplets, ${\cal M}_3\,^{2,\hat{2}}+{\cal M}_6\,^{\hat{2}}$,
in fact corresponds to ${\cal M}_4\,^{3,\hat{2}}$.  This exposes some multiplet alchemy
which might have an amusing mathematical context.
The supersymmetry transformation rules in terms of the natural components (\ref{retmcoz}) are
shown below in Section \ref{confy}.

\subsection{The $SU(2)_A\times SU(2)_B$ Automorphism of the DSDQ}
Using the DSDQ transformation rules and using the algorithm described above, we find
the following unique solution for the hidden $SU(2)_B$ automorphism of the DSDQ.
Under $SU(2)_B$, the level-1 bosons are singlets,
 \brr \delta_B\,H_i\,^{\hat{\imath}} &=& 0 \,,
\err
corresponding to a ${\bf 1}\oplus{\bf 1}$ representation.
The level-2 fermions transform as
\brr  \delta_C\,\beta^{\hat{\imath}} &=&
      \fr13\,\Theta_C^I\,\ve^{ij}\,(\,T_I\,)_i\,^k\,\Omega^*_{jk}\,^{\hat{\imath}}
      \nonumber\\[.1in]
       \delta_C\,\Omega^*_{ij}\,^{\hat{\imath}} &=&
      -\fr12\,\Theta_C^I\,(\,T_I\,)_{(i}\,^k\,\bpl\,
      3\,\ve_{j)k}\,\beta^{\hat{\imath}}
      +2\,\Omega^*_{j)k}\,^{\hat{\imath}}\,\bpr \,.
 \err
 This describes a ${\bf 2}\oplus{\bf 2}$ representation.
 The level-3 bosons transform as
 \brr  \delta_C\,S_i\,^{\hat{\imath}} &=&
      -\fr13\,\Theta_C^I\,(\,T_I\,)_l\,^j\,\bpl
      4\,\delta_i\,^l\,S_j\,^{\hat{\imath}}
     -\ve^{lk}\,R^*_{ijk}\,^{\hat{\imath}}\,\bpr
     \nonumber\\[.1in]
     \delta_C\,R^*_{ijk}\,^{\hat{\imath}} &=&
     -2\,\Theta_C^I\,(\,T_I\,)_{(i}\,^l\,\bpl\,
     S_{j}\,^{\hat{\imath}}\,\ve_{k)l}
     +R^*_{jk)l}\,^{\hat{\imath}}\,\bpr \,,
\err
which is a ${\bf 3}\oplus{\bf 3}$ representation, and
\brr  \delta_C\,U_{a\,i}\,^{\hat{\imath}} &=&
      -\fr13\,\Theta_C^I\,(\,T_I\,)_l\,^j\,\bpl\,
      4\,\delta_i\,^l\,U_{a\,j}\,^{\hat{\imath}}
     +\ve^{lk}\,T_{a\,ijk}\,^{\hat{\imath}}\,\bpr
     \nonumber\\[.1in]
     \delta_C\,T_{a\,ijk}\,^{\hat{\imath}} &=& 2\,\Theta_C^I\,(\,T_I\,)_{(i}\,^l\,\bpl\,
     U_{a\,j}\,^{\hat{\imath}}\,\ve_{k)l}
     -T_{a\,jk)l}\,^{\hat{\imath}}\,\bpr \,.
\err
which is also a ${\bf 3}\oplus{\bf 3}$ representation.
The level-4 fermions transform as
 \brr \delta_C\,\varphi_{ij}\,^{\hat{\imath}} &=&
      -\fr16\,\Theta_C^I\,(\,T_I\,)_m\,^k\,\bpl
      15\,\delta_{(i}\,^m\,\varphi_{j)k}\,^{\hat{\imath}}
      -\ve^{ml}\,\eta^*_{ijkl}\,^{\hat{\imath}}\,\bpr
      \nonumber\\[.1in]
      \delta_C\,\eta^*_{ijkl}\,^{\hat{\imath}} &=&
      -\fr32\,\Theta_C^I\,(\,T_I\,)_{(i}\,^m\,\bpl\,
      3\,\varphi_{jk}\,^{\hat{\imath}}\,\ve_{l)m}
      +2\,\eta^*_{jkl)m}\,^{\hat{\imath}}\,\bpr \,,
 \err
 which is a ${\bf 4}\oplus{\bf 4}$ representation.
 Finally, the level-5 bosons transform as
 \brr \delta_C\,d_{ijk}\,^{\hat{\imath}} &=&
     -\Theta_C^I\,(\,T_I\,)_m\,^l\,\bpl\,
     \fr{18}{5}\,\delta_{(i}\,^m\,d_{jk)l}\,^{\hat{\imath}}
     +\fr16\,\ve^{mn}\,Y_{ijkln}\,^{\hat{\imath}}\,\bpr
     \nonumber\\[.1in]
     \delta_C\,Y_{ijklm}\,^{\hat{\imath}} &=&
     \fr45\,\Theta_C^I\,(\,T_I\,)_{(i}\,^n\,\bpl\,
     6\,d_{jkl}\,^{\hat{\imath}}\,\ve_{m)n}
     -5\,Y_{jklm)n}\,^{\hat{\imath}}\,\bpr
\err
which is a ${\bf 5}\oplus{\bf 5}$ representation.  The specified representations are
determined using the method explained after equation (\ref{dbxi}).
The relationship between the isometries $SU(2)_A$, $SU(2)_B$ and $SU(2)_C$ are similar for
the DSDQ as in the RETM, described above.  Namely, $SU(2)_C$ is the diagonal subgroup of $SU(2)_A\times SU(2)_B$.

Since each of the $SU(2)_B$ representations spanned by DSDQ components are triple sums, and since
$SU(2)_A$ commutes with $SU(2)_B$, we conclude that the
$SU(2)_A\times SU(2)_B$ representations, and their branching
under $SU(2)_A\times SU(2)_B\to SU(2)_C$ are
\brr {\bf 5}\otimes{\bf 2} &\to& {\bf 4}\oplus{\bf 6}
     \hspace{.5in}
     (\,d_{ijk}\,^{\hat{\imath}}\,,\,Y_{ijklm}\,^{\hat{\imath}}\,)
     \nonumber\\[.1in]
     {\bf 4}\otimes{\bf 2} &\to& {\bf 3}\oplus{\bf 5}
     \hspace{.5in}
     (\,\varphi_{ij}\,^{\hat{\imath}}\,,\,\eta_{ijkl}\,^{\hat{\imath}}\,)
     \nonumber\\[.1in]
     {\bf 3}\otimes{\bf 2} &\to& {\bf 4}\oplus{\bf 2}
     \hspace{.5in}
     (\,R_{ijk}\,^{\hat{\imath}}\,,\,S_i\,^{\hat{\imath}}\,) \,\,{\rm and}\,\,
     (\,T_{a\,ijk}\,^{\hat{\imath}}\,,\,U_{a\,i}\,^{\hat{\imath}}\,)
     \nonumber\\[.1in]
     {\bf 2}\otimes{\bf 2} &\to& {\bf 3}\oplus{\bf 1}
     \hspace{.5in}
     (\,\Omega_{ij}\,^{\hat{\imath}} \,,\,\beta^{\hat{\imath}}\,)
     \nonumber\\[.1in]
     {\bf 1}\otimes{\bf 2} &\to& {\bf 2}
     \hspace{.8in}
     H_i\,^{\hat{\imath}}
\label{dbranch}\err
where the relevant components are identified to the right, in the frame where
the $SU(2)_C$ representation is manifest but the $SU(2)_A\times SU(2)_B$ representation is
not.  With (\ref{rbranch}) as a guide, we can diagrammatically
re-configure the DSDQ as
\begin{center}
\includegraphics[width=2in]{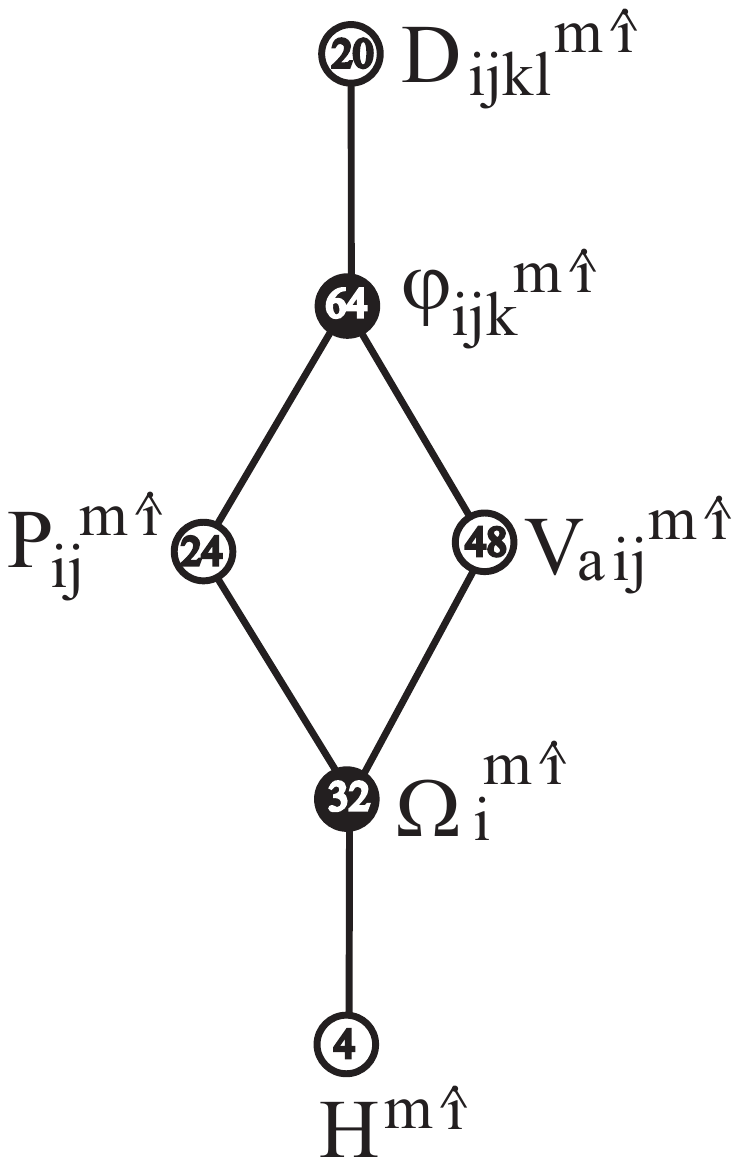}
\label{dsdqcoz}\end{center}
where, as above $SU(2)_A$ acts on the lower indices and $SU(2)_B$ acts on the upper
unadorned indices.  $SU(2)_C$ acts on all unadorned indices.

The relationship between the new DSDQ guise an the old DSDQ guise is clarified
by the following decompositions
 \brr H^{i\hat{\imath}} &=& \ve^{ij}\,H_j\,^{\hat{\imath}}
      \nonumber\\[.1in]
      \Omega^i\,_{j\hat{\imath}} &=&
      \fr32\,\delta^i\,_j\,\beta_{\hat{\imath}}
      -\ve_{jk}\,\Omega^{*\,ik}\,_{\hat{\imath}}
      \nonumber\\[.1in]
      P^{ij}\,_{k\hat{\imath}} &=&
      \delta^{(i}\,_k\,S^{j)}\,_{\hat{\imath}}
      -\fr12\,\ve_{km}\,R^{*\,mij}\,_{\hat{\imath}}
      \nonumber\\[.1in]
      V^{a\,ij}\,_{k\hat{\imath}} &=&
      -2\,\delta^{(i}\,_k\,U^{a\,j)}\,_{\hat{\imath}}
      -\ve_{km}\,T^{a\,mij}\,_{\hat{\imath}}
      \nonumber\\[.1in]
      \varphi^{ijk}\,_{l\hat{\imath}} &=&
      \fr94\,\delta^{(i}\,_l\,\varphi^{jk)}\,_{\hat{\imath}}
      -\fr12\,\ve_{lm}\,\eta^{*\,mijk}\,_{\hat{\imath}}
      \nonumber\\[.1in]
      D^{ijkl}\,_{m\hat{\imath}} &=&
      \fr{12}{5}\,\delta^{(i}\,_m\,d^{jkl)}\,_{\hat{\imath}}
      +\fr12\,\ve_{mn}\,Y^{nijk}\,_{\hat{\imath}} \,.
\label{dsdqdecom}\err
The natural tensor structures exposed by this analysis indicate that the DSDQ,
which we originally constructed as a
tethered combination of two multiplets, $\widetilde{\cal M}_6\,^{2}+\widetilde{\cal M}_4\,^{\hat{2}}$,
in fact corresponds to $\widetilde{\cal M}_5\,^{2,\hat{2}}$.
The supersymmetry transformation rules in terms of the natural components (\ref{dsdqcoz}) are
shown below in Section \ref{confy}.

It is possible to form a DSDQ from the components of the RETM, by making the following
identifications,
\brr H^{i\hat{\imath}} &=& \fr12\,\ve^{ij}\,C_{jkl}\,^{kl\hat{\imath}}
      \nonumber\\[.1in]
      \Omega^i\,_{j\hat{\imath}} &=&
      -\fr12\,\delta^i\,_j\,\rho^{mn}\,_{mn\hat{\imath}}
      +\fr13\,\rho^{im}\,_{jm\hat{\imath}}
      \nonumber\\[.1in]
      P^{ij}\,_{k\hat{\imath}} &=&
      -\fr{1}{12}\,\ve^{m(i}\,Y^{j)}\,_{km\hat{\imath}}
      -\fr14\,\delta_k\,^{(i}\,\ve^{j)m}\,Y^n\,_{mn\hat{\imath}}
      \nonumber\\[.1in]
      V^{a\,ij}\,_{k\hat{\imath}} &=&
      \fr12\,\delta_k\,^{(i}\,\ve^{j)m}\,T^{a\,n}\,_{mn\hat{\imath}}
      +\fr16\,\ve^{m(i}\,T^{a\,j)}\,_{km\hat{\imath}}
      +\fr13\,\delta_k\,^{(i}\,\der^a C^{j)mn}\,_{mn\hat{\imath}}
      +\fr49\,\der^a C^{ijm}\,_{km\hat{\imath}}
      \nonumber\\[.1in]
      \varphi^{ijk}\,_{l\hat{\imath}} &=&
      \fr14\,\ve^{m(i}\,\ve^{|n|j}\,\delta_l\,^{k)}\,\Lambda^*_{mn\hat{\imath}}
      +\fr{5}{12}\,\ve^{m(i}\,\delta_l\,^j\,\rho^{k)n}\,_{mn\hat{\imath}}
      -\fr16\,\dslash\Sigma^{*\,mijk}\,_{\hat{\imath}}
      \nonumber\\[.1in]
      D^{ijkl}\,_{m\hat{\imath}} &=&
      \fr13\,\ve^{p(i}\,\Box\,C^{jkl)}\,_{mq\hat{\imath}}
      -\ve^{p(i}\,\delta_m\,^j\,\Box\,C^{kl)q}\,_{pq\hat{\imath}}
      +\ve^{p(i}\,\ve^{|q|j}\,\delta_m\,^k\,\der_a T^{a\,l)}\,_{pq\hat{\imath}}
 \label{rfromd}\err
 It is not possible to define a DSDQ using the components of a RETM, since the
 RETM is ``shorter" than the DSDQ in the sense that its components span fewer
 distinct dimensions.

\subsection{The globally supersymmetric action}
The supersymmetry transformation rules for the RETM and DSDQ can be derived from the diagrams
(\ref{retmcoz}) and (\ref{dsdqcoz}).
Using these, we find that the following tensor transforms under global $Q$ supersymmetry into a total derivative,
\footnote{The precise transformation rules are presented in the following section
with extra details related to superconfomal generalizations. For the simpler case of global
supersymmetry, we should replace the covariant deriviatives $D_\mu$ with
ordinary derivatives $\der_\mu$, ignore all terms involving
background supergravity matter fields $(\,T_{ab}\,,\,\chi^i\,,\,D\,)$, and also
ignore the $S$ supersymmetry transformations parameterized by $\eta^i$.}
\brr {\cal I}^{ij} &=&
      -\fr12\,D^{ikln}\,_{m\hat{\imath}}\,C_{kln}\,^{jm\hat{\imath}}
      -\fr12\,\ve^{im}\,T_{a\,m}\,^{jn\hat{\imath}}\,\der^a H_{n\hat{\imath}}
      +\fr14\,\ve^{im}\,V^{a\,kl}\,_{n\hat{\imath}}\,\der_a C_{mkl}\,^{jn\hat{\imath}}
      \nonumber\\[.1in]
      & &
      -\fr12\,P^{im}\,_{n\hat{\imath}}\,Y_{m}\,^{jn\hat{\imath}}
      -\fr12\,P^{*\,im}\,_{n\hat{\imath}}\,Y^*_{m}\,^{jn\hat{\imath}}
      -\fr12\,V^{a\,im}\,_{m\hat{\imath}}\,T_{a\,n}\,^{jn\hat{\imath}}
      \nonumber\\[.1in]
      & &
      +\bpl\,i\,\bar{\varphi}^{imn}\,_{l\hat{\imath}}\,\rho_{mn}\,^{jl\hat{\imath}}
      +\fr12\,i\,\bar{\Omega}^{i}\,_{m\hat{\imath}}\,\Lambda^{*\,jm\hat{\imath}}
      +\fr13\,i\,\ve^{im}\,\bar{\Omega}^{n}\,_{l\hat{\imath}}\,\dslash \rho_{mn}\,^{jl\hat{\imath}}
      +{\rm h.c.}\,\bpr \,.
 \label{iij}\err
 Thus, we can form a globally supersymmetric action by writing
 \brr S &=& \int d^4x\,\ve_{ij}\,{\cal I}^{ij} \,.
 \label{glosac}\err
 Notice that the first index on ${\cal I}^{ij}$ corresponds to $SU(2)_A$ and the second corresponds
 to $SU(2)_B$, while both indices correspond to $SU(2)_C$.
 Thus, ${\cal I}^{ij}$ transforms as
 a vector {\bf 2} under each of $SU(2)_A$ and $SU(2)_B$ but as a rank two tensor ${\bf 2}\otimes{\bf 2}={\bf 1}\oplus{\bf 3}$ under $SU(2)_C$.
 The contraction $\ve_{ij}\,{\cal I}^{ij}$ extracts the $SU(2)_C$ singlet, so that the action is, in fact
 $SU(2)_C$ invariant.  In the context of global supersymmetry it is not problematic that the multiplets possess a larger global isometry
 group $SU(2)_A\times SU(2)_B$ than the symmetry exhibited by the action $SU(2)_C\subset SU(2)_A\times SU(2)_B$.  In fact, using the
 decompositions (\ref{retmdecom}) and (\ref{dsdqdecom}), the action (\ref{glosac}) is identical to the final six lines of (\ref{retmdsdqaction}).
 As explained in \ref{hpac}, this yields, using particular field re-definitions, the component action (\ref{theaction}) discussed above.
 Based on our previous discussions, we conclude that (\ref{glosac}) suffers from a saddle-point vacuum instability and ghost-like
 propagators.  To resolve these, we proceed, using the philosophy discussed above, to couple this model to conformal supergravity
 as a means of inoculating the ghosts.

\setcounter{equation}{0}
\section{The Conformal Hyperplet}
\label{confy}
An efficient way to couple ${\cal N}=2$ gauge and matter fields to Poincar{\'e} supergravity
is by generalizing these to represent the larger ${\cal N}=2$
superconformal algebra, and then using select multiplets as compensators to fix the
extra generators.  A key proposal in this paper is that the Conformal Hyperplet, given by
superconformal analogs of the RETM and DSDQ provides a novel superconformal compensator,
enabling a new off-shell description of ${\cal N}=2$ supergravity.
By rendering the Hyperplet superconformally, the ghostlike sector, which seems problematic
in the globally supersymmetric theory,
becomes pure gauge, leaving exactly one propagating hypermultiplet.  This theory
has no off-shell central charge, by construction.

In earlier sections we assigned the dimensions of the RETM and the DSDQ multiplets to unity based considerations of our
action, which was quadratic in component fields.  Since our constructions were not conformal, we
had the freedom to adjust those dimensions unencumbered by Jacobi identities involving local dilatations.
In the superconformal theory we need to be more careful.
In fact, in the absence of extra background fields, the scaling weights of the superconformal RETM and DSDQ turn out to be three
and minus two, respectively. The requirement of an appropriate dimension four supersymmetric density function requires that we introduce
Vector multiplets, which may gauge the $U(1)_R$ and/or the $SU(2)_B$ isometries, to provide new dimension one
scalar fields to adjust multiplet weights via multiplication.  (These also provide candidates for a second conformal compensator.)

As a first step, we augment the RETM and DSDQ transformation rules by adding all possible $S$ supersymmetry transformations
consistent with Lorentz and $U(1)_R\times SU(2)_A\times SU(2)_B$ covariance, leaving coefficients and Weyl weights as variables to
be determined.  Then we impose the $[\delta_Q,\delta_S]$
part of the supergravity algebra, shown in \ref{n2algebra}.  Imposing the algebra determines the $S$ transformations
and the Weyl weights.

For example, suppose the Weyl weight of
the RETM  is ${\red w}$, meaning that its lowest component transforms under a dilatation as
$\delta_D\,C_{ijk}\,^{lm\,\hat{\imath}}={\red w}\,\Lambda_D\,C_{ijk}\,^{lm\,\hat{\imath}}$.  Since that component
is a real Lorentz scalar, it follows that $\delta_D\,C_{ijk}\,^{lm\,\hat{\imath}}=\delta_{\cal A}\,C_{ijk}\,^{lm\,\hat{\imath}}=0$.
Also, since the $SU(2)_{\cal R}$ symmetry must correspond to $SU(2)_A$ as explained above, we can use
the first equation in (\ref{abtrans}), along with the algebra requirements in (\ref{sg1sq}), and the third identity in
(\ref{su2ids}) to determine that the supergravity algebra imposes
\brr [\,\delta_S\,,\,\delta_Q\,]\,C_{ijk}\,^{lm\,\hat{\imath}} &=&
     i\,\bpl\,(\,{\red w}+3\,)\,\bar{\e}^m\,\eta_m+(\,{\red w}-3\,)\,\bar{\e}_m\,\eta^m\,\bpr\,C_{ijk}\,^{lm\,\hat{\imath}}
     \nonumber\\[.1in]
     & & +6\,i\,\bpl\,\bar{\e}_{(i}\,\eta^m-i\,\bar{\e}^m\,\eta_{(i}\,\bpr\,C_{jk)m}\,^{lm\,\hat{\imath}} \,.
\err
Given the $Q$ supersymmetry transformation of $C_{ijk}\,^{lm\,\hat{\imath}}$ and the only possible covariant $S$ supersymmetry transformation
rule for $\rho_{ij}\,^{lm\,\hat{\imath}}$, a small computation tells us that ${\red w}=3$ and
$\delta_S\,\rho_{ij}\,^{lm\,\hat{\imath}}=6\,C_{ijk}\,^{lm\,\hat{\imath}}\,\eta^k$.

Using similar considerations on all of the RETM component fields, we determine the following
$Q$ and $S$ transformations
\brr \delta\,C_{ijk}\,^{lm\hat{\imath}} &=&
      i\,\bar{\e}_{(i}\,\rho_{jk)}\,^{lm\hat{\imath}}
      -i\,\ve_{p(i}\,\bar{\e}^p\,\rho^*_{jk)}\,^{lm\hat{\imath}}
      \nonumber\\[.1in]
      \delta\,\rho_{ij}\,^{lm\hat{\imath}} &=&
      {\Dslash} C_{ijk}\,^{lm\hat{\imath}}\,\e^k
      +\fr12\,Y_{(i}\,^{lm\hat{\imath}}\,\e_{j)}
      +\fr12\,\ve_{k(i}\,T_{a\,j)}\,^{lm\hat{\imath}}\,\gamma^a\,\e^k
      +6\,C_{ijk}\,^{lm\hat{\imath}}\,\eta^k
      \nonumber\\[.1in]
      \delta\,Y_{i}\,^{lm\hat{\imath}} &=&
      2\,i\,\bar{\e}^j\,{\Dslash}\rho_{ij}\,^{lm\hat{\imath}}
      +i\,\ve_{ij}\,\bar{\e}^j\,\Lambda^{*\,lm\hat{\imath}}
      -8\,i\,\bar{\eta}^j\,\rho_{ij}\,^{lm\hat{\imath}}
      \nonumber\\[.1in]
      \delta\,T_{a\,i}\,^{lm\hat{\imath}} &=&
      -\fr13\,i\,\ve^{jk}\,\bar{\e}_j\,(\,3\,{\Dslash}\gamma_a
      +2\,{D}_a\,)\,\rho_{ik}\,^{lm\hat{\imath}}
      -\fr12\,i\,\ve_{ij}\,\bar{\e}^j\,\gamma_a\,\Lambda^{lm\hat{\imath}}
      \nonumber\\[.1in]
      & &
      -\fr13\,i\,\bar{\e}^k\,
      (\,3\,{\Dslash}\gamma_a+2\,{D}_a\,)\,\rho^*_{ik}\,^{lm\hat{\imath}}
      +\fr12\,i\,\bar{\e}_i\,\gamma_a\,\Lambda^{*\,lm\hat{\imath}}
      \nonumber\\[.1in]
      & &
      -\fr{16}{3}\,i\,\ve^{jk}\,\bar{\eta}_j\,\gamma_a\,\rho_{ik}\,^{lm\hat{\imath}}
      -\fr{16}{3}\,i\,\bar{\eta}^j\,\gamma_a\,\rho^*_{ij}\,^{lm\hat{\imath}}
      \nonumber\\[.1in]
      \delta\,\Lambda^{lm\hat{\imath}} &=&
      -\fr12\,\ve^{jk}\,(\,{\Dslash}\gamma^a+4\,{D}^a\,)\,T_{a\,j}\,^{lm\hat{\imath}}\,\e_k
      +\fr12\,{\Dslash} Y^*_i\,^{lm\hat{\imath}}\,\e^i
      \nonumber\\[.1in]
      & &
      +3\,\ve^{jk}\,T_{a\,j}\,^{lm\hat{\imath}}\,\gamma^a\,\eta_k
      +4\,Y^*_i\,^{lm\hat{\imath}}\,\eta^i \,,
\label{cretm}\err
where the derivatives $D_a$ are covariant with respect to the superconformal algebra.
These represent the complete supersymmetry transformations of the Conformal RETM except for
terms involving the supergravity matter fields $(\,T_{ab}\,,\,\chi^i\,,\,D)$ which are
straightforward to derive using the algebra exhibited in \ref{n2algebra}.

Using considerations similar to those employed for the RETM, we determine that the conformal weight of the DSDQ is $-2$
and that its $Q$ and $S$ transformation rules are
 \brr \delta_Q\,H_{i\,\hat{\imath}} &=&
      i\,\ve_{jk}\,\bar{\e}^j\,\Omega^k\,_{i\hat{\imath}}
      +i\,\bar{\e}_j\,\Omega^{*\,j}\,_{i\hat{\imath}}
      \nonumber\\[.1in]
      \delta_Q\,\Omega^{i}\,_{j\hat{\imath}} &=&
      -\fr12\,\ve^{ik}\,\Dslash H_{j\hat{\imath}}\,\e_k
      -\fr12\,V^{a\,ik}\,_{j\hat{\imath}}\,\gamma_a\,\e_k
      +\ve_{kl}\,P^{ik}\,_{j\hat{\imath}}\,\e^l
      +2\,\ve^{ik}\,H_{j\hat{\imath}}\,\eta_k
      \nonumber\\[.1in]
      \delta_Q\,P^{ij}\,_{k\hat{\imath}} &=&
      -\fr23\,i\,\ve^{l(i}\,\bar{\e}_l\,\Dslash\Omega^{j)}\,_{k\hat{\imath}}
      +i\,\bar{\e}_l\,\varphi^{ijl}\,_{k\hat{\imath}}
      -4\,i\,\ve^{m(i}\,\bar{\eta}_m\,\Omega^{j)}\,_{k\hat{\imath}}
      \nonumber\\[.1in]
      \delta_Q\,V^{a\,ij}\,_{k\hat{\imath}} &=&
      \fr23\,i\,\bar{\e}^{(i}\,(\,\Dslash\gamma^a-D^a\,)\,\Omega^{j)}\,_{k\hat{\imath}}
      -i\,\ve_{mn}\,\bar{\e}^m\,\gamma^a\,\varphi^{ijn}\,_{k\hat{\imath}}
      +2\,i\,\bar{\eta}^{(i}\,\gamma^a\,\Omega^{j)}\,_{k\hat{\imath}}
      \nonumber\\[.1in]
      & &
      -\fr23\,i\,\ve^{m(i}\,\bar{\e}_{m}\,(\,\Dslash\gamma^a-D^a\,)\,\Omega^{*\,j)}\,_{k\hat{\imath}}
      -i\,\bar{\e}_l\,\gamma^a\,\varphi^{*\,ijl}\,_{k\hat{\imath}}
      -2\,i\,\ve^{m(i}\,\bar{\eta}_m\,\gamma^a\,\Omega^{*\,j)}\,_{k\hat{\imath}}
      \nonumber\\[.1in]
      \delta_Q\,\varphi^{ijk}\,_{l\hat{\imath}} &=&
      \Dslash P^{(ij}\,_{l\hat{\imath}}\,\e^{k)}
      -\fr14\,\ve^{m(i}\,(\,2\,\Dslash\gamma_a+3\,D_a\,)\,
      V^{a\,jk)}\,_{l\hat{\imath}}\,\e_m
      +\fr12\,D^{ijkm}\,_{l\hat{\imath}}\,\e_m
      \nonumber\\[.1in]
      & &
      -2\,P^{(ij}\,_{l\hat{\imath}}\,\eta^{k)}
      -2\,\ve^{m(i}\,V^{a\,jk)}\,_{l\hat{\imath}}\,\gamma_a\,\eta_m
      \nonumber\\[.1in]
      \delta_Q\,D^{ijkl}\,_{m\hat{\imath}} &=&
      2\,i\,\bar{\e}^{(i}\,\Dslash\varphi^{jkl)}\,_{m\hat{\imath}}
      -4\,i\,\bar{\eta}^{(i}\,\varphi^{jkl)}\,_{m\hat{\imath}}
      \nonumber\\[.1in]
      & &
      -2\,i\,\ve^{n(i}\,\bar{\e}_{n}\,\Dslash\varphi^{*\,jkl)}\,_{m\hat{\imath}}
      +4\,i\,\ve^{p(i}\,\bar{\eta}_p\,\varphi^{*\,jkl)}\,_{m\hat{\imath}}
      \,,
 \label{cdsdq}\err
where the derivatives $D_a$ are covariant with respect to the superconformal algebra.
These represent the complete supersymmetry transformations of the Conformal DSDQ except for
terms involving the supergravity matter fields $(\,T_{ab}\,,\,\chi^i\,,\,D)$ which are
straightforward to derive using the algebra exhibited in \ref{n2algebra}.

The conformal RETM and the conformal DSDQ each describe 96+96 off-shell degrees of freedom.
The isometry $SU(2)_A$ is gauged by the field $V_\mu^I$ in the Weyl multiplet.  We can also gauge
the isometry $SU(2)_B$ by coupling a Vector multiplet.
\footnote{In fact, we have already worked out
the fully-gauged RETM and DSDQ multiplets, in which the transformation rules
include terms involving the components of the background.  We will exhibit those results in a subsequent paper.}
Taken together, the Weyl multiplet and the Vector multiplet serve to gauge the
isometry $U(1)_{\cal A}\times SU(2)_A\times SU(2)_B$.

Using the dilatation and $SU(2)_A$ isometries, both gauged by the Weyl multiplet, and the $SU(2)_B$
isometry gauged by the background Vector multiplet, one linear combination of the
non-auxiliary quaternions $\phi_i\,^{\hat{\imath}}$ and $H_i\,^{\hat{\imath}}$ is pure gauge.  Similarly,
using $S$ supersymmetry, gauged by its composite connection, one linear combination of the
non-auxiliary fermions $\psi^{\hat{\imath}}$ and $\beta^{\hat{\imath}}$
is pure gauge.  We can fix the gauge degrees of freedom
by coordinating a residual field-dependent $S$ transformation with $Q$ transformations which would otherwise spoil the gauge
choice.
This inextricably connects the Hyperplet off-shell to the supergravity background.
Since the Hyperplet also includes
the propagating hypermultiplet, this explains a crucial difference between the model we are promulgating in this
paper as compared to earlier off-shell hypermultiplets, which exist without supergravity.

It is necessary that we identify an action coupling the Hyperplet to
supergravity. This would comprise the local analog of the
global action explained in (\ref{iij}) and (\ref{glosac}).  We relegate the development
of that action to a future manuscript.  But we will end this paper with a brief speculation about
its structure.

We might presume that the tensor density ${\cal I}^{ij}$ forms the highest-weight component of a multiplet that we can couple
supersymmetrically to the Vector multiplet which gauges $SU(2)_B$.  For example, consider
a Linear Multiplet with components $(\,L_{ij}\,^{mn}\,|\,\varphi_i\,^{mn}\,|\,G^{mn}\,,\,E_a\,^{mn}\,)$,
where lower indices correspond to $SU(2)_A$ and upper indices correspond to $SU(2)_B$, using the
convention described above.  Consider as well the $SU(2)_B$ Vector multiplet with components
$(\,W_a^I\,,\,X^I\,|\,\widehat{\Omega}_i^I\,|\,\widehat{Y}_{ij}^I\,)$.  There is a unique
globally supersymmetric Lagrangian density which couples these two, given by
${\cal L}=\bar{X}^I\,G^{mn}\,(\,T_I\,\ve\,)_{mn}+\cdots$ where most terms have been suppressed.
In principle, the component $G^{mn}$, or an analog with hybrid $SU(2)_A\times SU(2)_B$ indices, can be
formed from the components of the RETM, the DSDQ, and the Vector multiplet that gauges $SU(2)_B$.
In this case, the the globally supersymmetric density could organize, after fixing the
superconformal gauge, as ${\cal L}\propto |X|\,\ve_{ij}\,{\cal I}^{ij}+\cdots$,
forming a direct analog of (\ref{glosac}).  In this case, the field $X^I$ would properly adjust the scaling weight.
A Noethering procedure would then expose the Poincar{\'e} supergravity action.  There are interesting considerations, however,
involving the interplay between $SU(2)_A$ and $SU(2)_B$ tensor structures which need to be resolved.
We might expect the ultimate theory to possess the global symmetry $SU(2)_C\subset SU(2)_A\times SU(2)_B$.
So the gauge fixing may implement a ``twist" which identifies $SU(2)_A$ with $SU(2)_B$.
These speculations motivate further work, to be addressed in a future paper.

\setcounter{equation}{0}
\section{Conclusions and Outlook}
The two superconformal multiplets given in (\ref{cretm}) and (\ref{cdsdq})
provide an existence proof that any argument based on off-shell state counting
which implies that hypermultiplets cannot be realized off-shell with finite degrees of
freedom or without an off-shell central charge possesses a loophole, related to
a role played by couplings to supergravity.
In this sense we have resolved the off-shell hypermultiplet problem at the level of state counting.
This is the primary result of this paper.
This provides hope that similar considerations
might resolve the off-shell problem associated with ${\cal N}=4$ Super Yang-Mills.

The fields of the Conformal Hyperplet assemble as
\brr
\includegraphics[width=3in]{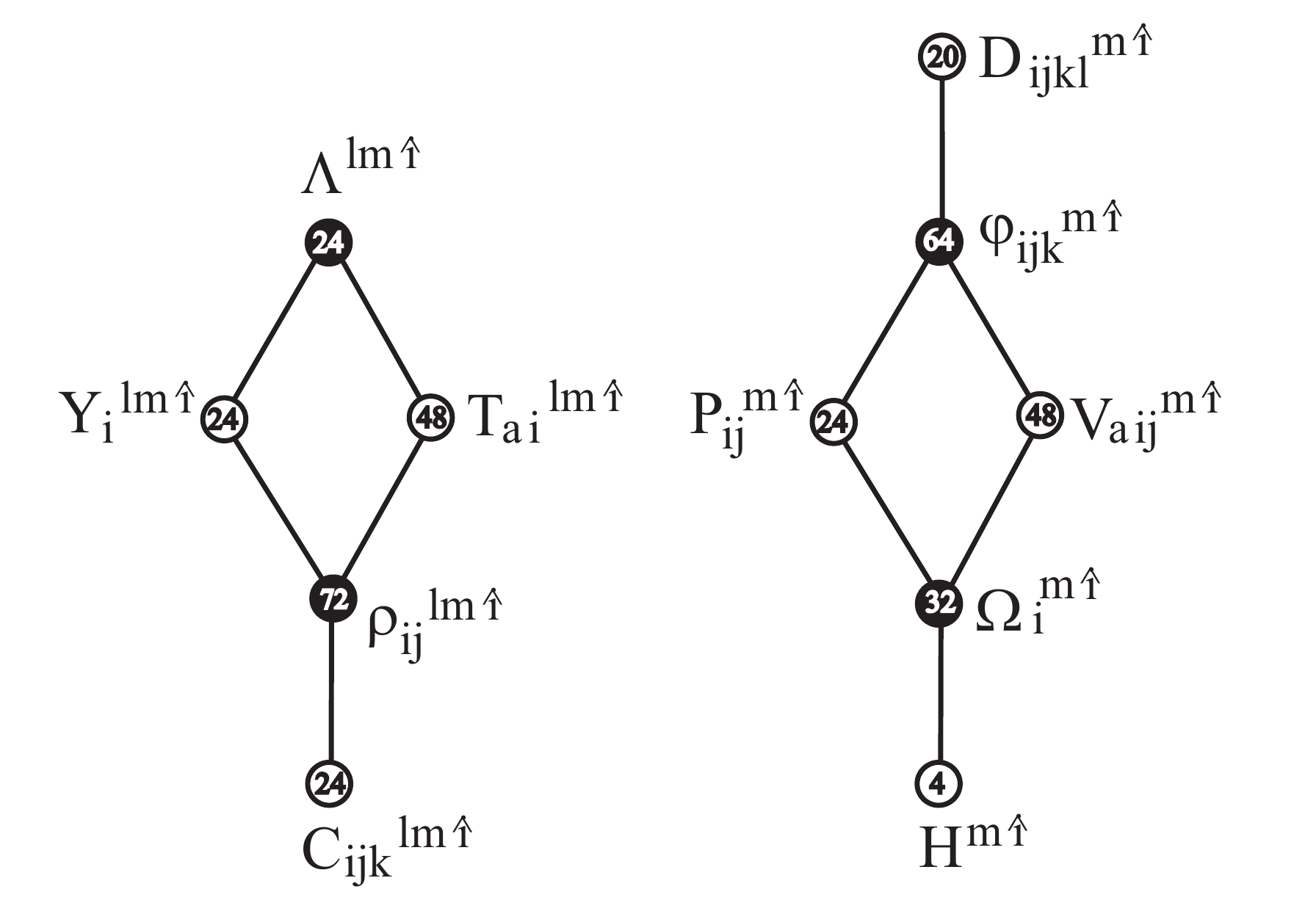}
\label{finalform}\err
where the left multiplet is the RETM and the right multiplet is the DSDQ.
The component fields transform as tensors under the isometry
$SU(2)_A\times SU(2)_B\times SU(2)_H$. The $SU(2)_H$ indices are hatted, $SU(2)_B$ indices are placed at the same
level (up or down) as the $SU(2)_H$ indices, and $SU(2)_A$ indices are placed
at the level opposite the $SU(2)_H$ indices.  Strings of indices are always symmetric.
Thus, for example, $C_{ijk}\,^{lm\,\hat{\imath}}$ transforms as a symmetric rank three $SU(2)_A$
tensor, a symmetric rank two $SU(2)_B$ tensor, and as a vector under $SU(2)_H$.  We use the
Weyl multiplet component $V_\mu^I$ to gauge $SU(2)_A$, and we use the component $W_\mu^I$
of a background Vector multiplet to gauge $SU(2)_B$.  The diagonal subgroup $SU(2)_C\subset SU(2)_A\times SU(2)_B$
is gauged (in tandem) by $V_\mu^I$ and $W_\mu^I$.

The components of (\ref{finalform}) can be expressed in terms of $SU(2)_C$ irreps as
\brr \includegraphics[width=5in]{RETMDSDQ.pdf}
\label{retmdsdq}\err
Under dilatations and $SU(2)_A\times SU(2)_B$ transformations, $\phi_i\,^{\hat{\imath}}$
is pure gauge and the scalars $X^I$ of the $SU(2)_B$ Vector multiplet are also
pure gauge.  Under $S$-supersymmetry,  $\psi^{\hat{\imath}}$
is pure gauge.  The remaining fields are all auxiliary except for $H_i\,^{\hat{\imath}}$
and $\beta^{\hat{\imath}}$.
Thus, the only propagating fields are those of a single hypermultiplet:
under $SU(2)_{\cal R}\times SU(2)_H$ the physical bosons
transform as ${\bf 2}\times {\bf 2}$ and the physical fermions transform as ${\bf 1}\times {\bf 2}$.

As a next step, we aim to construct supergravity actions associated with the
Hyperplet transformation rules presented above.  This should involve a superconformal
Hyperplet coupled to the ${\cal N}=2$ Weyl multiplet and to the Vector multiplet used to gauge
the $SU(2)_B$ isometry, the latter allowing us to suitably adjust conformal weights.

We are interested in describing additional hypermultiplets
in generalized off-shell Hyperplet models.  Presumably these would include
equal numbers of RETM and DSDQ multiplets. One might expect that this would
involve equal numbers of ``healthy" and ghost-like hypermultiplet sectors,
suggesting that there might be insufficient symmetry to gauge away the ghosts.
But this is not necessarily so.  For example, a given RETM can couple to more
than one DSDQ, and vice-versa.  In principle, this can be arranged so that the
kinetic sector involves $N_H$ healthy eigenvalues and only one ghost-like eigenvalue.
The latter self-identifies as the relevant conformal compensator.  We intend to develop such
theories, and to investigate the corresponding restrictions on sigma model geometry.

Off-shell ${\cal N}=2$ theories have proved
invaluable in elucidating the structure of
${\cal N}=2$ supergravity, and have demonstrated their worth in many other ways as well, including
providing contexts for special geometry \cite{SpecG,SpecG1,SpecG2} and Seiberg-Witten theory \cite{SWitt1,SWitt2}, to name a couple of
examples.  It might be interesting to re-evaluate these using the Hyperplet introduced in this paper.
Since Hyperplets do not have an off-shell central charge, these might
admit superspace descriptions which could complement other constructions,
such as harmonic superspace \cite{Harm1}.

\appendix
\renewcommand{\thesection}{Appendix \Alph{section}}
\renewcommand{\theequation}{\Alph{section}.\arabic{equation}}

\setcounter{equation}{0}
\section{Elements of ${\cal N}=4$ SYM}
\label{n4elements}
In this Appendix we review the basic elements of the standard presentation of ${\cal N}=4$ SYM theory,
in its well-known on-shell guise.  These details are included for reference purposes and to aid
the presentation in the main text.

The ${\cal N}=4$ Vector multiplet involves
gauge potentials $A_a^I$,
massless chiral gaugino fields $\chi_i^I=-\gamma_5\,\chi_i^I$, and
massless complex scalars $\phi_{ij}^I=\phi_{[ij]}^I$, subject to
\brr \phi_{ij}^I &=& \fr12\,\ve_{ijkl}\,\phi^{kl\,I} \,,
\err
where $\ve_{ijkl}$ is the four-dimensional Levi-Civita tensor.
Mid-alphabet Latin indices $i$ and $j$ are fundamental
${\bf 4}$ indices under $SU(4)_R$, while the index $a$ is an adjoint index under the
gauge group ${\cal G}$. For each value of $I$ the
field $\phi_{ij}^I$ describes six degrees of freedom and $\chi_i^I$ describes
eight degrees of freedom {\it on-shell}.

The anti-hermitian generators $T_I$ of the gauge group ${\cal G}$ respect the commutation relationship
$[\,T_I\,,\,T_J\,]=f_{IJ}\,^K\,T_K$.  The component fields
$(\,\phi_{ij}^I\,,\,\chi_i^I\,,\,F_{\mu\nu}^I\,)$ transform in the
adjoint representation of ${\cal G}$.

A supersymmetry transformation acts as
\brr \delta_Q\,\phi_{ij}^I &=& 2\,i\,\bar{\e}_{[i}\,\chi_{j]}^I
     +i\,\ve_{ijkl}\,\bar{\e}^k\,\chi^{l\,I}
     \nonumber\\[.1in]
     \delta_Q\,\chi_i^I &=&
     -\Dslash\phi_{ij}^I\,\e^j
     +\fr14\,F_{ab}^I\,\gamma^{ab}\,\e_i
     +f_{JK}\,^I\,\phi_{ik}^J\,\phi^{kj\,K}\,\e_j
     \nonumber\\[.1in]
     \delta_Q\,F_{\mu\nu}^I &=&
     -2\,i\,\bar{\e}_i\,\gamma_{[\mu}\,D_{\nu]}\chi^{i\,I}
     -2\,i\,\bar{\e}^i\,\gamma_{[\mu}\,D_{\nu]}\chi_i^I \,.
\label{n4tran}\err
where the field strengths, $F_{\mu\nu}^I=2\,\der_{[\mu}A_{\nu]}^I-f_{JK}\,^I\,A_\mu^J\,A_\nu^K$,
satisfy the Bianchi identity $D_{[\mu}F_{\nu\lambda]}^I=0$.
The gauge potentials transform as
\brr \delta_Q\,A_a^I &=&
     i\,\bar{\e}_i\,\gamma_\mu\,\chi^{i\,I}
     +i\,\bar{\e}^i\,\gamma_\mu\,\chi_i^I
\err
The dual field strength is $\tilde{F}^I_{\mu\nu}=\fr12\,\ve_{\mu\nu}\,^{\lambda\sigma}\,F^I_{\lambda\sigma}$,
and the complex chiral field strengths are $F_{\mu\nu}^{(\pm)\,I}=\fr12\,(\,F_{\mu\nu}^I\pm i\,\tilde{F}^I_{\mu\nu}\,)$.
It follows that
\brr \delta_Q\,F_{\mu\nu}^{(-)\,I} &=&
     \fr12\,i\,\bar{\e}_i\,\gamma_{\mu\nu}\Dslash\,\chi^{I\,i}
     -\fr12\,i\,\bar{\e}^i\,\Dslash\gamma_{\mu\nu}\,\chi^I_i
\err
and a similar expression for $\delta_Q\,F_{\mu\nu}^{(+)\,I}$ obtained by toggling
the up/down placement of $SU(4)_R$ indices.

The supersymmetry transformations in (\ref{n4tran}) satisfy the following
commutation relationship on the bosonic fields
$\phi_{ij}^I$ and $F_{\mu\nu}^I$,
\brr [\,\delta_Q(\e_1)\,,\,\delta_Q(\e_2)\,]
     &=& 2\,i\,\bar{\e}_{[2}^i\,\gamma^a\,\e_{1]\,i}\,D_a+\delta_g(\tilde{\theta})
\label{qqcom}\err
where
\brr \tilde{\theta}^I &=& 2\,i\,\bar{\e}_{2\,i}\,\e_{1\,j}\,\phi^{ij\,I}
      +2\,i\,\bar{\e}_2^i\,\e_1^j\,\phi_{ij}^I
\err
parameterizes a field-dependent gauge transformation.  When acting on the
fermions $\chi_i^I$, two susy transformations obey (\ref{qqcom})
plus new terms proportional to the field equations associated with the
supersymmetric action given below.

The action
\brr S &=& \int d^4x\,\bpl
     -\fr14\,F_{\mu\nu}^I\,F^{\mu\nu\,I}
     +\fr12\,D_\mu\,\phi_{ij}^I\,D^\mu\phi^{ij\,I}
     -2\,i\,\bar{\chi}^{i\,I}\,\Dslash\chi_i^I
     \nonumber\\[.1in]
     & &\hspace{.6in}
     +2\,i\,f_{IJK}\,(\,\phi_{ij}^I\,\bar{\chi}^{i\,J}\,\chi^{j\,K}
     +\phi^{ij\,I}\,\bar{\chi}_i^J\,\chi_j^K\,)
     \nonumber\\[.1in]
     & &\hspace{.6in}
     +\fr14\,(\,f_{JKI}\,\phi_{ij}^J\,\phi_{kl}^K\,)\,
     (\,f_{LM}\,^I\,\phi^{ij\,L}\,\phi^{kl\,M}\,)\,\bpr \,,
\label{ac4}\err
is invariant under (\ref{n4tran}).  An important generalization
involves a topological term
proportional to $\theta\,F_{\mu\nu}^I\,\tilde{F}^{\mu\nu\,I}$, which
plays a central role in $S$-duality.

\setcounter{equation}{0}
\section{A proof about $SU(4)$ representations}
\label{pcount}
In this appendix we prove that any $SU(4)$ representation structured as an odd-rank tensor
has a dimensionality which is an integer multiple of four.

Any $SU(N)$ representation can be written as a tensor $\Phi_{i_1\cdots i_k}$, where the indices are
fundamental ${\bf N}$ indices. For example, the symmetric tensor $\Phi_{(i_1\cdots i_k)}$
has dimension
\brr D_S(N,k) &=& {N+k-1}\choose{k}
\label{symrep}\err
and the antisymmetric tensor $\Phi_{[i_1\cdots i_k]}$ has dimension
\brr D_A(N,k) &=& {N\choose{k}}
     \,.
\err
Other representations are described by tensors with mixed index symmetry.  The different possibilities
are characterized by Young tableaux.

A given tableau has $n$ rows of left-justified boxes, where $1\le n\le N$, and where no row is longer than the previous row.
Each box corresponds to a tensor index.  Horizontal juxtaposition indicates index symmetrization while vertical juxtaposition
indicates index anti-symmetrization.
Any $N$-row tableau is equivalent to another tableau with fewer
rows.  (This is because an $N$-box column corresponds to the Levi-Civita tensor.) As a consequence, appending or removing
any number of $N$-box columns to or from the left side of any tableau produces an equivalent representation.
Thus, we can study all $SU(4)$ representations by restricting attention to tableaux with one, two, or
three rows.

One way to determine the dimension of a representation is to compute the ratio of
two integers determined using the following simple algorithm:  For the numerator, start with a blank tableau,
then place an entry in each box, so that the $r$-th row is filled, left to right,
with consecutive integers, starting with $N+1-r$.  Then multiply all the
entries in the entire tableau together.  For the denominator, start with another blank tableau,
fill each box with its hook length, and then multiply
all of the entries together.  (Hook length is defined as the number of boxes
to the right, plus the number of boxes beneath, plus one for the box itself.)

If an $SU(4)$ tableau has $a$ boxes in the first row, $b$ boxes in the second row, and $c$ boxes in the
third row then, using a well-known result \cite{Georgi}, the dimension of the corresponding representation is
\brr D_{a,b,c} &=&
     \fr{1}{12}\,(\,a+3\,)(\,b+2\,)(\,c+1\,)(\,a-b+1\,)(\,a-c+2\,)(\,b-c+1\,) \,.
\label{wdfk}\err
This is equivalent to the Weyl dimension formula \cite{Weyl}, and is readily derived using the algorithm
described above.
The number of boxes in a given tableau, namely $k=a+b+c$, is equivalent to the rank of the corresponding tensor.

Since (\ref{wdfk}) must be an integer, it follows that the factors in that formula excluding the $\fr{1}{12}$ pre-factor
must be a multiple of three.  This may be verified as follows.  The factor $(b+2)$ is a multiple of three whenever $b$ is 1 mod 3.  The factor
$(c+1)$ is a multiple of three whenever $c$ is 2 mod 3.  And the factor $(1+b-c)$ is
a multiple of three in all remaining cases except when $(\,b\,,\,c\,)$ is
$(\,0\,,\,0\,)$ or $(\,2\,,\,1\,)$ where all integers are expressed mod 3.  In both of the
exceptional cases the factors $(a+3)(a-b+1)(a-c+2)$ involve three consecutive integers, one of which
therefore must be a multiple of three.

Given that there is a factor of three in its numerator, (\ref{wdfk}) is an integral multiple
of four provided this contains either 1) four even factors, or 2) two even factors and another doubly-even factor.
We proceed to show that this criterion is met for for any tableau with odd $k$.
If $k=a+b+c$ is odd, then there are four possibilities: either all three
of $a$, $b$, and $c$ are odd, or two of these are odd and one of these is even.

Consider the first case, where $a$, $b$, and $c$ are each odd. It follows that $(\,a+3\,)(\,c+1\,)(\,a-b+1\,)(\,b-c+1\,)$ is the product
of four even factors, proving, based on the above discussion, that (\ref{wdfk}) is an integral multiple of 4.

Consider the second case, where $a$ and $b$ are even while $c$ is odd, so that $c=1$ mod 4, or $c=3$ mod 4.  If $c=1$ mod 4 then,
using mod 4 arithmetic, we have $(b+2)(c+1)(b-c+1)=(b+2)\cdot 2\cdot b$.  Since at least one of the
factors $(b+2)$ or $b$ is doubly even, it follows that this product involves two even factors
and one doubly even factor.  If, on the other hand, $c=3$ mod 4 then, using mod 4 arithmetic, the same product is $(b+2)\cdot 4\cdot b$,
which again includes two even factors and the explicit doubly-even middle factor.

Consider the third case, where $b$ and $c$ are even while $a$ is odd, so that $a=1$ mod 4, or $a=3$ mod 4.  If $a=1$ mod 4 then,
using mod 4 arithmetic, we have $(a+3)(b+2)(a-b+1)=4\cdot (2+b)(2-b)$, where the first factor is manifestly doubly-even while the
second and third factors are even.  If, on the other hand, $a=3$ mod 4 then $(a+3)(b+2)(a-b+1)=2\cdot (b+2)\cdot b$.  In this case
the second and third factors are consecutive even integers, one of which is necessarily doubly-even.

Consider the remaining case, where $c$ and $a$ are even while $b$ is odd, so that $b=1$ mod 4, or $b=3$ mod 4.  If $b=1$ mod 4 then,
using mod 4 arithmetic,
$(a-b+1)(a-c+2)(b-c+1)=a\,(a-c+2)(2-c)$.  In this case if $a=0$ mod 4 then the first of these factors is doubly-even while the other two
are even, while if $a=2$ mod 4 then the first factor is even, and the other two are consecutive even integers, one of which is necessarily
doubly-even.  If, on the other hand, $b=3$ mod 4 then $(a-b+1)(a-c+2)(b-c+1)=(a+2)(a-c+2)\cdot c$.  In this case
if $a=0$ mod 4 then the first factor is even while the other two are consecutive even integers, one of which is necessarily doubly-even,
while if $a=2$ mod 4 then the first factor is doubly-even while the other two are even.

We have shown that in every possible case (\ref{wdfk}) is one-twelfth of a factor of 3 and a factor of 16.  Thus,
the dimension of any representation for odd $k$, meaning any representation corresponding to an odd rank tensor, is
necessarily an integer multiple of four.  Q.E.D.

 \setcounter{equation}{0}
\section{Counting Argument}
\label{adinkrareview}
This Appendix provides a cursory overview of contexts and results from one-dimensional supersymmetry representation theory
\cite{adinkras,DFGHIL_codes,DFGHIL_topology,DFGHIL_graph}
pertinent to the conclusion that any off-shell 4D ${\cal N}=4$ theory must involve an integral multiple of 128+128 degrees
of freedom and any off-shell 4D ${\cal N}=2$ theory must involve an integral multiple of 8+8 degrees of freedom.  These conclusions follow
from an analysis of (shadow) theories obtained by switching off any field dependence
on the three spatial coordinates, thereby restricting to a time-like one-dimensional sub-manifold of spacetime.
\footnote{A complementary approach to the classification of 1D supersymmetry is provided in \cite{Toppan,Toppan_01,Toppan_02}.}

In $N$-extended supersymmetry in one-dimension (for example on a world-line or on a 0-brane) an unconstrained superfield has $2^N$ real components.
These identify one-to-one with the vertices of an $N$ dimensional hypercube.  The $N$ supercharges
identify one of $N$ distinct ``colors" with each edge.  The supercharges also impart a
directional sense and a kind of parity to each edge.  Each vertex represents one degree of freedom.  These come in two
sorts: bosons are represented by black vertices and fermions are represented by white vertices.
All data about the corresponding supermultiplet lies encoded in the hypercube augmented with the data described
above.  The set of vertices and edges define a bipartite graph equivalent to the
hypercube.  The embellishments corresponding to black and white vertices, edge color, edge directionality, and edge
parity, are applied to the graph as well.  The embellished graph is called an Adinkra.

Irreducible multiplets are delineated by imposing differential constraints on unconstrained
superfields.  These translate as restrictions on Adinkras, generated by consistent involutions,
implemented by pairwise identification of vertices along with corresponding
identification of edges.  Consistency requires that each edge identifies with another edge possessing the same color (supercharge),
the same sense, and the same edge parity.  Each such operation, called a folding, defines a new Adinkra which is smaller (by half) compared to the original Adinkra.
The topology of any Adinkra along with the embellishments described above define its ``chromotopology".
The number of possible iterated foldings grows with $N$.

The mathematical problem of classifying all possible Adinkra chromotopologies
corresponds to the classification of doubly-even binary codes as described in \cite{DFGHIL_codes}.
For the case $N=16$, the principal result states that there are a maximum of $\aleph(16)=8$ independent consistent involutions, so that the number of component fields in the smallest possible multiplet is $2^{N-\aleph(N)}=2^{16-8}=256$. Half of these are bosons and half are fermions.  So the smallest
$N=16$ multiplet has 128+128 degrees of freedom.
In the case of $N=8$, the principal result of \cite{DFGHIL_codes} states that there are a maximum of $\aleph(8)=4$ independent consistent
involutions, so that the number of component fields in the smallest possible $N=8$ multiplet is $2^{N-\aleph(N)}=2^{8-4}=16$.
Half of these are bosons and half are fermions.  So the smallest $N=8$ multiplet has 8+8 degrees of freedom.

Information about the engineering dimensions, or the mass units, associated with fields is related to the
sense data associated with the edges.  The dimension of the fields defines a ``height" for the corresponding vertex.
If an Adinkra admits consistent vertex height assignments it is called ``engineerable".
There are a variety of distinct height assignments, called hangings, for a given chromotopology.
If all edges are directed from a boson toward a fermion the Adinkra spans only
two distinct heights.  An Adinkra with this feature is called a Valise Adinkra.  All engineerable Adinkras
can be obtained from a Valise by a sequence of ``unpacking" or raising operations.

Every irreducible and engineerable supermultiplet in one dimension corresponds to an Adinkra obtained using the method
described above.  Since there are no $N=16$ Adinkras with fewer than 128+128 vertices, and since any Adinkra is obtained
by iterated half-wise restrictions on maximal Adinkra with $2^{N-1}+2^{N+1}$ vertices, it follows that
any $N=16$ multiplet must have $128\,m$ bosons, where $m$ is a positive integer, and a like number of fermions.
Any off-shell 4D ${\cal N}=4$ multiplet dimensionally reduces to a (shadow)
1D $N=16$ multiplet.  Therefore, any off-shell 4D ${\cal N}=4$ supermultiplet must involve
$128\,m$ bosonic and $128\,m$ fermionic degrees of freedom, regardless of whether the corresponding degrees of freedom
are physical, gauge, or auxiliary.  Analogous considerations tell us that any off-shell 4D ${\cal N}=2$  multiplet
must involve $8\,m$ bosonic and $8\,m$ fermionic degrees of freedom.

\setcounter{equation}{0}
\section{On a class of ${\cal N}=2$ supermultiplets}
\label{multclass}
For every non-negative odd integer $\alpha\ge 5$, there is an ${\cal N}=2$ supermultiplet with a trivial
central charge off-shell, given by
\brr {\cal M}_\alpha &=& (\,\Phi_{(\alpha-1)}\,|\,\Psi_{(\alpha-2)}\,|\,C_{(\alpha-3)}\,,\,V_{a\,(\alpha-3)}\,|\,\Lambda_{(\alpha-4)}\,|\,E_{(\alpha-5)}\,) \,,
\label{mnu1}\err
where ${\cal M}_\alpha$ is a moniker, and the right-hand side lists the component fields
as follows.  The fields $\Phi$, $C$, $V_a$, and $E$ are a real scalar,
a complex scalar, a real Lorentz vector, and another real scalar, respectively, where the concept of
``real" is clarified below.  The fields $\Psi$ and $\Lambda$ are
Weyl spinors.
The vertical bars delineate field cells with different scaling dimension, which increases by
one-half unit as we move from cell-to-cell, left-to-right.  (Adjacent cells correspond to adjacent ``component levels" in superspace,
where a given level corresponds to a certain number of fermionic $\theta$ coordinates.)
  The fields transform as symmetric tensors
under $SU(2)_R$, and the subscripts indicate the tensor rank.
For example, the lowest component $\Phi_{(\alpha-1)}$ transforms as a $\alpha$-dimensional representation
of $SU(2)_R$, structured as a symmetric tensor $\Phi_{\i_1\cdots i_{\alpha-1}}=\Phi_{(\i_1\cdots i_{\alpha-1})}$,
where the indices are fundamental ${\bf 2}$ indices.

The degrees of freedom for the component fields are listed in Table \ref{mnudofs}.
\begin{table}
\begin{center}
\begin{tabular}{|c||c|c|cc|c|c|}
\hline
&&&&&&\\[-.1in]
field & $\Phi_{(\alpha-1)}$ &
$\Psi_{(\alpha-2)}$ &
$C_{(\alpha-3)}$ &
$V_{a\,(\alpha-3)}$ &
$\Lambda_{(\alpha-4)}$ &
$E_{(\alpha-5)}$ \\[.1in]
\hline
&&&&&&\\[-.1in]
d.o.f.s & $\alpha$ &
$4\,(\,\alpha-1)$ &
$2\,(\,\alpha-2)$ &
$4\,(\,\alpha-2\,)$ &
$4\,(\,\alpha-3)$ &
$\alpha-4$ \\[.1in]
\hline
\end{tabular}
\caption{The degrees of freedom in the component fields of the multiplet ${\cal M}_\alpha$
specified in (\ref{mnu1}). The coefficients 2 and 4 on the dimensions for $C$ and $V_a$
indicate that $C$ is complex and $V_a$ is a Lorentz vector.  The leading coefficients 4 on the
dimensions for $\Psi$ and $\Lambda$ reflect that these are Weyl spinors with four degrees
of freedom per choice of $SU(2)_R$ index.}
\label{mnudofs}
\end{center}
\end{table}
Adding these up, we compute $N_B$ bosonic degrees of freedom and $N_F$ fermionic
degrees of freedom, where
\brr N_B &=& \alpha+2\,(\,\alpha-2\,)+4\,(\,\alpha-2)+(\,\alpha-4\,)
     \nonumber\\[.1in]
     N_F &=& 4(\,\alpha-1\,)+4\,(\,\alpha-3\,)
\err
so that $N_B=N_F=8\,(\,\alpha-2\,)$, showing that there is a proper balance. The transformation rules are
\brr \delta_Q\,\Phi_{i_1\cdots i_{\alpha-1}}  &=&
     i\,\bar{\e}_{(i_1}\,\Psi_{i_2\cdots i_{\alpha-1})}
     -i\,\ve_{m(i_1}\,\bar{\e}^m\,\Psi^{*}_{i_2\cdots i_{\alpha-1})}
     \nonumber\\[.1in]
     \delta_Q\,\Psi_{i_1\cdots i_{\alpha-2}}  &=&
     \dslash\Phi_{i_1\cdots i_{\alpha-2}k}\,\e^k
     +\fr12\,C_{(i_1\cdots i_{\alpha-3}}\,\e_{i_{\alpha-2})}
     +\fr12\,\ve_{k(i_1}\,V_{a\,i_2\cdots i_{\alpha-2})}\,\gamma^a\,\e^k
     \nonumber\\[.1in]
     \delta_Q\,C_{i_1\cdots i_{\alpha-3}}  &=&
     2\,i\,\bar{\e}^k\,\dslash\Psi_{k i_1\cdots i_{\alpha-3}}
     -i\,\ve_{k(i_1}\,\bar{\e}^k\,\Lambda_{i_2\cdots i_{\alpha-3})}^{*}
     \nonumber\\[.1in]
     \delta_Q\,V_{a\,i_1\cdots i_{\alpha-3}}  &=&
     -i\,\ve^{mn}\,\bar{\e}_m\,\bpl\,\dslash\,\gamma_a
     +\frac{2}{\alpha-1}\,\der_a\,\bpr\,\Psi_{i_1\cdots i_{\alpha-3}n}
     +\fr12\,i\,\ve_{k(i_1}\,\bar{\e}^k\,\gamma_a\,\Lambda_{i_2\cdots i_{\alpha-3})}
     +{\rm h.c.}
     \nonumber\\[.1in]
     \delta_Q\,\Lambda_{i_1\cdots i_{\alpha-4}}   &=&
     \bpl\,\frac{\alpha-3}{\alpha-2}\,\bpr\,\dslash C_{i_1\cdots i_{\alpha-4}k}^{*}\,\e^k
     -\ve^{mn}\,\bpl\,\bpl\,\frac{\alpha-3}{\alpha-2}\,\bpr\,\dslash\gamma^a
     +2\,\der^a\,\bpr\,V_{a\,i_1\cdots i_{\alpha-4}m}\,\e_n
     \nonumber\\[.1in]
     & & +\bpl\,\frac{\alpha-4}{\alpha-3}\,\bpr\,E_{(i_1\cdots i_{\alpha-5}}\,\e_{i_{\alpha-4})}
     \nonumber\\[.1in]
     \delta_Q\,E_{i_1\cdots i_{\alpha-5}} &=&
     i\,\bar{\e}^k\,\dslash\Lambda_{k i_1\cdots i_{\alpha-5}}
     -i\,\ve^{mn}\,\bar{\e}_m\,\dslash\Lambda^{*}_{n i_1\cdots i_{\alpha-5}}
\label{nurules}\err
The reality constraint satisfied by the fields $\Phi_{(\alpha-1)}$, $V_{a\,(\alpha-3)}$, and $\Lambda_{(\alpha-5)}$
are typified by
\brr \Phi_{i_1\cdots i_{\alpha-1}} &=& \ve_{i_1j_1}\cdots\ve_{i_{\alpha-1}j_{\alpha-1}}\,\Phi^{j_1\cdots j_{\alpha-1}} \,.
\label{rconp}\err
We reiterate that raising or lowering indices {\it en-masse} corresponds to complex conjugation.
The constraint (\ref{rconp}) is consistent only for even-rank tensors, owing
to the anti-symmery of the Levi-Civita tensor $\ve_{ij}$.  This is what restricts $\alpha$ to odd values.

The choice $\alpha=3$ was excluded in the above analysis.  Indeed, that choice would render the tensor rank of the
two highest components $\Lambda_{(\alpha-4)}$ and $E_{(\alpha-5)}$ as negative.  However, there {\it is} a multiplet which
fits into the ${\cal M}_\alpha$ hierarchy for $\alpha=3$, although there is an extra subtlety in that case.  The multiplet
${\cal M}_3$ does not have fields corresponding to $\Lambda$ and $E$.  Instead, the components are
${\cal M}_3=(\,\Phi_{ij}\,|\,\Psi_i\,|\,C\,,\,V_a\,)$, where we have used the explicit index notation,
$\Phi_{ij}$ and $\Psi_i$, rather than $\Phi_{(2)}$ and $\Psi_{(1)}$.  In this case
the superficial state counting does not balance.  Instead, we seem to have 3+2(1)+4(1)=9 bosonic degrees of freedom,
and only 4(2)=8 fermionic degrees of freedom.  This is reconciled by use of an extra constraint,
applied to the vector field, namely $\der^a V_a=0$.
In fact, ${\cal M}_3$ is the well-known ${\cal N}=2$ Tensor multiplet, which plays a central role in the construction
of ${\cal N}=2$ supergravity theories. Note that the vector constraint $\der^a V_a$ implies
that $V_a=\ve_a\,^{bcd}\,\der_b B_{cd}$, where $B_{ab}$ is a two-form tensor potential.
\footnote{In analogous versions of ${\cal M}_3$ involving gauged symmetries,
including the superconformal case, the constraint generalizes to a more complicated version involving a covariant
derivative $D^a V_a$ and includes contributions from other background fields.}
Thus, the Tensor multiplet sits at the bottom of a semi-infinite hierarchy of multiplets ${\cal M}_\alpha$, and represents
the only element with an extra constraint.

For each multiplet ${\cal M}_\alpha$ there is a separate ``dual" multiplet
\brr \widetilde{\cal M}_\alpha &=&
     (\,\widetilde{\Phi}_{(\alpha-5)}\,|\,\widetilde{\Psi}_{(\alpha-4)}\,|\,\widetilde{C}_{(\alpha-3)}\,,\,
     \widetilde{V}_{a\,(\alpha-3)}\,|\,\widetilde{\Lambda}_{(\alpha-2)}\,|\,\widetilde{E}_{(\alpha-1)}\,)
\label{dualm}\err
in which the $SU(2)_R$ tensor ranks are reversed relative to the scaling dimensions of the component fields
as compared to ${\cal M}_\alpha$.  (In the dual multiplet the tensor ranks increase rather than decrease as
the scaling dimension increases.)  The transformation rules are
\brr \delta_Q\,\widetilde{\Phi}_{i_1\cdots i_{\alpha-1}} &=&
     i\,\ve^{mn}\,\bar{\e}_m\,\widetilde{\Psi}_{n i_1\cdots i_{\alpha-1}}
     +i\,\bar{\e}^m\,\widetilde{\Psi}^*_{m i_1\cdots i_{\alpha-1}}
     \nonumber\\[.1in]
     \delta_Q\,\widetilde{\Psi}_{i_1\cdots i_\alpha} &=&
     \bpl\,\frac{\alpha}{\alpha+1}\,\bpr\,\ve_{k(i_1}\,\dslash\widetilde{\Phi}_{i_2\cdots i_\alpha)}\,\e^k
     -\fr12\,\widetilde{V}_{a\,k i_1\cdots i_{\alpha}}\,\gamma^a\,\e^k
     +\ve^{mn}\,\widetilde{C}_{m i_1\cdots i_{\alpha}}\,\e_n
     \nonumber\\[.1in]
     \delta_Q\,\widetilde{C}_{i_1\cdots i_{\alpha+1}} &=&
     -\bpl\,\frac{\alpha+1}{\alpha+2}\,\bpr\,i\,\ve_{k(i_1}\,\bar{\e}^k\,\dslash\widetilde{\Psi}_{i_2\cdots i_{\alpha+1})}
     +i\,\bar{\e}^k\,\widetilde{\Lambda}_{k i_1\cdots i_{\alpha+1}}
     \nonumber\\[.1in]
     \delta_Q\,\widetilde{V}_{a\,i_1\cdots i_{\alpha+1}} &=&
     i\,\bar{\e}_{(i_1}\,\bpl\,\bpl\,\frac{\alpha+1}{\alpha+2}\,\bpr\,\dslash\gamma_a
     -\bpl\,\frac{2}{\alpha+2}\,\bpr\,\der_a\,\bpr\,\widetilde{\Psi}_{i_2\cdots i_{\alpha+1})}
     -i\,\ve^{mn}\,\bar{\e}_m\,\gamma_a\,\widetilde{\Lambda}_{n i_1\cdots i_{\alpha+1}}
     +{\rm h.c.}
     \nonumber\\[.1in]
     \delta_Q\,\widetilde{\Lambda}_{i_1\cdots i_{\alpha+2}} &=&
     \dslash \widetilde{C}_{(i_1\cdots i_{\alpha+1}}\,\e_{i_{\alpha+2})}
     -\ve_{k(i_1}\,\bpl\,\fr12\,\dslash\gamma_a
     +\bpl\,\frac{\alpha+2}{\alpha+3}\,\bpr\,\der_a\,\bpr\,\widetilde{V}_{a\,i_2\cdots i_{\alpha+2})}\,\e^k
     +\widetilde{E}_{i_1\cdots i_{\alpha+2}k}\,\e^k
     \nonumber\\[.1in]
     \delta_Q\,\widetilde{E}_{i_1\cdots i_{\alpha+3}} &=&
     i\,\bar{\e}_{(i_1}\,\dslash\widetilde{\Lambda}_{i_2\cdots i_{\alpha+3})}
     -i\,\ve_{k(i_1}\,\bar{\e}^k\,\dslash\widetilde{\Lambda}^*_{i_2\cdots i_{\alpha+3})}
\err
By coupling a multiplet with its corresponding dual multiplet, one can build an $SU(2)_R$-invariant supersymmetric density.
Also, one can define the components of a dual multiplet in terms of the components of the original multiplet,
enabling self-couplings.

Given a multiplet ${\cal M}_{\alpha}$ or $\widetilde{\cal M}_\alpha$, there exist more multiplets
obtained by (formally) tensoring a common set of extra indices to every component field.  For example,
\brr {\cal M}_5\,^2 &=&
     (\,\Phi_{(4)}\,^{(1)}\,|\,\Psi_{(3)}\,^{(1)}\,|\,C_{(2)}\,^{(1)}\,,\,V_{a\,(2)}\,^{(1)}\,|\,\Lambda_{(1)}\,^{(1)}\,|\,
     E_{(0)}\,^{(1)}\,) \,,
\err
where, say, $V_{a\,(2)}\,^{(1)}=V_{a\,ij}\,^{m}$ transforms as ${\bf 3}\otimes {\bf 2}$ under $SU(2)_R$.
Similarly, $\Phi_{(4)}\,^{(1)}=\Phi_{ijkl}\,^m$ transforms as ${\bf 5}\otimes{\bf 2}$.  The real fields in this case
satisfy reality constraints which involve the new indices, for example
\brr \Phi_{ijkl}\,^{m} &=&
     \ve_{in}\,\ve_{jp}\,\ve_{kq}\,\ve_{lr}\,\ve^{ms}\,\Phi^{npqr}\,_{s} \,.
\err
The transformation rules for these generalized multiplets involve terms similar to those
exhibited in (\ref{nurules}), but with modified coefficients.  One can also tensor other index structures corresponding to
additional $Sp(2\,n)$ groups, distinct from $SU(2)_R$, for any positive integer $n$. For example, if we tensor on
a single $Sp(2)_H=SU(2)_H$ index, we can define the multiplet
\brr {\cal M}_5\,^{3,\hat{2}} &=&
     (\,\Phi_{(4)}\,^{(2)(\hat{1})}\,|\,\Psi_{(3)}\,^{(2)(\hat{1})}\,|\,
     C_{(2)}\,^{(2)(\hat{1})}\,,\,V_{a\,(2)}\,^{(2)(\hat{1})}\,|\,\Lambda_{(1)}\,^{(2)(\hat{1})}\,|\,
     E_{(0)}\,^{(2)(\hat{1})}\,) \,,
     \nonumber\\
\err
which has as its lowest component $\Phi_{ijkl}\,^{mn\,\hat{i}}$, which transforms under
$SU(2)_R\times SU(2)_H$ as $(\,{\bf 5}\otimes{\bf 3}\,)\times{\bf 2}$.

In the generalized case, the consistency of the reality constraints on the real scalar fields requires that
the total tensor rank of the lowest component be even.  Thus, multiplets such as ${\cal M}_4\,^{\hat{2}}$,
which has lowest component $\Phi_{ijk}\,^{\hat{\imath}}$ exists, even though ${\cal M}_4$ does not
exist by itself.  Multiplets of the sort described in the Appendix play a central role in the main text,
above.  In particular, we make use of the following named multiplets:
\brr  {\cal M}_3\,^{2,\hat{2}} &=& {\rm Extended\,\,Tensor\,\,Multiplet}
      \nonumber\\[.1in]
      {\cal M}_4\,^{\hat{2}} &=& {\rm Quadruplet\,\,Multiplet}
      \nonumber\\[.1in]
      {\cal M}_6\,^{\hat{2}} &=& {\rm Sextet\,\,Multiplet}
\err
along with the duals of the second two of these.  The first of these is called the Extended Tensor Multiplet
because this is obtained from the Tensor multiplet ${\cal M}_3$ by tensoring on the index structure
$(\cdot)^{i\hat{\imath}}$, thereby ``extending" the components.

\setcounter{equation}{0}
\section{Diagrammatics}
\label{diappen}
Diagrams helpfully depict the field content of supermultiplets.  For example, the
Extended Tensor multiplet may be depicted thusly,
\brr
\includegraphics[width=1.7in]{ETM.pdf} \,.
\label{etmdia}
\err
In such a diagram white circles represent bosonic fields and black circles represent fermionic fields.
The degrees of freedom described by a field are indicated by the numeral inside its corresponding
circle. In (\ref{etmdia}) the $SU(2)_R\times SU(2)_H$ representations are specified by the tensor indices on the field
names.  The height of a field on the diagram correlates with its scaling dimension:
adjacent levels are separated by one-half mass unit, and the field dimension increases moving upward
on the diagram.
All field dimensions are fixed by the dimension of the lowest level.
\footnote{For the diagram  (\ref{etmdia})
the lowest level, with fields $\phi_i\,^{\hat{\imath}}$ and $u_{ijk}\,^{\hat{\imath}}$,
has dimension one.   Thus, the second level, with
fields $\psi^{\hat{\imath}}$ and $\lambda_{ij}\,^{\hat{\imath}}$, has dimension three-halfs, and so forth.}
When two circles (fields) are connected by a dark line, this means that there are terms in the supersymmetry transformation
rules for each of those fields involving the other.  For example, the vertical black line connecting
$\phi_i\,^{\hat{\imath}}$ with $\xi^{\hat{\imath}}$ indicates a term or terms in $\delta_Q\,\phi_i\,^{\hat{\imath}}$
proportional to $\xi^{\hat{\imath}}$ and a term or terms in $\delta_Q\,\xi^{\hat{\imath}}$ proportional to
$\der_a\phi_i\,^{\hat{\imath}}$.

A double-circle, such as the one associated with the field $E_{a\,i}\,^{\hat{\imath}}$ in (\ref{etmdia})
indicates that the field is subject to a constraint.  In the simplest cases, if the field is a vector, such as $E_a$,
then a double-circle implies the field is divergence-free, $\der^a E_a=0$; if the field is an antisymmetric tensor,
then a double-circle implies a Bianchi identity, {\it e.g.} $\der_{[a}F_{bc]}=0$ .  (More generally, the constraints
involve covariant derivatives and/or terms involving background fields.)

In the context of $N$-extended {\it one-dimensional} supersymmetry, similar diagrams yield precise mathematical characterization
relevant to 1D (worldline) supersymmetry representation theory \cite{adinkras,DFGHIL_graph,doubly,frames,Toppan,Toppan_01,Toppan_02}.
The one-dimensional diagrams
are the Adinkras described in \ref{adinkrareview}.  The 4D analogs, such as (\ref{etmdia}) are not as mathematically
developed, although we refer to these as Adinkras too.  An Adinkra is a graph, and the circles and lines are, respectively,
its vertices and edges.

It is straightforward, using Lorentz covariance and other information
such as reality constraints, to determine the precise sorts of terms represented by the edges in
an Adinkra. The corresponding numerical coefficients are fixed by the supersymmetry algebra, and can
be worked out readily.  Such diagrams are useful for organizing
and arranging multiplets, for guiding thoughts about how these can couple to each other,
and for ascertaining whether mutiplets are reducible \cite{prepotentials}.

For some multiplets, is possible to choose a ``frame", by redefining fields, so that transformation rules
occur ``one-way", meaning that a field transforms into another field with a higher mass dimension, while
the higher--dimensional field does not transform ``back" into the lower-dimensional field.  Transformation
rules like this, which are not paired, are represented on the 4D Adinkras as grey edges rather than as black edges.
For example the multiplet $\widetilde{\cal M}_5\,^{\hat{1}}$ can be expressed as
\brr \includegraphics[width=3in]{DSDQy.pdf}
\err
When an Adinkra admits a frame in which otherwise disconnected sub-graphs are connected only by grey lines,
this means that the multiplet is reducible, such that the sub-graphs represent independent multiplets
in their own right.  The connections implied by grey lines represent a consistent way to ``tether"
multiplets together to form larger multiplets.  This also provides a means to identify gauge structures in
multiplets, as explained in \cite{frames}.

\setcounter{equation}{0}
\section{Transformation rules and action}
\label{hpac}
In this Appendix we exhibit the transformation rules for the global RETM and DSDQ
using the presentment associated with (\ref{retmdsdq}), in which the components
are manifest $SU(2)_A\times SU(2)_H$ tensors, but the isometry $SU(2)_B$ is
not manifest.  We also exhibit a supersymmetric action, and a field redefinition
(shift) which renders the action diagonal in terms of components.
This construction was described, as an epilog, in \cite{RETM}.  The presentation in this
Appendix includes details not shown in that paper, however.

The supersymmetry transformation rules for the RETM are
\brr \delta_Q\,\phi_i\,^{\hat{\imath}} &=&
      i\,\bar{\e}_i\,\psi^{\hat{\imath}}
      +i\,\ve^{mn}\,\bar{\e}_m\,
      \lambda_{ni}\,^{\hat{\imath}}
      +i\,\ve_{ij}\,\ve^{\hat{\imath}\hat{\jmath}}\,\bar{\e}^j\,\psi_{\hat{\jmath}}
      +i\,\bar{\e}^m\,\lambda^*_{im}\,^{\hat{\imath}}
      \nonumber\\[.1in]
      \delta_Q\,u_{ijk}\,^{\hat{\imath}} &=&
      i\,\bar{\e}_{(i}\,\lambda_{jk)}\,^{\hat{\imath}}
      -\fr{16}{15}\,i\,\bar{\e}^l\,\Sigma_{ijkl}\,^{\hat{\imath}}
      -i\,\ve_{m(i}\,\bar{\e}^{m}\,\lambda^*_{jk)}\,^{\hat{\imath}}
      +\fr{16}{15}\,i\,\ve^{mn}\,\bar{\e}_m\,\Sigma^*_{ijkn}\,^{\hat{\imath}}
      \nonumber\\[.1in]
      \delta_Q\,Z_{ijklm}\,^{\hat{\imath}} &=&
      i\,\ve_{n(i}\,\bar{\e}^n\,\Sigma_{jklm)}\,^{\hat{\imath}}
      +i\,\bar{\e}_{(i}\,\Sigma^*_{jklm)}\,^{\hat{\imath}}
      \nonumber\\[.1in]
      \delta_Q\,\psi^{\hat{\imath}} &=&
      \fr23\,\dslash\,\phi_i\,^{\hat{\imath}}\,\e^i
      +\fr12\,A_{a\,i}\,^{\hat{\imath}}\,\gamma^a\,\e^i
      +\fr12\,\ve^{mn}\,N_m\,^{\hat{\imath}}\,\e_n
      \nonumber\\[.1in]
      \delta_Q\,\lambda_{ij}\,^{\hat{\imath}} &=&
      \fr29\,\ve_{k(i}\,\dslash\,\phi_{j)}\,^{\hat{\imath}}\,\e^k
      +\fr56\,\dslash u_{ijk}\,^{\hat{\imath}}\,\e^k
      -\fr13\,N_{(i}\,^{\hat{\imath}}\,\e_{j)}
      -\fr23\,\ve^{mn}\,P^*_{ijm}\,^{\hat{\imath}}\,\e_n
      \nonumber\\[.1in]
      & &
      -\fr13\,\ve_{k(i}\,A_{a\,j)}\,^{\hat{\imath}}\,\gamma^a\,\e^k
      +\fr23\,K_{a\,ijm}\,^{\hat{\imath}}\,\gamma^a\,\e^m
      \nonumber\\[.1in]
      \delta_Q\,\Sigma_{ijkl}\,^{\hat{\imath}} &=&
      -\fr18\,\dslash u_{(ijk}\,^{\hat{\imath}}\,\e_{l)}
      -\ve^{mn}\,\dslash\,Z_{ijklm}\,^{\hat{\imath}}\,\e_n
      +\fr12\,\ve_{m(i}\,P_{jkl)}\,^{\hat{\imath}}\,\e^m
      +\fr12\,K_{a\,(ijk}\,^{\hat{\imath}}\,\gamma^a\,\e_{l)}
      \nonumber\\[.1in]
      \delta_Q\,N_i\,^{\hat{\imath}} &=&
      i\,\ve_{ij}\,\bar{\e}^j\,\dslash\,\psi^{\hat{\imath}}
      -2\,i\,\bar{\e}^j\,\dslash\,\lambda_{ij}\,^{\hat{\imath}}
      -\fr83\,i\,\bar{\e}^m\,\xi^*_{mi}\,^{\hat{\imath}}
       \nonumber\\[.1in]
      \delta_Q\,P_{ijk}\,^{\hat{\imath}} &=&
      \fr38\,i\,\bar{\e}_{(i}\,\dslash\lambda^*_{jk)}\,^{\hat{\imath}}
      +2\,i\,\ve^{mn}\,\bar{\e}_m\,\dslash\,\Sigma_{nijk}\,^{\hat{\imath}}
      +i\,\bar{\e}_{(i}\,\xi_{jk)}\,^{\hat{\imath}}
      \nonumber\\[.1in]
      \delta_Q\,A_{a\,i}\,^{\hat{\imath}} &=&
      -\fr16\,\,i\,\bar{\e}_i\,(\,3\,\dslash\gamma_a+2\,\der_a\,)\,\psi^{\hat{\imath}}
      +\fr13\,i\,\ve^{mn}\,\bar{\e}_m\,(\,3\,\dslash\gamma_a+2\,\der_a\,)\,\lambda_{ni}\,^{\hat{\imath}}
      +\fr43\,i\,\bar{\e}^m\,\gamma_a\,\xi_{mi}\,^{\hat{\imath}}
      \nonumber\\[.1in]
      & &
      -\fr16\,\,i\,\ve_{ij}\,\ve^{\hat{\imath}\hat{\jmath}}\,\bar{\e}^j\,(\,3\,\dslash\gamma_a+2\,\der_a\,)\,\psi_{\hat{\jmath}}
      +\fr13\,i\,\bar{\e}^m\,(\,3\,\dslash\gamma_a+2\,\der_a\,)\,\lambda^*_{mi}\,^{\hat{\imath}}
      -\fr43\,i\,\ve^{mn}\,\bar{\e}_m\,\gamma_a\,\xi^*_{ni}\,^{\hat{\imath}}
      \nonumber\\[.1in]
      \delta_Q\,K_{a\,ijk}\,^{\hat{\imath}} &=&
      \fr{1}{16}\,i\,\ve_{m(i}\,\bar{\e}^m\,(\,3\,\dslash\gamma_a+2\,\der_a\,)\,\lambda^*_{jk)}\,^{\hat{\imath}}
      -\fr13\,i\,\bar{\e}^m\,(\,3\,\dslash\gamma_a+2\,\der_a\,)\,\Sigma_{mijk}\,^{\hat{\imath}}
      -\fr12\,i\,\ve_{m(i}\,\bar{\e}^m\,\gamma_a\,\xi_{jk)}\,^{\hat{\imath}}
      \nonumber\\[.1in]
      & &
      -\fr{1}{16}\,i\,\bar{\e}_{(i}\,(\,3\,\dslash\gamma_a+2\,\der_a\,)\,\lambda_{jk)}\,^{\hat{\imath}}
      +\fr13\,i\,\ve^{mn}\,\bar{\e}_m\,
      (\,3\,\dslash\gamma_a+2\,\der_a\,)\,\Sigma^*_{nijk}\,^{\hat{\imath}}
      -\fr12\,i\,\bar{\e}_{(i}\,\gamma_a\,\xi^*_{jk)}\,^{\hat{\imath}}
      \nonumber\\[.1in]
      \delta_Q\,\xi_{ij}\,^{\hat{\imath}} &=&
      -\fr18\,\ve_{k(i}\,\dslash N^*_{j)}\,^{\hat{\imath}}\,\e^k
      +\fr12\,\dslash P_{ijm}\,^{\hat{\imath}}\,\e^m
      +\fr18\,(\,\dslash\gamma^a+4\,\der^a\,)\,A_{a\,(i}\,^{\hat{\imath}}\,\e_{j)}
      \nonumber\\[.1in]
      & &
      +\fr12\,\ve^{mn}\,(\,\dslash\gamma^a+4\,\der^a\,)\,K_{a\,ijm}\,^{\hat{\imath}}\,\e_n
       \,.
\label{retmrules}\err
The supersymmetry transformation rules for the DSDQ are
\brr \delta_Q\,H_i\,^{\hat{\imath}} &=&
      i\,\bar{\e}^j\,\Omega_{ij}\,^{\hat{\imath}}
      -i\,\ve^{mn}\,\bar{\e}_m\,\Omega^*_{ni}\,^{\hat{\imath}}
      -\fr32\,i\,\bar{\e}_i\,\beta^{\hat{\imath}}
      -\fr32\,i\,\ve_{ij}\,\ve^{\hat{\imath}\hat{\jmath}}\,\bar{\e}^j\,\beta_{\hat{\jmath}}
      \nonumber\\[.1in]
      \delta_Q\,\beta^{\hat{\imath}} &=&
      -\fr16\,\dslash H_i\,^{\hat{\imath}}\,\e^i
      +\fr12\,\ve^{ij}\,S_i\,^{\hat{\imath}}\,\e_j
      +\fr12\,U_{a\,i}\,^{\hat{\imath}}\,\gamma^a\,\e^i
      \nonumber\\[.1in]
      \delta_Q\,\Omega_{ij}\,^{\hat{\imath}} &=&
      \fr12\,\dslash H_{(i}\,^{\hat{\imath}}\,\e_{j)}
      +\fr12\,\ve^{mn}\,T_{a\,ijm}\,^{\hat{\imath}}\,\gamma^a\,\e_n
      +\fr12\,R_{ijm}\,^{\hat{\imath}}\,\e^m
      -\fr12\,\ve_{k(i}\,S^*_{j)}\,^{\hat{\imath}}\,\e^k
      +\fr12\,U_{a\,(i}\,^{\hat{\imath}}\,\gamma^a\,\e_{j)}
      \nonumber\\[.1in]
      \delta_Q\,S_i\,^{\hat{\imath}} &=&
      i\,\ve_{ij}\,\bar{\e}^j\,\dslash\beta^{\hat{\imath}}
      -\fr29\,i\,\bar{\e}^j\,\dslash\Omega^*_{ij}\,^{\hat{\imath}}
      +2\,i\,\bar{\e}^j\,\varphi_{ij}\,^{\hat{\imath}}
      \nonumber\\[.1in]
      \delta_Q\,R_{ijk}\,^{\hat{\imath}} &=&
      \fr43\,i\,\bar{\e}_{(i}\,\dslash\,\Omega_{jk)}\,^{\hat{\imath}}
      -\fr32\,i\,\bar{\e}_{(i}\,\varphi^*_{jk)}\,^{\hat{\imath}}
      +i\,\ve^{mn}\,\bar{\e}_m\,\eta_{nijk}\,^{\hat{\imath}}
      \nonumber\\[.1in]
      \delta_Q\,U_{a\,i}\,^{\hat{\imath}} &=&
      -\fr12\,i\,\bar{\e}_i\,(\,\dslash\gamma_a-\der_a\,)\,\beta^{\hat{\imath}}
      +\fr19\,i\,\ve^{mn}\,\bar{\e}_m\,(\,\dslash\gamma_a-\der_a\,)\,\Omega^*_{ni}\,^{\hat{\imath}}
      +i\,\ve^{jk}\,\bar{\e}_j\,\gamma_a\,\varphi_{ik}\,^{\hat{\imath}}
      \nonumber\\[.1in]
      & &
      -\fr12\,i\,\ve_{ij}\,\ve^{\hat{\imath}\hat{\jmath}}\,\bar{\e}^j\,(\,
      \dslash\gamma_a-\der_a\,)\,\beta_{\hat{\jmath}}
      -\fr19\,i\,\bar{\e}^m\,(\,\dslash\gamma_a-\der_a\,)\,\Omega^*_{mi}\,^{\hat{\imath}}
      +i\,\bar{\e}^m\,\gamma_a\,\varphi^*_{mi}\,^{\hat{\imath}}
      \nonumber\\[.1in]
      \delta_Q\,T_{a\,ijk}\,^{\hat{\imath}} &=&
      \fr23\,i\,\ve_{m(i}\,\bar{\e}^m(\,\dslash\gamma_a-\der_a\,)\,\Omega_{jk)}\,^{\hat{\imath}}
       +\fr34\,i\,\ve_{m(i}\,\bar{\e}^m\,\gamma_a\,\varphi^*_{jk)}\,^{\hat{\imath}}
      +\fr12\,i\,\bar{\e}^m\,\gamma_a\,\eta_{mijk}\,^{\hat{\imath}}
      \nonumber\\[.1in]
      & &
      +\fr23\,i\,\bar{\e}_{(i}\,(\,\dslash\gamma_a-\der_a\,)\,\Omega^*_{jk)}\,^{\hat{\imath}}
       -\fr34\,i\,\bar{\e}_{(i}\,\gamma_a\,\varphi_{jk)}\,^{\hat{\imath}}
      -\fr12\,i\,\ve^{mn}\,\bar{\e}_m\,\gamma_a\,\eta^*_{nijk}\,^{\hat{\imath}}
       \nonumber\\[.1in]
      \delta_Q\,\varphi_{ij}\,^{\hat{\imath}} &=&
      \fr49\,\dslash S_{(i}\,^{\hat{\imath}}\,\e_{j)}
      +\fr{1}{18}\,\ve^{mn}\,\dslash R^*_{ijm}\,^{\hat{\imath}}\,\e_n
      +\fr29\,\ve_{k(i}\,(\,2\,\dslash\gamma^a+3\,\der^a\,)\,U_{a\,j)}\,^{\hat{\imath}}\,\e^k
      \nonumber\\[.1in]
      & &
      -\fr{1}{36}\,(\,2\,\dslash\gamma^a+3\,\der^a\,)\,T_{a\,ijk}\,^{\hat{\imath}}\,\e^k
      +\fr12\,d_{ijk}\,^{\hat{\imath}}\,\e^k
      \nonumber\\[.1in]
      \delta_Q\,\eta_{ijkl}\,^{\hat{\imath}} &=&
      \ve_{m(i}\,\dslash R_{jkl)}\,^{\hat{\imath}}\,\e^m
      +\fr12\,(\,2\,\dslash\gamma^a+3\,\der^a\,)\,T_{a\,(ijk}\,^{\hat{\imath}}\,\e_{l)}
      +\fr12\,\ve^{mn}\,Y_{ijklm}\,^{\hat{\imath}}\,\e_n
      +\fr35\,d_{(ijk}\,^{\hat{\imath}}\,\e_{l)}
       \nonumber\\[.1in]
      \delta_Q\,d_{ijk}\,^{\hat{\imath}} &=&
      \fr{15}{8}\,i\,\bar{\e}_{(i}\,\dslash\,\varphi_{jk)}\,^{\hat{\imath}}
      +\fr{1}{12}\,i\,\bar{\e}^m\,\dslash\eta_{mijk}\,^{\hat{\imath}}
      -\fr{15}{8}\,i\,\ve_{l(i}\,\bar{\e}^l\,\dslash\,\varphi^*_{jk)}\,^{\hat{\imath}}
      -\fr{1}{12}\,i\,\ve^{mn}\,\bar{\e}_m\,\dslash\eta^*_{nijk}\,^{\hat{\imath}}
      \nonumber\\[.1in]
      \delta_Q\,Y_{ijklm}\,^{\hat{\imath}} &=&
      -2\,i\,\ve_{n(i}\,\bar{\e}^n\,\dslash\eta_{jklm)}\,^{\hat{\imath}}
      -2\,i\,\bar{\e}_{(i}\,\dslash\eta^*_{jklm)}\,^{\hat{\imath}} \,.
 \label{dsdqrules}\err
A supersymmetric action involving the components of these two multiplets is given by
 \brr S &=& \int d^4x\,\bpl\,
      \fr{5}{18}\,\phi_i\,^{\hat{\imath}}\,\Box\,\phi^i\,_{\hat{\imath}}
       +\fr23\,\phi^i\,_{\hat{\imath}}\,\der_a A^a_i\,^{\hat{\imath}}
       +\fr12\,A_{a\,i}\,^{\hat{\imath}}\,A^{a\,i}\,_{\hat{\imath}}
       \nonumber\\[.1in]
       & & \hspace{.6in}
      +\fr{5}{12}\,u_{ijk}\,^{\hat{\imath}}\,\Box\,u^{ijk}\,_{\hat{\imath}}
      +\fr{20}{3}\,u^{ijk}\,_{\hat{\imath}}\,\der_a K^a_{ijk}\,^{\hat{\imath}}
      -\fr83\,K_a^{ijk}\,_{\hat{\imath}}\,K_{a\,ijk}\,^{\hat{\imath}}
      \nonumber\\[.1in]
      & & \hspace{.6in}
      -\fr{16}{3}\,Z^{ijklm}\,_{\hat{\imath}}\,\Box\,Z_{ijklm}\,^{\hat{\imath}}
      -\fr12\,N_i\,^{\hat{\imath}}\,N^i\,_{\hat{\imath}}
      +\fr83\,P^{ijk}\,_{\hat{\imath}}\,P_{ijk}\,^{\hat{\imath}}
      \nonumber\\[.1in]
      & & \hspace{.6in}
      +i\,\bar{\psi}^{\hat{\imath}}\dslash \psi_{\hat{\imath}}
      +\fr32\,i\,\bar{\lambda}_{ij}\,^{\hat{\imath}}\,\dslash\lambda^{ij}\,_{\hat{\imath}}
      -\fr{32}{3}\,i\,\bar{\Sigma}^{ijkl}\,_{\hat{\imath}}\,\dslash\Sigma_{ijkl}\,^{\hat{\imath}}
      \nonumber\\[.1in]
      & & \hspace{.6in}
      +4\,i\,\bar{\lambda}^{ij}\,_{\hat{\imath}}\,\xi^*_{ij}\,^{\hat{\imath}}
      +4\,i\,\bar{\lambda}_{ij}\,^{\hat{\imath}}\,\xi^{*\,ij}\,_{\hat{\imath}}
      \nonumber\\[.1in]
      & & \hspace{.6in}
      +\fr34\,H_i\,^{\hat{\imath}}\,\der_a A^{a\,i}\,_{\hat{\imath}}
       -\fr12\,\phi_i\,^{\hat{\imath}}\,\der_a U^{a\,i}\,_{\hat{\imath}}
      -\fr{5}{16}\,u_{ijk}\,^{\hat{\imath}}\,\der_a T^{a\,ijk}\,_{\hat{\imath}}
       \nonumber\\[.1in]
      & & \hspace{.6in}
      +P_{ijk}\,^{\hat{\imath}}\,R^{ijk}\,_{\hat{\imath}}
      +P^{ijk}\,_{\hat{\imath}}\,R_{ijk}\,^{\hat{\imath}}
      +\fr38\,S^i\,_{\hat{\imath}}\,N_i\,^{\hat{\imath}}
      +\fr38\,S_i\,^{\hat{\imath}}\,N^i\,_{\hat{\imath}}
      \nonumber\\[.1in]
      & & \hspace{.6in}
      -\fr34\,U^{a\,i}\,_{\hat{\imath}}\,A_{a\,i}\,^{\hat{\imath}}
      -2\,K_a^{ijk}\,_{\hat{\imath}}\,T^a_{ijk}\,^{\hat{\imath}}
      -\fr98\,d^{ijk}\,_{\hat{\imath}}\,u_{ijk}\,^{\hat{\imath}}
      -Z_{ijklm}\,^{\hat{\imath}}\,Y^{ijklm}\,_{\hat{\imath}}
      \nonumber\\[.1in]
      & &\hspace{.6in}
      -\fr34\,i\,\bar{\beta}^{\hat{\imath}}\,\dslash\psi_{\hat{\imath}}
      -\fr34\,i\,\bar{\beta}_{\hat{\imath}}\,\dslash\psi^{\hat{\imath}}
      +\fr12\,i\,\bar{\lambda}^{ij}\,_{\hat{\imath}}\,\dslash\Omega^*_{ij}\,^{\hat{\imath}}
      +\fr12\,i\,\bar{\lambda}_{ij}\,^{\hat{\imath}}\,\dslash\Omega^{*\,ij}\,_{\hat{\imath}}
      \nonumber\\[.1in]
      & & \hspace{.6in}
      +\fr94\,i\,\bar{\varphi}^{ij}\,_{\hat{\imath}}\,\lambda_{ij}\,^{\hat{\imath}}
      +\fr94\,i\,\bar{\varphi}_{ij}\,^{\hat{\imath}}\,\lambda^{ij}\,_{\hat{\imath}}
      -2\,i\,\bar{\eta}^{ijkl}\,_{\hat{\imath}}\,\Sigma_{ijkl}\,^{\hat{\imath}}
      -2\,i\,\bar{\eta}_{ijkl}\,^{\hat{\imath}}\,\Sigma^{ijkl}\,_{\hat{\imath}}
      \nonumber\\[.1in]
      & & \hspace{.6in}
      -2\,i\,\bar{\xi}^{ij}\,_{\hat{\imath}}\,\Omega_{ij}\,^{\hat{\imath}}
      -2\,i\,\bar{\xi}_{ij}\,^{\hat{\imath}}\,\Omega^{ij}\,_{\hat{\imath}}
      \,\bpr\,.
 \label{retmdsdqaction}\err
 In fact, the first five lines and the final six lines of this action are independently supersymmetric.
 The first five lines represent a self-coupling of the RETM to itself, while the final six lines
 represent a coupling of the RETM to the DSDQ.  The latter includes the auxiliary field couplings
 needed to inoculate the non-hypermultiplet degrees of freedom.

 Starting with (\ref{retmdsdqaction}), there is a particular basis choice for which the propagating fields appear
 only squared, with no cross terms.  This is obtained from the above
 via
 \brr \phi_i\,^{\hat{\imath}} &\to&
      \phi_i\,^{\hat{\imath}}-\fr12\,H_i\,^{\hat{\imath}}
      \nonumber\\[.1in]
      A_{a\,i}\,^{\hat{\imath}} &\to&
      A_{a\,i}\,^{\hat{\imath}}
      +\fr23\,\der_a\phi_i\,^{\hat{\imath}}
      -\fr13\,\der_a H_i\,^{\hat{\imath}}
      \nonumber\\[.1in]
      U_{a\,i}\,^{\hat{\imath}} &\to&
      U_{a\,i}\,^{\hat{\imath}}
      +\fr23\,A_{a\,i}\,^{\hat{\imath}}
      -\der_a H_i\,^{\hat{\imath}}
      \nonumber\\[.1in]
      T_{a\,ijk}\,^{\hat{\imath}} &\to&
      T_{a\,ijk}\,^{\hat{\imath}}
      -\fr43\,K_{a\,ijk}\,^{\hat{\imath}}
      \nonumber\\[.1in]
      N_i\,^{\hat{\imath}} &\to&
      N_i\,^{\hat{\imath}}
      +\fr34\,S_i\,^{\hat{\imath}}
      \nonumber\\[.1in]
      P_{ijk}\,^{\hat{\imath}} &\to&
      P_{ijk}\,^{\hat{\imath}}
      -\fr38\,R_{ijk}\,^{\hat{\imath}}
      \nonumber\\[.1in]
      d_{ijk}\,^{\hat{\imath}} &\to&
      d_{ijk}\,^{\hat{\imath}}
      -\fr{5}{18}\,\der_a T_{a\,ijk}\,^{\hat{\imath}}
      +\fr{160}{27}\,\der^a K_{a\,ijk}\,^{\hat{\imath}}
      +\fr{10}{27}\,\Box\,u_{ijk}\,^{\hat{\imath}}
      \nonumber\\[.1in]
      Y_{ijklm}\,^{\hat{\imath}} &\to&
      Y_{ijklm}\,^{\hat{\imath}}
      -\fr{16}{3}\,\Box\,Z_{ijklm}\,^{\hat{\imath}}
      \nonumber\\[.1in]
      \psi^{\hat{\imath}} &\to&
      \psi^{\hat{\imath}}
      +\fr34\,\beta^{\hat{\imath}}
      \nonumber\\[.1in]
      \varphi_{ij}\,^{\hat{\imath}} &\to&
      \varphi_{ij}\,^{\hat{\imath}}
      -\fr{16}{9}\,\xi^*_{ij}\,^{\hat{\imath}}
      -\fr13\,\dslash\lambda_{ij}\,^{\hat{\imath}}
      \nonumber\\[.1in]
      \xi_{ij}\,^{\hat{\imath}} &\to&
      \xi_{ij}\,^{\hat{\imath}}
      -\fr14\,\dslash\lambda^*_{ij}\,^{\hat{\imath}}
      \nonumber\\[.1in]
      \eta_{ijkl}\,^{\hat{\imath}} &\to&
      \eta_{ijkl}\,^{\hat{\imath}}
      -\fr{8}{3}\,\dslash\Sigma_{ijkl}\,^{\hat{\imath}}
 \err
 After performing these shifts, the action becomes
  \brr S &=& \int d^4x\,\bpl
      \fr12\,\phi_i\,^{\hat{\imath}}\,\Box\,\phi^i\,_{\hat{\imath}}
      -\fr18\,H_i\,^{\hat{\imath}}\,\Box\,H^i\,_{\hat{\imath}}
      +i\,\bar{\psi}^{\hat{\imath}}\dslash \psi_{\hat{\imath}}
      -\fr{9}{16}\,i\,\bar{\beta}^{\hat{\imath}}\,\dslash\beta_{\hat{\imath}}
      \nonumber\\[.1in]
      & &
      \hspace{.6in}
      -\fr34\,A_{a\,i}\,^{\hat{\imath}}\,U^{a\,i}\,_{\hat{\imath}}
       -2\,K_a^{ijk}\,_{\hat{\imath}}\,T^a_{ijk}\,^{\hat{\imath}}
       -\fr98\,d^{ijk}\,_{\hat{\imath}}\,u_{ijk}\,^{\hat{\imath}}
       -Z_{ijklm}\,^{\hat{\imath}}\,Y^{ijklm}\,_{\hat{\imath}}
      \nonumber\\[.1in]
      & & \hspace{.6in}
      -\fr12\,N_i\,^{\hat{\imath}}\,N^i\,_{\hat{\imath}}
      +\fr{9}{32}\,S^i\,_{\hat{\imath}}\,S_i\,^{\hat{\imath}}
      +\fr83\,P_{ijk}\,^{\hat{\imath}}\,P^{ijk}\,_{\hat{\imath}}
      -\fr32\,R^{ijk}\,_{\hat{\imath}}\,R_{ijk}\,^{\hat{\imath}}
      \nonumber\\[.1in]
      & & \hspace{.6in}
       +\fr94\,i\,\bar{\varphi}^{ij}\,_{\hat{\imath}}\,\lambda_{ij}\,^{\hat{\imath}}
      +\fr94\,i\,\bar{\varphi}_{ij}\,^{\hat{\imath}}\,\lambda^{ij}\,_{\hat{\imath}}
      -2\,i\,\bar{\eta}^{ijkl}\,_{\hat{\imath}}\,\Sigma_{ijkl}\,^{\hat{\imath}}
      -2\,i\,\bar{\eta}_{ijkl}\,^{\hat{\imath}}\,\Sigma^{ijkl}\,_{\hat{\imath}}
       \nonumber\\[.1in]
      & & \hspace{.6in}
      -2\,i\,\bar{\xi}^{ij}\,_{\hat{\imath}}\,\Omega_{ij}\,^{\hat{\imath}}
      -2\,i\,\bar{\xi}_{ij}\,^{\hat{\imath}}\,\Omega^{ij}\,_{\hat{\imath}}
       \,\bpr\,.
 \label{otmfinal}\err
 In fact, one could obtain exactly this same action by starting with only the final six lines of
 (\ref{retmdsdqaction}) and performing a different field redefinition.  Thus, in the limited context
 of this presentation, the first five lines of (\ref{retmdsdq}) are superfluous.  We exhibit them here, however because
 these exhibit an explicit RETM self-coupled action, and also exhibits that the DSDQ is dual to the
 RETM, in the sense that these two can be quadratically coupled supersymmetrically.

 In section \ref{hidis} we explain how these constructions can be re-packaged in a more concise and
 ``natural" manner in which the hidden $SU(2)_B$ isometry is made manifest.

 \setcounter{equation}{0}
\section{The ${\cal N}=2$ Supergravity Algebra}
\label{n2algebra}
In this Appendix we review superficial basics of
${\cal N}=2$ supergravity, to the minimal extent relevant to the discussion in the main text.

The gauge fields of ${\cal N}=2$ conformal supergravity are found in the ${\cal N}=2$ Weyl multiplet,
expressed diagrammatically by
\footnote{The degrees of freedom indicated in the vertices of (\ref{weyl2ad}) reflect inherent gauge
structures.  For example, the vierbein $e_\mu^a$ has $4\times 4=16$ dofs minus eleven dofs
associated with coordinate transformations, Lorentz transformations, and dilatations.  Similarly, the
dilatational gauge field $b_\mu$ is itself pure gauge under special conformal transformations, and therefore
describes no physical degrees of freedom.}
\brr \includegraphics[width=2.5in]{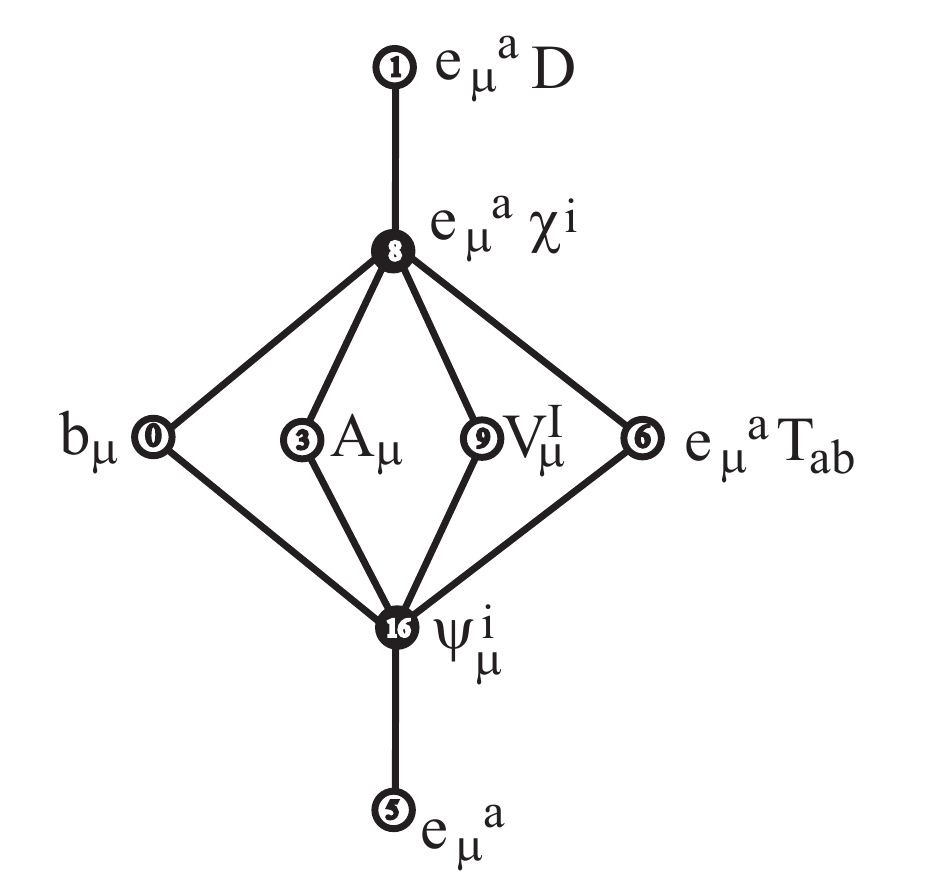}
\label{weyl2ad}\err
The vierbien $e_\mu\,^a$ and the gravitino field $\psi_\mu^i$ have scaling dimensions
minus one and minus one-half, respectively. The latter of these is the gauge field for $Q$ supersymmetry.  The
gauge fields for dilatations, $U(1)_A$ transformations,
and $SU(2)_A$ transformations are supplied, respectively, by $(\,b_\mu\,,\,A_\mu\,,\,V_\mu^I\,)$.
The remaining fields $(\,T_{ab}\,,\,\chi^i\,,\,D\,)$
are ``supergravity matter" fields needed to close the supersymmetry algebra off shell.
These appear multiplied or contracted with the vierbien in (\ref{weyl2ad}) to render the
field dimensions commensurate with the diagrammatic presentation.  For example, the two-form $T_{ab}$ has
dimension one, so that $e_\mu\,^a\,T_{ab}$ has dimension zero.

The remaining superconformal generators comprise local Lorentz transformations, $S$-supersymmetry transformations,
and special conformal transformations.  The corresponding respective gauge fields
$(\,\omega_\mu^{ab}\,,\,\phi_\mu^i\,,\,f_\mu^a\,)$ are composites built from
the component fields in (\ref{weyl2ad}).

The Weyl multiplet represents a distorted version of the ${\cal N}=2$ superconformal algebra organized to include
covariant general coordinate transformations as a subset of the generators.  The commutator of two $Q$ supersymmetry transformations
is
\brr [\,\delta_Q(\e)\,,\,\delta_Q(\e_2)\,] &=&
     \delta^{({\rm cov})}(\xi)
     +\delta_M(\theta)
     +\delta_K(\Lambda_K)
     +\delta_S(\eta)
\label{sg1}\err
where $\delta^{({\rm cov})}$ is a covariant coordinate transformation and $M$, $K$, and $S$ describe
a Lorentz transformation, a Special Conformal transformation, and an $S$ supersymmetry transformation,
respectively. The parameters of the transformations on the right-hand side of (\ref{sg1}) are
given by
\brr \xi^\mu &=& 2\,i\,\bar{\e}_{[2}^i\,\gamma^\mu\,\e_{1]\,i}
     \nonumber\\[.1in]
     \theta^{ab} &=& 4\,i\,\ve_{ij}\,\bar{\e}_2^i\,\e_1^j\,T^{(-)\,ab}+{\rm h.c.}
     \nonumber\\[.1in]
     \Lambda_K^a &=&
     \bpl\,\fr32\,i\,\ve^{ij}\,\bar{\e}_{2\,i}\,\e_{1\,j}\,D_b\,T^{(+)\,ab}+{\rm h.c.}\,\bpr
     -\fr12\,i\,\e_{[2}^i\,\gamma^a\,\e_{1]\,i}\,{\rm D}
     \nonumber\\[.1in]
     \eta^i &=& 2\,i\,\bar{\e}_{[2}^i\,\e_{1]}^j\,\chi_j \,.
\label{pa2}\err
The commutator of an $S$ supersymmetry transformation and a $Q$ supersymmetry transformation is
\brr [\,\delta_S(\eta)\,,\,\delta_Q(\e)\,] &=&
     \delta_M(\hat{\theta})+\delta_D(\Lambda_D)+\delta_{\cal A}(\alpha)
     +\delta_G(\Theta)
\label{sg1sq}\err
where $D$ is a dilatation, ${\cal A}$ is a $U(1)_R$ transformation, and $G$ is an $SU(2)_{\cal R}\cong SU(2)_A$
transformation.   The parameters are
\brr \hat{\theta}^{ab} &=&
     i\,\bar{\e}^i\,\gamma^{ab}\,\eta_i
     +i\,\bar{\e}_i\,\gamma^{ab}\,\eta^i
     \nonumber\\[.1in]
     \Lambda_D &=&
     i\,\bar{\e}^i\,\eta_i
     +i\,\bar{\e}_i\,\eta^i
     \nonumber\\[.1in]
     \alpha &=&
     \bar{\eta}^i\,\e_i
     -\bar{\eta}_i\,\e^i
     \nonumber\\[.1in]
     \Theta^I &=&
     4\,(\,T^I\,)_i\,^j\,(\,
     i\,\bar{\eta}^i\,\e_j
     -i\,\bar{\eta}_j\,\e^i\,) \,.
\label{pa2SQ}\err
The hat is placed on the parameter $\hat{\theta}^{ab}$ merely to distinguish this from the parameter of the Lorentz
transformation appearing in the $\delta_Q^2$ commutator (\ref{sg1sq})

A special conformal transformation is determined by the commutator of two $S$ supersymmetry transformations,
$[\,\delta_S(\eta_1)\,,\,\delta_S(\eta_2)\,]=\delta_K(\Lambda_K)$,
where
$\Lambda_K^a=2\,i\,\bar{\eta}_{[1}^i\,\gamma^a\,\eta_{2]\,i}$
is the parameter.

It is noteworthy that the $N=2$ supergravity algebra has non-trivial structure functions, appearing in (\ref{pa2}),
which depend on the matter fields $(\,T_{ab}\,,\,\chi^i\,,\,D\,)$.

\setcounter{equation}{0}
\section{Some $SU(2)$ Identities}
\label{su2rules}
The anti-hermitian $SU(2)$ generators $T_I$ satisfy the algebra
$[\,T_I\,,\,T_J\,]=\ve_{IJ}\,^K\,T_K$. In the fundamental ${\bf 2}$ representation
these are $(\,T_I\,)_i\,^j=-\fr12\,i\,(\,\sigma_I\,)_i\,^j$, where $\sigma_{1,2,3}$ are
the Pauli matrices.  The group acts on a vector $\phi_i$ as
\brr \phi_i &\to& (\,e^{-\theta^I\,T_I}\,)_i\,^j\,\phi_j \,.
\err
Accordingly, $\delta\,\phi_i=-\theta^I\,(\,T_I\,)_i\,^j$. The complex conjugate $\phi^i$
transforms as
\brr \phi^i &\to& \phi^j\,(\,e^{\theta^I\,T_I}\,)_j\,^i \,.
\err
so that $\delta\,\phi^i=\theta^I\,\phi^j\,(\,T_I\,)_j\,^i$.
Two infinitesimal
transformations obey the following commutation relationship,
\brr [\,\delta(\theta_1)\,,\,\delta(\theta_2)\,] &=& \delta(\tilde{\theta})
\err
where $\tilde{\theta}^I=\ve^I\,_{JK}\,\theta_1^J\,\theta_2^K$.  A rank $k$ tensor
with down indices transforms as
\brr \delta\,\Phi_{i_1\cdots i_k} &=& -k\,\Theta^I\,(\,T_I\,)_{(i_1}\,^j\,\Phi_{i_2\cdots i_k)j}
\err
and a rank $k$ tensor with up indices transforms as
\brr \delta\,\Phi^{i_1\cdots i_k} &=& k\,\Theta^I\,(\,T_I\,)_j\,^{(i_1}\,\Phi^{i_2\cdots i_k)j}
\err
Tensors with mixed indices involve a combination of these two rules.

The fundamental generators satisfy the following relationships
\brr [\,T_I\,,\,T_J\,]_i\,^j &=& \ve_{IJ}\,^K\,(\,T_K\,)_i\,^j
     \nonumber\\[.1in]
     \{\,T_I\,,\,T_J\,\}_i\,^j &=& -\fr12\,\delta_{IJ}\,\delta_i\,^j
     \nonumber\\[.1in]
     (\,T^I\,)_i\,^k\,(\,T_I\,)_j\,^l &=&
     \fr14\,(\,\delta_i\,^k\,\delta_j\,^l
     -2\,\delta_i\,^l\,\delta_j\,^k\,)
     \nonumber\\[.1in]
     (\,T_{[I}\,)_i\,^j\,(\,T_{J]}\,)_m\,^n
     &=& \fr14\,\ve_{IJ}\,^K\,\bpl\,
     \delta_m\,^j\,(\,T_K\,)_i\,^n
     -\delta_i\,^n\,(\,T_K\,)_m\,^j\,\bpr
     \nonumber\\[.1in]
     {\rm Tr}\,(\,T_I\,T_J\,) &=& -\fr12\,\delta_{IJ} \,.
\label{su2ids}\err
These prove useful in deriving many results in the main text.

\pagebreak

\noindent
{\bf\large Acknowledgements}\\[.1in]
\noindent
The global Hyperplet model, reviewed in Section 4 of this paper,
was created collaboratively with Jim Gates, who provided useful insight and helpful
suggestions. Several mathematical quandaries, such as the elucidation of the hidden
representation structures described in Section 5, were worked out with invaluable help
from Kevin Iga, Greg Landweber, and Chuck Doran. Much of this work was done at
the Slovak Institute for Basic Research, in Podvazie, Slovakia.

\end{document}